\documentclass[%
 reprint,
 amsmath,amssymb,
pra,
]{revtex4-1}

\usepackage{graphicx}
\usepackage{dcolumn}
\usepackage{bm}
\usepackage{subfigure}
\usepackage{physics}
\usepackage[space]{grffile}
\usepackage{color}
\usepackage{empheq}
\usepackage{multirow}

\begin{document}


\title{Long time dynamics of a single particle extended quantum walk on a one dimensional lattice with complex hoppings: a generalized hydrodynamic description}

\author{Hemlata Bhandari}

\author{P. Durganandini}

\affiliation{%
Department of Physics, Savitribai Phule Pune University, Pune 411007, India\\
}%

\date{\today}

\begin{abstract}
We study the continuous time quantum walk of a single particle (initially localized at a single site) on a one-dimensional spatial lattice with complex nearest neighbour and next-nearest neighbour hopping amplitudes. Complex couplings lead to chiral propagation and a causal cone structure asymmetric about the origin. We provide a hydrodynamic description for quantum walk dynamics in large space time limit. We find a global "quasi-stationary state" which can be described in terms of the local quasi-particle densities satisfying Euler type of hydrodynamic equation and is characterized by an infinite set of conservation laws satisfied by scaled cumulative position moments. Further, we show that there is anomalous sub-diffusive scaling near the extremal fronts, which can be described by higher order hydrodynamic equations. The long time behaviour for any complex next-nearest neighbour hopping with a non-zero real component is similar to that of purely real hopping (apart from asymmetric distribution). There is a critical coupling strength at which there is a Lifshitz transition where the topology of the causal structure changes from a regime with one causal cone to a regime with two nested causal cones. On the other hand, for purely imaginary next-nearest neighbour hopping, there is a transition from one causal cone to a regime with two partially overlapping cones due to the existence of degenerate maximal fronts (moving with the same maximal velocity). The nature of the Lifshitz transition and the scaling behaviour (both) at the critical coupling strength is different in the two cases.
\end{abstract}

\maketitle

\section{\label{sec:level1}Introduction}
 Quantum notions of superposition, interference and coherence often  lead  to very different behaviour for a quantum walk as compared to that for a classical random walk. Quantum walks have been the focus of intense study in recent years due to their relevance in a wide variety of fields like condensed matter physics, quantum optics, astronomy, quantum information and computer science,  mathematics, biology, etc \citep{Aharonov,Kempe,VAndraca}. Various physical processes like quantum transport \citep{longhi,mulken}, Anderson localization \citep{QW_Anderson_localization} and topological phases \citep{QW_Topological_phases},etc,  have been modelled using quantum walks.  They have been used to build quantum algorithms~\citep{shor, Grover,Childs_Conf_Proc_2003, Childs_PRA.70.022314}. Experimentally, they have been realized using photons in waveguides \cite{QW_photons}, trapped ions \citep{QW_trapped_ion, PRA.65.032310,PRL.103.090504}, atoms in optical lattices \citep{QW_optical_lattice, QW_trapped_atom, QW_BEC}, etc.  

                   There has been particular interest in the study of continuous time quantum walks~\citep{Kempe, mulken, QW_cycle, QW_Krapivsky1, QW_Krapivsky2, Hemlata} because they provide  alternate ways to model and study many body lattice Hamiltonians. They have been shown to exhibit ballistic propagation instead of the diffusive behaviour expected of a classical random walk ~\citep{cuevas, Toro, QW_Krapivsky1, QW_Krapivsky2, Hemlata, schoenhammer}. In recent work~\citep{Hemlata}, we showed that interesting 'causal light-cone structures' appear in a single particle continuous time quantum walk with a finite range of hopping. In the bulk, the walk exhibits ballistic propagation of wave fronts. The wave front propagation is bounded by a maximal 'light-cone group velocity'; propagation of excitations with a velocity greater than the 'light-cone velocity' are suppressed. The wave fronts were characterized as ordinary or extremal; the latter are characterized at long times by a  $1/t^{2/k+2}$ probability scaling where $k~(k =1,2, 3, \cdots)$ is the order of front.  In a walk with nearest and next- nearest hopping, we showed that there is a transition from a regime with one causal cone to a regime with two nested causal cones. Further, we showed that the local probability densities exhibit anomalous subdiffusive scaling near extremal fronts and the nature of the scaling depends on the order of the front \citep{Hemlata}. We also connected the study to that of spin-chains where the existence of  such upper bounds on the spread  or Lieb-Robinson (LR) bounds for the speed with which information propagates has been earlier studied ~\citep{lieb-robinson, bravyi2006, cardy2006}. The analysis in the above work was however restricted to that of real hopping amplitudes.  In this work, we generalize the study to address the problem of the long time dynamics of a single particle walk on a one-dimensional spatial lattice with complex hopping amplitudes. It is of considerable interest to extend the study to the case of complex hopping amplitudes, since it is known that complex hopping strengths break time-reversal symmetry and can lead to novel effects like chiral propagation \citep{Chiral_QWs, TRS_quantum_transport}. The study of chiral walks is also of interest because they allow for controlled information transfer in quantum systems and directional biasing without a biased initial state \citep{Chiral_QWs, TRS_quantum_transport}.  Effect of time reversal symmetry breaking has been demonstrated in case of triangular chains showing enhancement of quantum transport whereas in case of loops there is complete suppression \citep{Chiral_QWs, TRS_quantum_transport}. Chiral quantum walks have also been considered to study exciton transport in naturally occurring several light harvesting complexes like Fenna-Matthew-Olsen complex (FMO) \citep{TRS_quantum_transport}. In the context of tight binding models employed for many body systems, it is well known that complex hopping amplitudes naturally occur in the presence of magnetic fields.

                   In this paper, we analytically study the problem using the stationary phase approximation and compare also with results obtained from exact numerical calculations. We show that for an initial localized state at the origin, a complex next-nearest neighbor (NNN) hopping leads to chiral propagation of the wave fronts. We show the existence of Lieb-Robinson bounds on the maximal velocities with which the wave fronts can propagate. Due to the chiral nature of the propagation, the left moving and right moving maximal velocities are different resulting in a causal region asymmetric about the origin. The skewness of the probability distribution which is a measure of the asymmetry in the distribution is zero for real hopping strength \citep{QW_Krapivsky2, Hemlata} and maximal for a purely imaginary NNN hopping.  At a certain critical strength of NNN hopping, the value of which depends on both the magnitude and phase of the NNN hopping, there is a transition from a regime with one causal cone to two causal cones. Such a transition  where the topology of the causal cone structure changes from one to two can be considered as a Lifshitz transition~\citep{Lifshitz60} in analogy with the usual definition of a Lifsthitz transition as a transition across which the topology of the Fermi surface changes. However, due to the breaking of inversion symmetry, the causal cones are no longer symmetric about the origin. We also find an interesting departure of the behaviour of the extremal fronts for a purely imaginary NNN hopping as compared to that for a real NNN hopping. For a purely imaginary NNN hopping, in the regime with two causal cones, there are three maximal fronts, two of which move with the same maximal velocity. This gives rise to partially overlapping causal cones. Even a small real component of NNN hopping breaks this degeneracy and one obtains asymmetric but completely nested cones. A symmetrically located nested cone structure is obtained only for  purely real NNN hopping.  We also find very different behaviour at the critical coupling for purely imaginary NNN hopping as compared to that for a purely real NNN hopping.  Exactly, at the critical coupling, for purely imaginary NNN hopping, the phase is characterised by a single causal cone with the two maximal fronts being of different order; one is a first order front while the other is a third order front. 
 We provide, at asymptotically long times and distances, a hydrodynamical description of the quantum walk dynamics. We show the existence of a  global "quasi-stationary state" which can be described in terms of the local density of quasi-particle excitations satisfying Euler type of hydrodynamic equations. The global quasi-stationary state is characterized by an infinite set of conservation laws satisfied by scaled cumulative position moments. Furthermore, we show that there is anomalous scaling behaviour in the vicinity of the extremal fronts which can be described in terms of higher order hydrodynamic equations. A generalized hydrodynamic framework with infinitely many conservation laws for quasi-particles has been used to study the non-equilibrium dynamics in integrable systems ~\citep{Fagotti_prl, Doyon_PhysRevX.6.041065,Fagotti, DOYON_2018, Agrawal_PhysRevB.99.174203}. For the model at hand, in the regimes $g<g_c$ and $g>g_c$, all extremal fronts are first order in nature and show a sub-diffusive $t^{1/3}$  Airy scaling and staircase structure, the two-fold multiplicity of the front (for $g>g_c$ ) is reflected in the area under the steps of the staircase or the quantization.  Exactly at the critical coupling $g=g_c$,  the different orders of the extremal fronts leads to different sub-diffusive $ t^{1/3}$ and $t^{1/5}$ scaling near the two front edges.  A local staircase structure and quantization is observed near both edges. This is in contrast to the case with real NNN hopping where at the critical coupling, there are three extremal fronts, two of which are first order and the other which is an internal front is a second order front.  At the second order front there is no local staircase structure~\citep{Hemlata}. Thus the nature of the Lifshitz transition is different in the two cases.  Similar to the case of real NNN hopping~\citep{Hemlata},  where we connected the  long time dynamics of a single particle (initially localized at the origin) quantum walk problem with the long time dynamics of domain wall propagation in spin chains ~\citep{antal_pre59, sasvari_pre69},  we suggest that the present study can be connected to the time evolution of a domain wall in a spin chain model with complex NNN spin hopping \citep{suzuki,pradeep2}.  
 
                The plan of the paper is as follows. In Sec \ref{sec:model} we introduce the continuous time quantum walk model on a one dimensional spatial  lattice with complex nearest neighbour(NN) and next-nearest neighbour(NNN) hopping amplitudes. We show that a complex NNN hopping amplitude leads to chiral propagation of wave fronts with unequal maximal left and right moving velocities. The consequent asymmetric causal structures, local probability and current density distributions are described for different NNN coupling strengths. In Sec.\ref{sec:Hydrodynamic}, we obtain, at asymptotically long times and distances, a hydrodynamic description of the quantum walk dynamics. Specifically, we describe the bulk scaling behaviour of the cumulative probability distribution and cumulative current density distribution and their dependence on the density of excitations using stationary phase approximation. We show that exact  numerical computations of the long time behaviour agree with the analytical results. Further, we show that higher order cumulative position moments satisfy  global scaling relations and obtain the conservation laws satisfied by them. In Sec. \ref{sec:local}, we discuss the nature of propagation near the extremal front edges at asymptotically large times and distances. We show the emergence of a local scaling behaviour which can be described by higher order hydrodynamic equations. The analytic results are compared with exact numerical results. Finally, we summarize our results in Sec \ref{sec:conclusions}. 

\section{\label{sec:model}Model}
The Hamiltonian for the continuous time quantum walk of a single particle on a one-dimensional spatial lattice with complex hopping amplitudes between nearest and next-nearest neighbour sites can be written in the second quantized form as:  
\begin{eqnarray}\label{eq:Hamiltonian}
H && = g_{1}\sum_{n=-N}^{N}(e^{i \phi_1} c_{n+1}^{\dagger}(t)c_n(t)+e^{-i\phi_1} c_{n}^{\dagger}(t)c_{n+1}(t))\nonumber\\
&& + g_{2} \sum_{n=-N}^{N}(e^{i \phi_2} c_{n+2}^{\dagger}(t)c_n(t)+e^{-i\phi_2} c_{n}^{\dagger}(t)c_{n+2}(t))
\end{eqnarray}
where $c_n(c^{\dagger}_n)$ denote the annihilation(creation) operators. $g_1$ and $g_2$ denote the magnitude of the NN and NNN hopping strengths while $\phi_1$ and $\phi_2$ denote the corresponding phases. We assume the lattice spacing to be $a$. 
We measure energy in units of the nearest neighbour coupling strength $g_1$ ($\hbar$ has been set to $1$). Time is measured in units of $1/g_1$,  length is measured in units of the lattice spacing $a$ and velocity is measured in units of $ag_1$.  We also define a dimensionless ratio $g=g_2/g_1$ and set without any loss of generality, $g_1=1$.  We restrict in the following to the case of a real NN hopping and complex NNN hopping. There is no loss of generality by doing this since the phase $\phi_1$ in Eq.~\ref{eq:Hamiltonian} can be eliminated by a gauge transformation, $c_n \rightarrow e^{i  n \phi_1} c_n$ and redefining $\phi_2$ as $\phi_2 -2 \phi_1$. We therefore set $\phi_1=0$ and $\phi_2 =\phi; \,\,\, 0 \leq \phi <2 \pi$  hereafter. The single particle wave function $\psi(n,t)$ at the $n$-th site and at time $t$ is obtained from the field operator $\Psi(t) = \sum_n c_n \mid n \rangle$ as $\psi(n,t) = \langle n \mid \Psi (t)\rangle$. Here $|n\rangle$ denotes the single particle position space eigen-basis vector.  The Heisenberg equation of motion for the single particle position-space wave function is obtained from Eq.~\ref{eq:Hamiltonian} as: 
\begin{eqnarray}
i\frac{\partial}{\partial t}{\psi(n,t)} = &&  [\psi(n,t),H] \nonumber \\
= &&  \,\, [\psi(n+1,t)+\psi(n-1,t)] \nonumber\\
&& +g[\psi(n+2,t)e^{i\phi}+\psi(n-2,t)e^{-i\phi}]
\end{eqnarray}
The  probability current density $j(n,t)$ can be similarly obtained from Heisenberg's equation of motion for the 
probability density  $p(n,t) = |\psi(n,t)|^2$, $i\frac{\partial }{\partial t} {p(n,t)}=[p(n,t),H]$~\cite{gd_mahan, current}  as:
\begin{eqnarray}\label{eq:prob_current}
j(n,t) &&=i[\psi^{*}(n-1,t)\psi(n,t)-\psi ^{*}(n,t)\psi(n-1,t)]\nonumber \\
   && + 2 i \, g \,[e^{i \phi}\psi^{*}(n-2,t)\psi(n,t)-e^{-i \phi}\psi^{*}(n,t)\psi(n-2,t)]\nonumber \\
&&    
\end{eqnarray}  
The first term in the above equation is the current at site $n$ due to  NN hopping while the second term gives the current due to NNN hopping.\\
		 The wave function at site $n$ and  time $t$ can be obtained by Fourier transforming to momentum space: 
\begin{equation}\label{eq:wavefunction_sum}
\psi(n,t) = \frac{1}{L}\sum_q  e^{i( nq- \omega (q)t)}\hat \psi(q,0);\qquad -\pi \leq q < \pi
\end{equation} 
The momentum space eigenfunctions are plane waves: $\psi_q(n) = e^{i q n}$ where $q$ is the wave vector measured in units of $1/a$ and the Fourier sum is performed over all wave-vectors $q$ lying in the first Brillouin zone,  $L=2Na$ and $\hat \psi(q,t=0)$ denotes the initial wave function in momentum space at time $t=0$.  The  single particle energies $\omega (q)$ obey  the dispersion relation:
\begin{equation}\label{eq:dispersion}
\omega (q, g, \phi)=2\cos q+ 2g\cos (2q+\phi)
\end{equation}  
While for real hopping amplitudes, $\omega(-q) =\omega(q)$,  for complex hopping amplitudes, the reflection symmetry of the dispersion relation is no longer present and, in general, $\omega(-q) \neq \omega(q)$.
From Eq.~\ref{eq:dispersion}, we can see that the dispersion relation satisfies the identities:
\begin{equation}
\omega(\pm q, g, \pi \pm \phi) = \omega (q, -g, \phi) 
\end{equation}
or in other words the analysis for the case $ \pi/2 < \phi \leq 2\pi$ can be obtained from $ 0 < \phi \leq \pi/2$ by appropriate transformation of $g$ and $q$ in the dispersion relation (Eq. \ref{eq:dispersion}) and subsequent analysis.
Hence,  it is sufficient to restrict $\phi$ to the range $ 0 \leq \phi \leq \pi/2$.
 In the limit of an infinite site lattice, the summation over the wave-vectors $q$ in Eq.~\ref{eq:wavefunction_sum} can be converted into an integral and the wave function at the $n$-th site is obtained as:
\begin{equation}\label{eq:fourier_integral}
\psi(n,t) = \int\limits_{-\pi}^{\pi} \frac{dq}{2 \pi} e^{i( nq- \omega(q)t)}\hat \psi(q,0)
\end{equation}
 In the rest of the work, we consider that the  particle is localized at the origin at time $t=0$, i.e., the wave-function of the particle is $\psi_n(0)=\delta_{n,0}/\sqrt L$ ($\hat \psi(q,0) =\sqrt L$).  For such an initial state $\psi_n(0)$ we have then in the infinite site lattice limit,  
\begin{equation}\label{eq:fourier_integral_1}
\psi_n(t) = \int\limits_{-\pi}^{\pi} \frac{dq}{2 \pi} e^{i( nq- w(q)t)}
\end{equation}

		 Although the Fourier integral in Eq. \ref{eq:fourier_integral_1} cannot, in general, be evaluated exactly, 
we can examine the long-time behaviour of the wave-function by evaluating the integral using  stationary phase  approximation~\citep{QW_Krapivsky2, Hemlata}.  To this end, the integral is expressed as:
   \begin{equation}\label{eq:saddle-point}
\psi(n,t) = \int\limits_{-\pi} ^{\pi} \frac{dq}{2 \pi} e^{i\varphi(n ,q) t}; \quad  \varphi(n,q)\equiv\frac{n}{t}q-\omega (q);
\end{equation}
 The dominant contribution to the above integral  comes, within the stationary phase approximation, from a small region of $q$ around the saddle point solutions $q^*$ satisfying the equation:
\begin{equation} \label{eq:saddle_pt_equation}
\frac{\partial}{\partial q} \varphi(n, q) = 0 \implies \omega '(q_n^*) =v(q_n^*)= \frac{n}{t} 
\end{equation}
The integral in Eq.~\ref{eq:saddle-point} is then performed by expanding  $\varphi(n,q)$ around the saddle point solutions $q_n^*$ and summing over all $q_n^*$.  It can be seen from Eq. (\ref{eq:saddle_pt_equation}) that the saddle point solutions describe ballistic propagation of wave fronts travelling with group velocity $v(q)= \omega '(q)$.  The group velocities $v(q)$ are bounded: $v_{lm}(q) \leq v(q) \leq v_{rm}(q)$; with  the maximal left and right moving velocities determined as: 
  \begin{equation}
 \quad v_{rm}(q) = \mbox{max}(\omega'(q)); \qquad  v_{lm}(q) = \mbox{min}(\omega'(q)) . 
  \end{equation} 
 The solutions $q_n^*$ of the saddle point equation\,(Eq.~\ref{eq:saddle_pt_equation}) are real for $n_{lm}(=v_{lm} t) <n<n_{rm}(=v_{rm} t)$, leading to oscillatory solutions for the wave function which decay with time as $1/t$.  When $n>n_{rm}$ or $n<n_{lm}$,  the solutions $q_n^*$ are imaginary, giving rise to exponentially decaying wave functions. Thus, the wave packet spreads with time with the propagation  bounded by the maximal left and right group velocities with which the fronts can travel. 
The maximal spread of the wave packet at any given instant of time is hence given by the length of the "allowed region",  $n_{rm} - n_{lm} = (v_{rm}-v_{lm}) t$.  In contrast to the case with real hopping amplitudes,  the Hamiltonian (Eq.~\ref{eq:Hamiltonian}) is not symmetric under the reflection transformation: $n\rightarrow -n$.  Equivalently, the single particle energies $\omega(q)$(Eq.~\ref{eq:dispersion}) are not symmetric under $q \rightarrow -q$, i.e., $ \omega(-q) \neq \omega (q)$.  Hence, there is asymmetric propagation of the wave fronts to the left and right of the origin and $\psi(-n,t) \neq \psi(n,t)$.  Also, in general, for complex hopping amplitudes, $|v_{lm}(q)|\neq |v_{rm}(q)|$,  hence the  "allowed region" is asymmetric about the origin.  Thus, we expect, for complex hopping amplitudes,  chiral propagation of the wave fronts and a consequent asymmetric probability distribution about the origin. 

         For the model at hand, the group velocities can be obtained from the dispersion relation (Eq. ~\ref{eq:dispersion}) as: 
\begin{equation}
v(q) =\omega'(q) =-2\sin q - 2g \sin (2q+\phi)
\end{equation}
\footnote {We previously)~\citep{Hemlata} classified fronts as an ordinary front if $v'(q) \neq 0$ while the $k$th ($k \geq 1$) order extremal front is that for which the first non-zero derivative of the group velocity is the $(k+1)^{th}$ derivative ($f^{(n)}(q) =\frac{d^nf(q)}{dq^n}$): $v^{(1)}(q) = v^{(2)}(q) = \cdots = v^{(k)}(q)=0;\quad v^{(k+1)}(q) \neq 0$ \vspace{0pt}}
The position and velocities of the extremal fronts are determined by setting $v'(q)=\omega''(q)=0$:
\begin{equation}
v'(q)= \omega ''(q)=-2 \cos q - 4 g\cos (2q+\phi)=0
\end{equation}
Defining $ y = \cos q$ and  $\alpha= \frac{\phi}{2}$, the above equation can be written as:
\begin{eqnarray}\label{eq:Solve}
64\, g^2 y^4 && +16 \,g  \cos  2 \alpha \, y^3 +(1+16\,g\mu\cos2\alpha-64\,g^2\sin^2 2\alpha) y^2 \nonumber \\
&& + \,\, 2 \mu y +\mu^2 =0 
\end{eqnarray}
Here $\mu=8\,g\sin^2\alpha-4 g$. \\

\begin{figure*}[htp]           
  \centering
  \subfigure[$\phi=0, g_c=0.25${\label{fig:front_a}}]{\includegraphics[width=5cm,height=4cm,keepaspectratio]{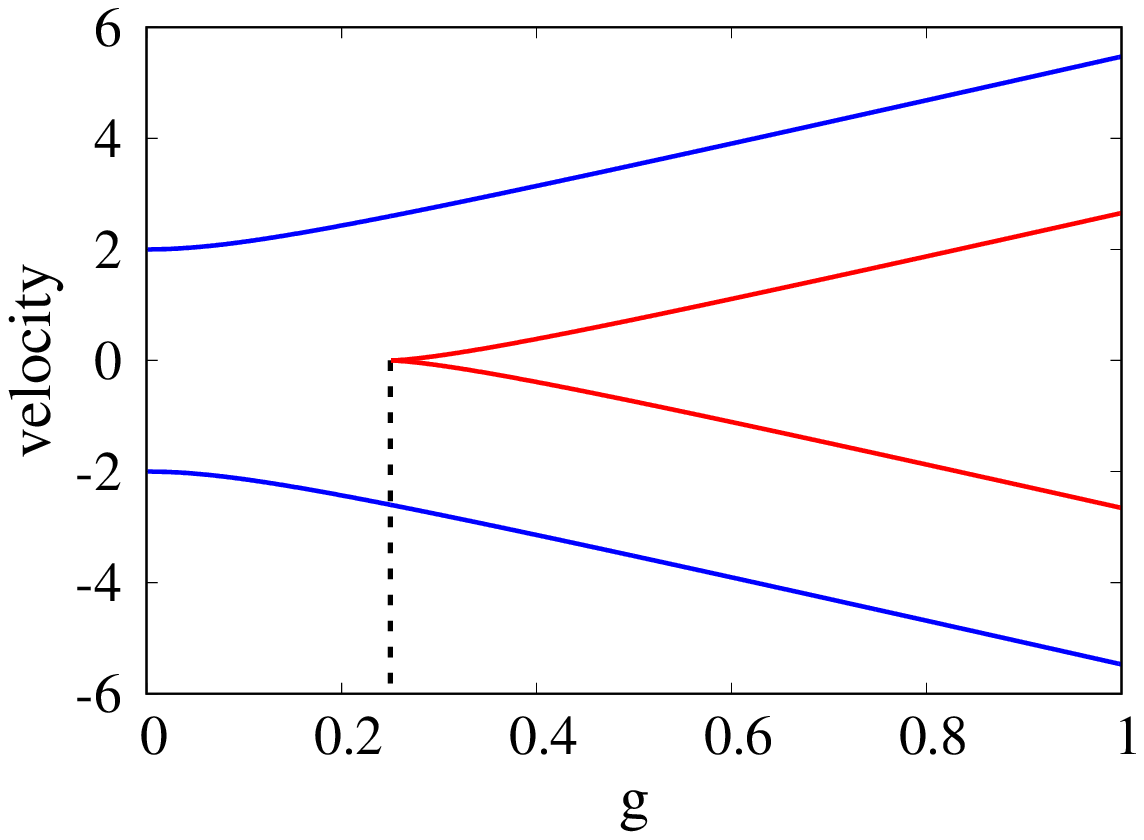}}\qquad 
  \subfigure[$\phi=\pi/6, g_c=0.239${\label{fig:front_b}}]{\includegraphics[width=5cm,height=4cm,keepaspectratio]{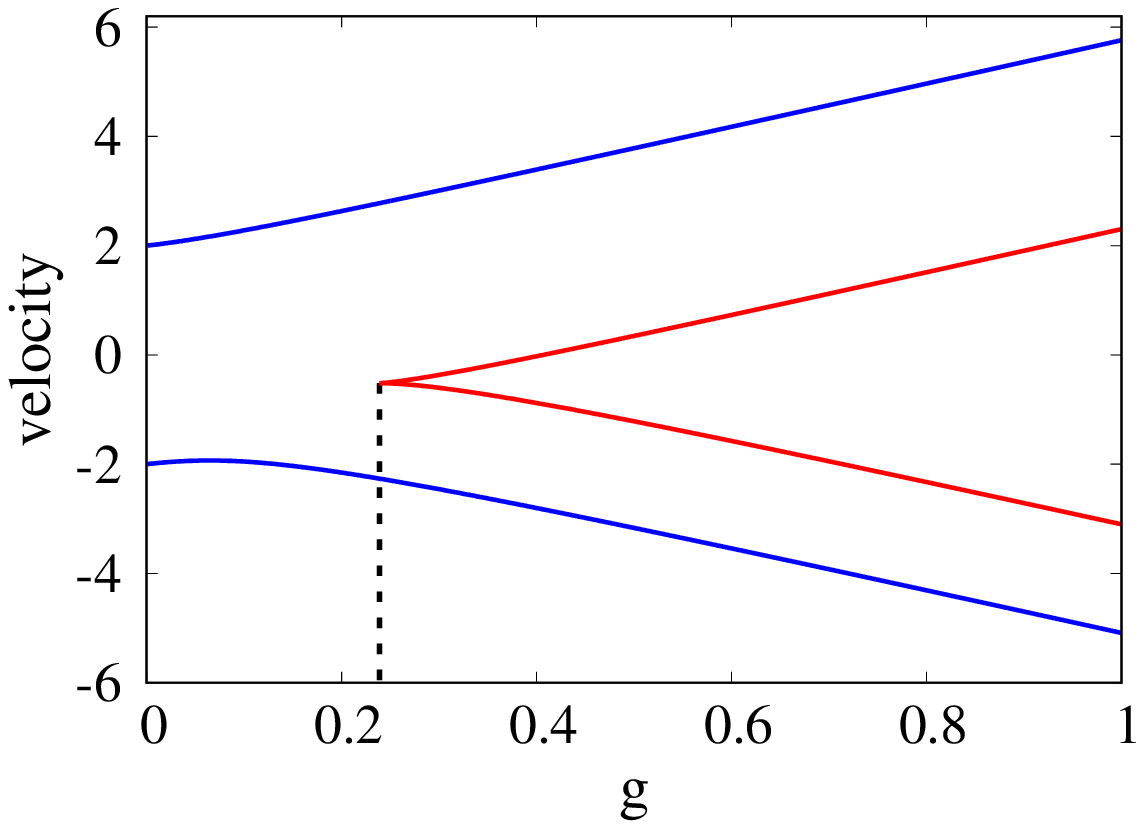}}\quad
\subfigure[$\phi=\pi/4, g_c=0.225${\label{fig:front_c}}]{\includegraphics[width=5cm,height=4cm,keepaspectratio]{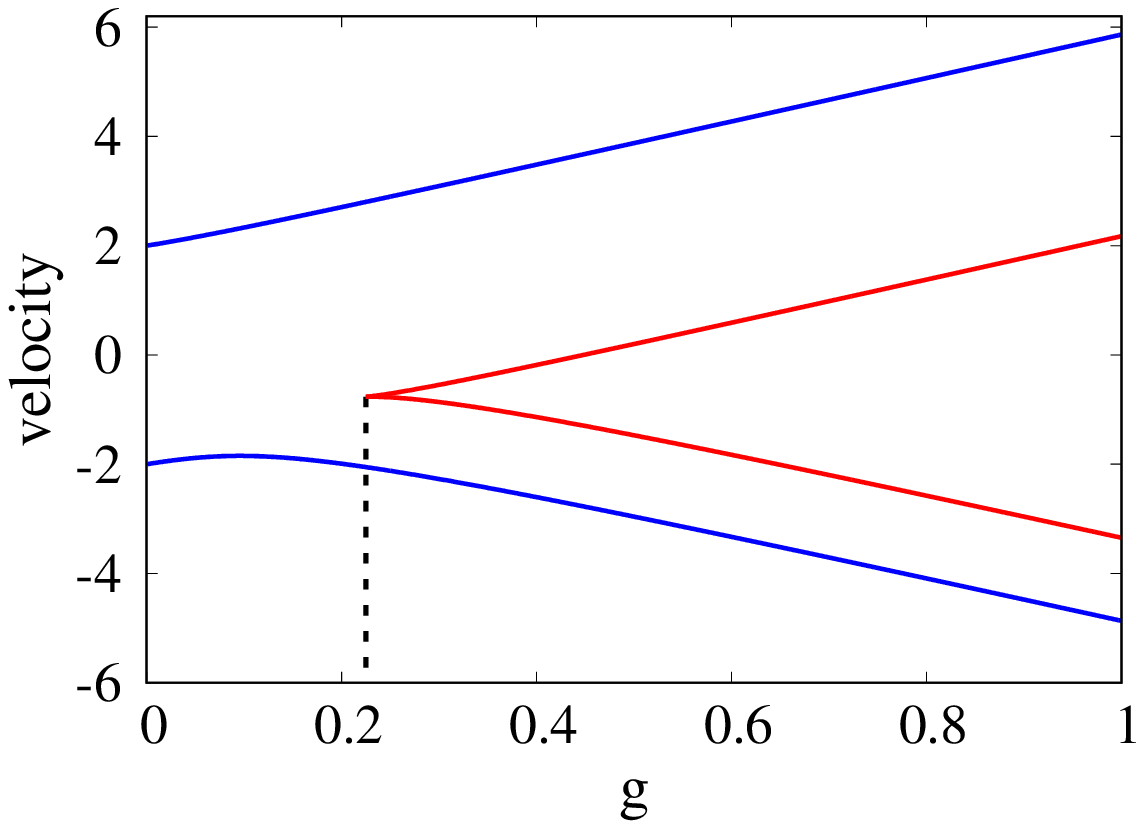}}\qquad

  \subfigure[$\phi=\pi/3, g_c=0.204${\label{fig:front_d}}]{\includegraphics[width=5cm,height=4cm,keepaspectratio]{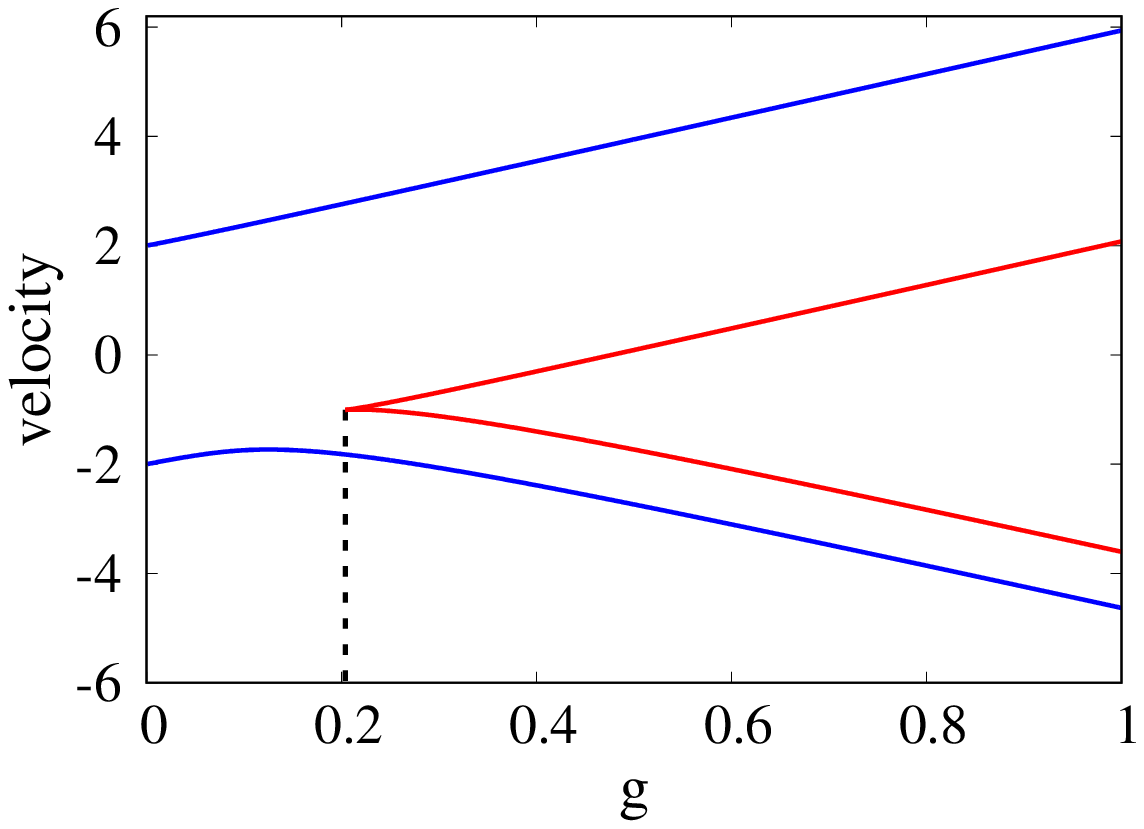}}\quad
  \subfigure[$\phi=5\pi/12, g_c=0.176${\label{fig:front_e}}]{\includegraphics[width=5cm,height=4cm,keepaspectratio]{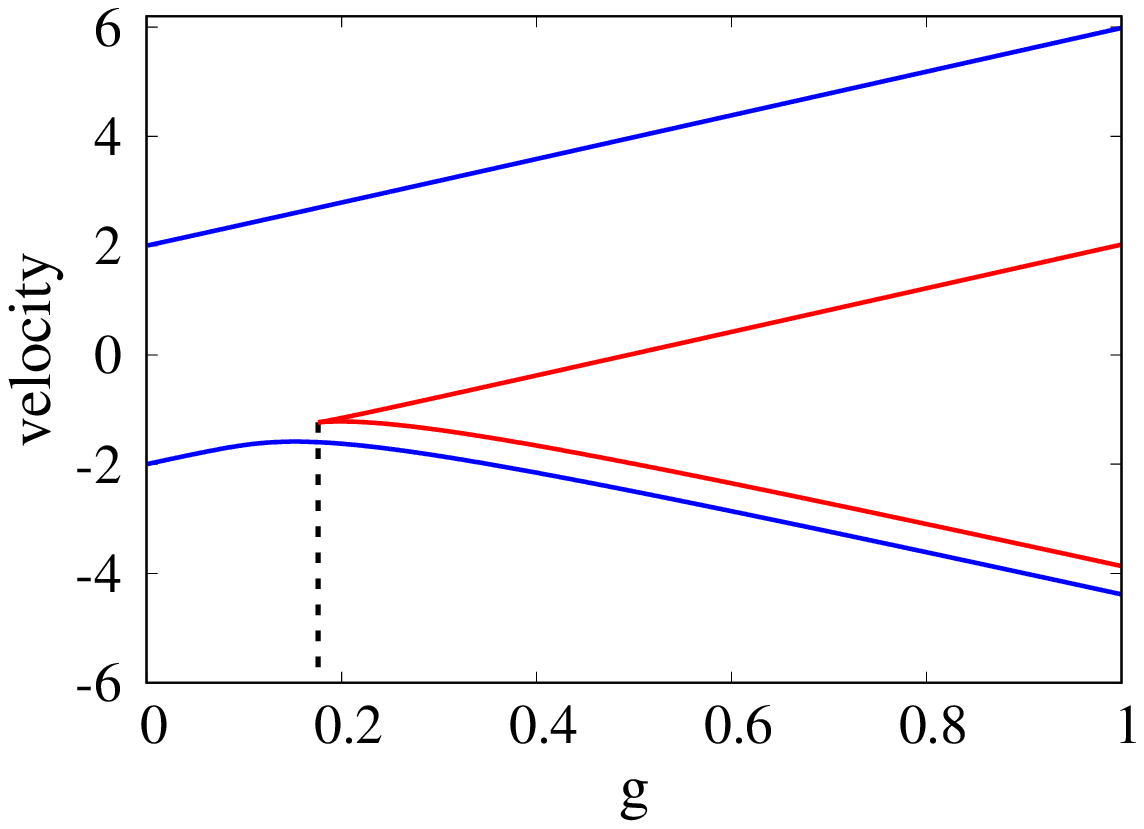}}\qquad
  \subfigure[$\phi=\pi/2, g_c=0.125${\label{fig:front_g}}]{\includegraphics[width=5cm,height=4cm,keepaspectratio]{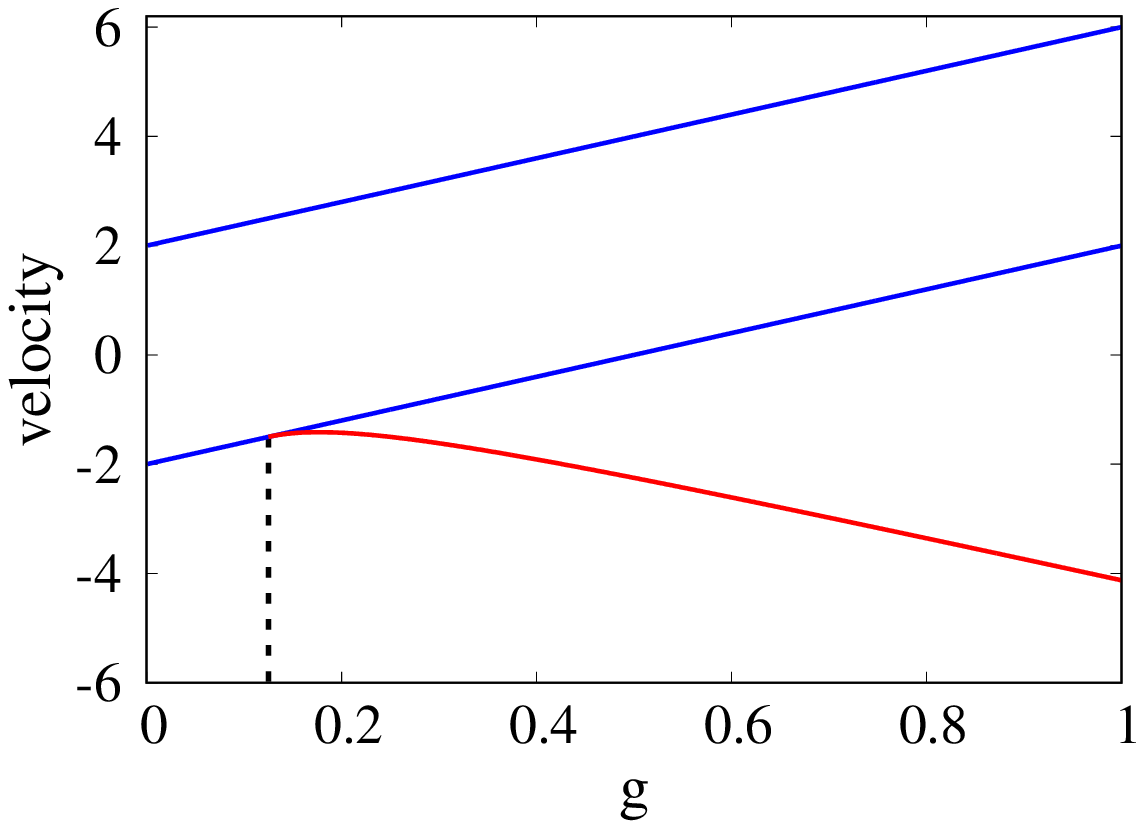}}\quad
    
\caption{Dependence of extremal front velocities on the NNN hopping strength g for different $\phi$ - values as we go from real to completely imaginary NNN hopping amplitude. The critical coupling strength $g_c$ (depends on phase and magnitude of NNN hopping strength) at which there is Lifshitz transition from regime with two extremal fronts to four extremal fronts is marked in the plots. Panel (a) shows the symmetrically placed extremal fronts about the origin for $\phi=0$ where NNN coupling is real. Panel (b)-(e) shows front velocities for increasing $\phi$-values where NNN hopping is complex (has both real and imaginary part). We see similar kind of Lifshitz transition taking place here as in case of $\phi=0$ with asymmetric distribution. We see that as the $\phi$ increases from zero towards $\pi/2$ the distance between two left moving fronts decreases leading to two-degenerate extremal fronts (moving with same velocity) at $\phi=\pi/2$ (purey imaginary NNN hopping) as shown in panel (f).  In panel (f), for $g<g_c$ blue lines represent extremal first order fronts moving with velocities $v_{lm},v_{rm}$. Critical value $g_c$ is marked in the plot where we have right first order front and left third order front. For $g>g_c$ blue lines represent two first order right moving fronts with velocities $v_{rm},v_{i}~ (v_{rm}>v_{i})$. Red line corresponds to degenerate first order extremal fronts moving with same velocity $v_{lm}$}\label{fig:front_phi}
\end{figure*} 


 We discuss the cases $0 \leq \phi <\pi/2$ and the case $\phi=\pi/2$ separately. \\
 
 \noindent (i) $ 0 \leq \phi <\pi/2$: \\
         From the solutions of Eq.~\ref{eq:Solve}, we find that for any  $ 0 \leq \phi  < \pi/2$, there is a critical coupling strength $g_c$ (which depends on both $g$ and $\phi$), at which a transition occurs from a regime ($g<g_c$) with two  first order extremal fronts  to a regime ($g>g_c$) with four first order extremal fronts.  Exactly at the critical coupling $g=g_c$, there are three extremal fronts: the two fronts with maximal left and right moving are first order extremal fronts while the internal extremal front is second order in nature.  When $\phi=0$, the extremal velocities are located symmetrically about the origin since $v(-q) = -v(q)$.  Even a small imaginary NNN coupling destroys the reflection symmetry: $v(-q) \neq -v(q)$.  Hence, the values of the left and right moving extremal velocities are not equal as can be seen from panels (a)-(e) of Fig.~\ref{fig:front_phi}, where we have shown the $g$ dependence of the extremal velocities for different $\phi$ values. It can also be seen from the figure that as $\phi$ increases from zero,  the distance between the two left moving extremal fronts decreases.   From the above analysis, we  expect that the long time quantum walk dynamics and scaling behaviour for any complex NNN  with even a small real component to remain similar to that for real NNN hopping~\citep{Hemlata} (albeit with asymmetric distributions). We also expect that the nature of the Lifshitz transition to remain the same as we go from real NNN coupling ($\phi=0$) to complex NNN hopping ($\phi < \pi/2$). We have checked this for some values of $\phi \neq \pi/2$.  \\

\noindent(ii) $\phi=\pi/2$:\\
         For purely imaginary NNN coupling, the behaviour is somewhat different. It can be seen from Eq.~\ref{eq:Solve} that for $\phi =\pi/2$, $\omega ''(q)$ vanishes at  $ q_{1,2}^*= \pm \pi/2$ for all values of $g$.  There exists a critical value of NNN hopping strength, $g=g_c=1/8$ beyond which two additional solutions occur at 
\begin{equation}
q_{3}^{*}= \sin^{-1}(1/8g); \,\,\,\text{and} \,\,\,q_4^* = \pi-q_3^{*}
\end{equation} 
The corresponding dispersion relations are related as $\omega(q_3^*) = -\omega (q_4^*)$.
For large $g$ ( $g \rightarrow \infty$), the extremal $q^*$-values are $\pm \pi/2, \,\,0,\,\, \pi$.
Thus, the number of extremal fronts changes from $2$ to $4$ at the  critical coupling strength, $g=g_c=1/8$.  
From the dispersion relation Eq.~\ref{eq:dispersion},  we can see that for $g < g_c$, the extremal fronts at $q*=\pm \pi/2$ are both first order maximal fronts ($v^{(2)} (q_{1(2)}^*) \neq 0$), moving with velocities $ v_{lm} = -2 + 4g $ and $v_{rm} = 2+ 4g$, while for $g>g_c$,  all the four extremal fronts are first order; the two extremal fronts at $ \pm \pi/2$ move with velocities $\mp 2+ 4g $ and the two additional first order fronts at $q_3^*$ and $q_4^*$ are both left moving with the same velocity: $v_3=v_4= -4g -\frac{1}{8g}$.
The maximal left and right velocities are  $v_{lm}=v_3 (=v_4)$ and $v_{rm}=v_1$.  Exactly at the critical coupling, $g=g_c=1/8$,  of the two extremal fronts, located at $q^*=\pm \pi/2$, the one corresponding to $q_1^* = \pi/2$ is a third order left moving front  while the extremal front corresponding to $q_2^* = -\pi/2$ is a first order right moving front. Hence, near $q_1^*$,  the group velocity shows a quartic $q$ dependence: 
\begin{equation}
v(q_1^*) \approx v (q_1^*) +  (q-q_1^*)^4 \frac{v^{(4)}(q_1^*)}{4!} =  -\frac{3}{2}+\frac{1}{4}(q-q_1^*)^4
\end{equation}
The maximal left and right velocities are $v_{lm}= -2 + 4g $ and $v_{rm}=2 + 4g$.

\begin{figure*}[htp]           
  \centering
  \subfigure{\label{fig:disp_g_0}}{\includegraphics[width=5.6cm,height=4.6cm,keepaspectratio]{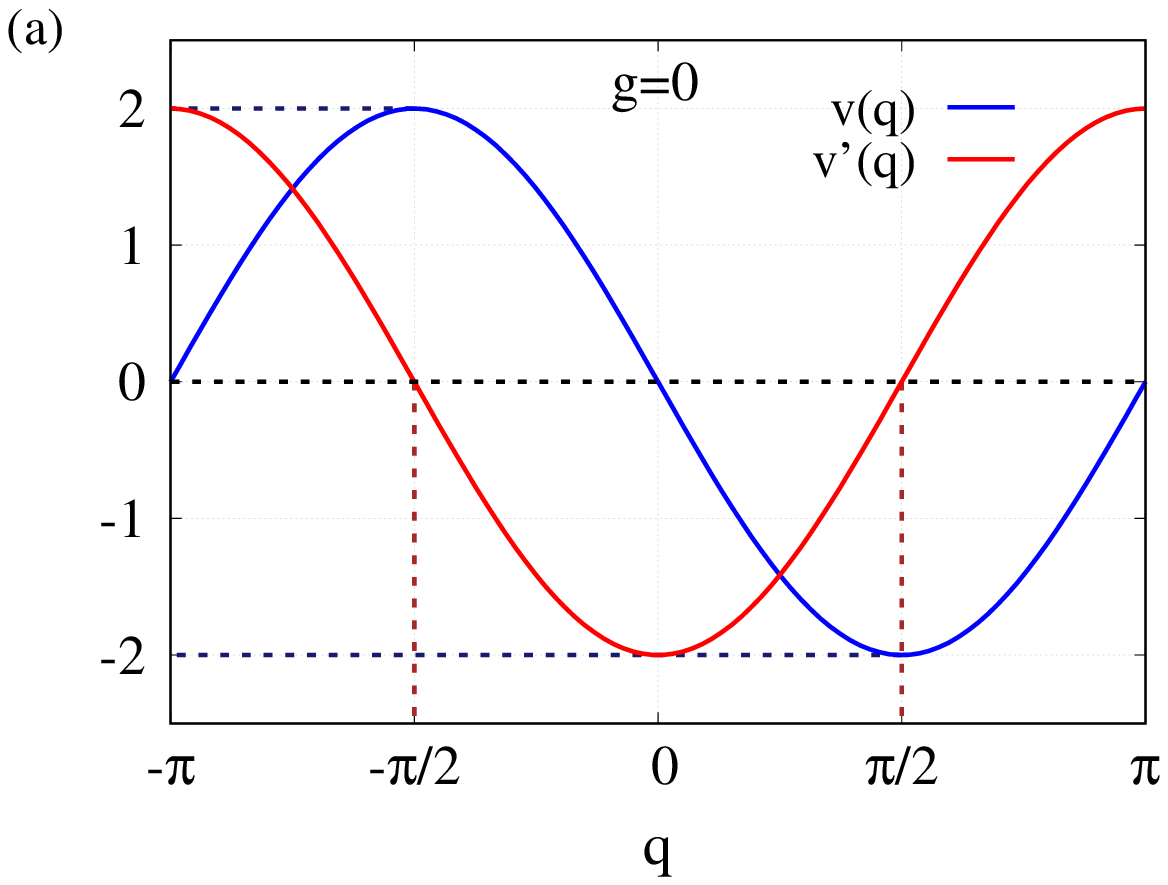}}\qquad 
\subfigure{\label{fig:pd_g_0}}{\includegraphics[width=5.6cm,height=4.6cm,keepaspectratio]{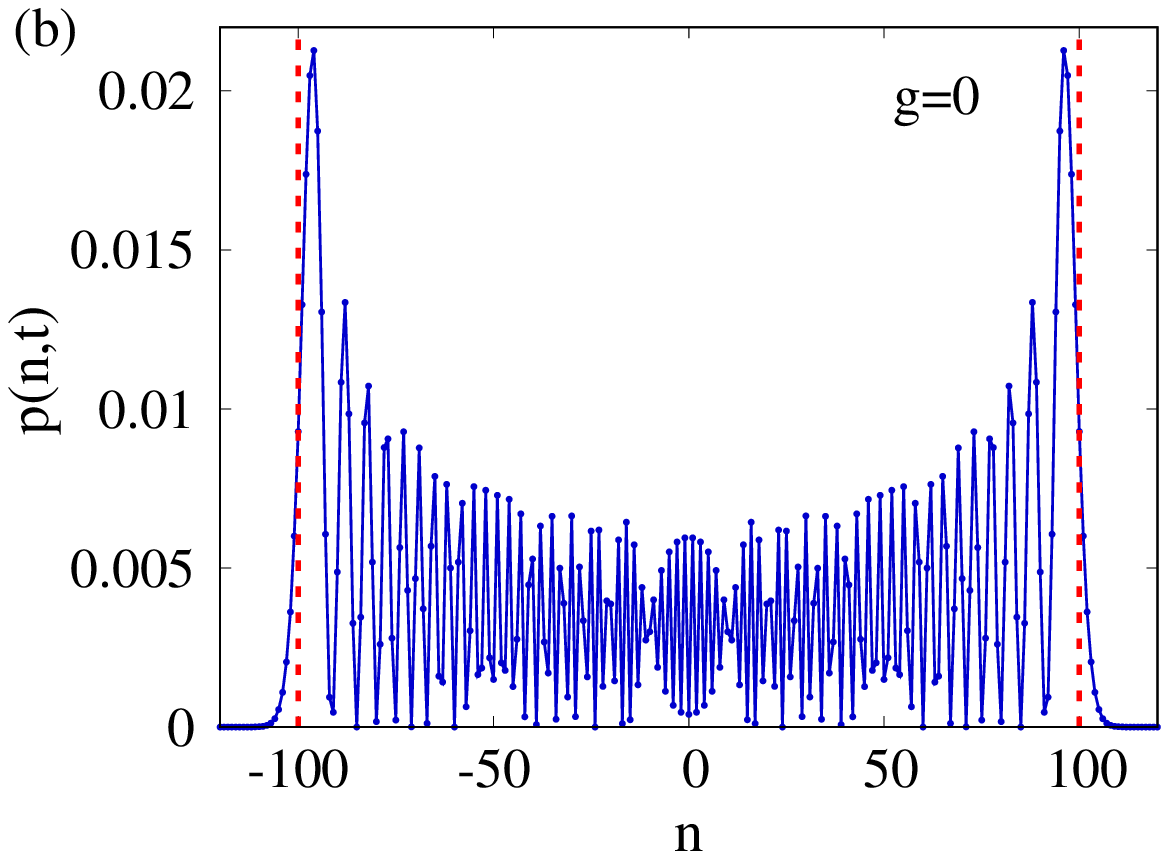}}\quad
\subfigure{\label{fig:pc_g_0}}{\includegraphics[width=5.6cm,height=4.6cm,keepaspectratio]{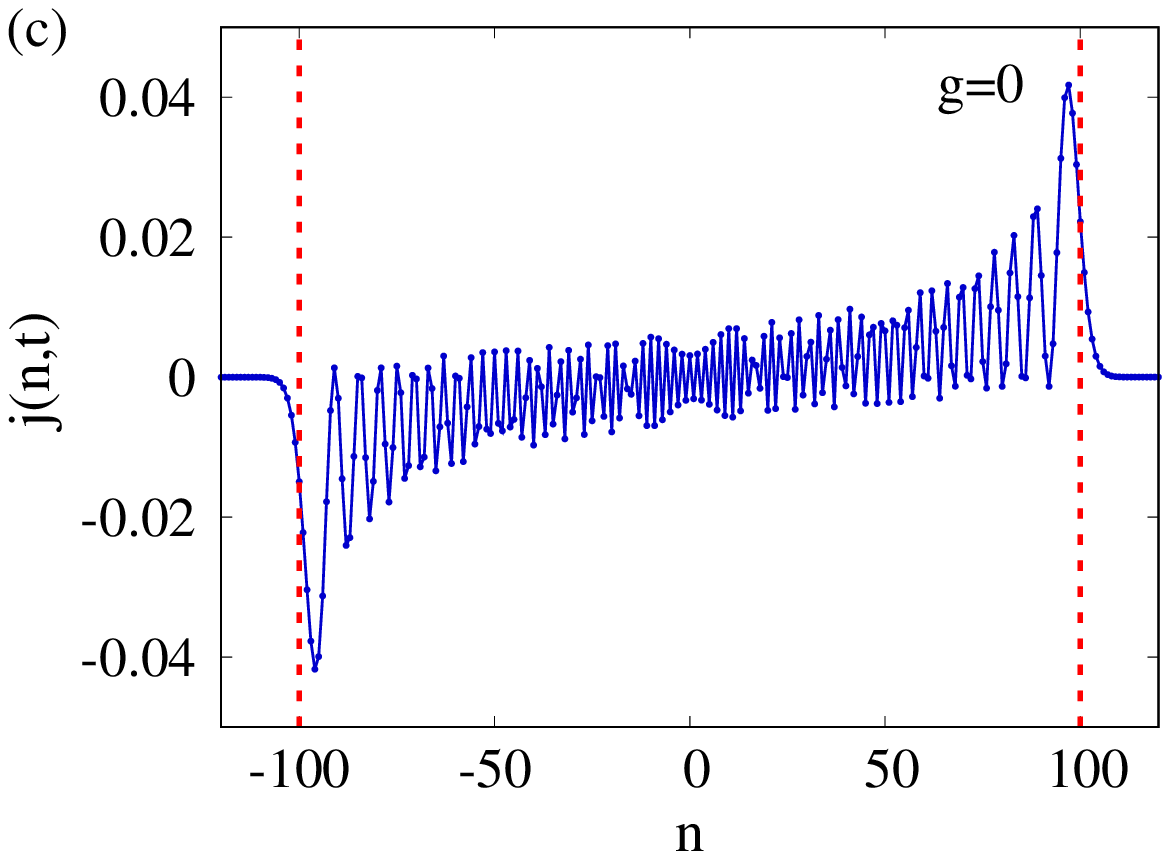}}\qquad

  \subfigure{\label{fig:disp_g_0p0625}}{\includegraphics[width=5.6cm,height=4.6cm,keepaspectratio]{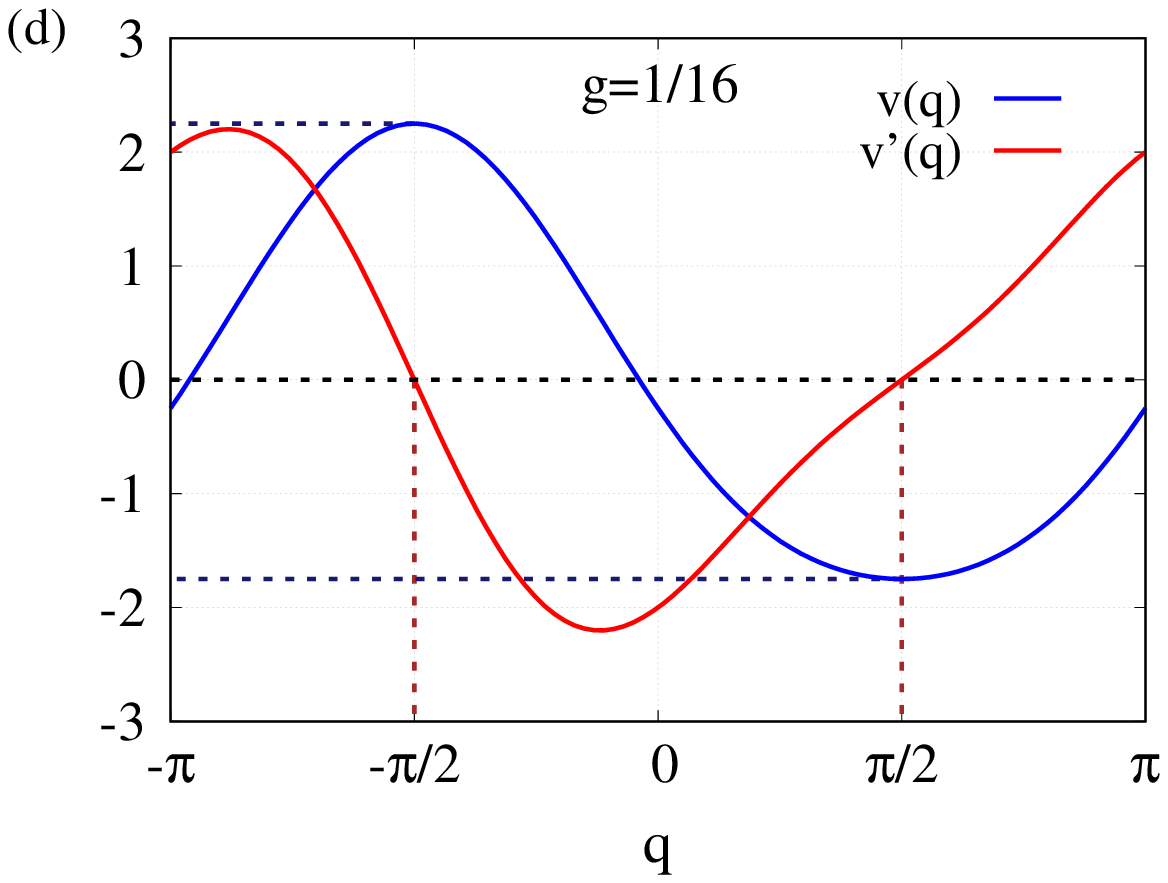}}\quad
  \subfigure{\label{fig:pd_g_1by16}}{\includegraphics[width=5.6cm,height=4.6cm,keepaspectratio]{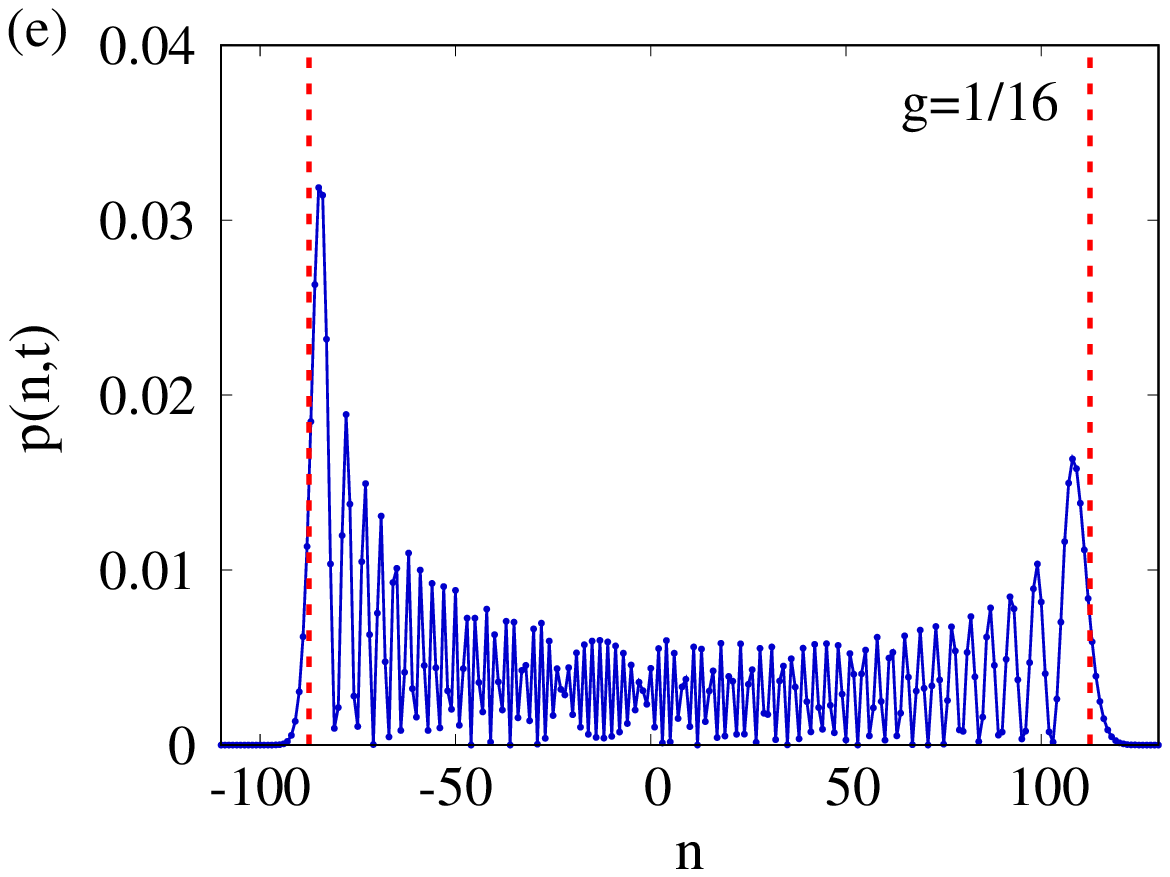}}\qquad
  \subfigure{\label{fig:pc_g_1by16}}{\includegraphics[width=5.6cm,height=4.6cm,keepaspectratio]{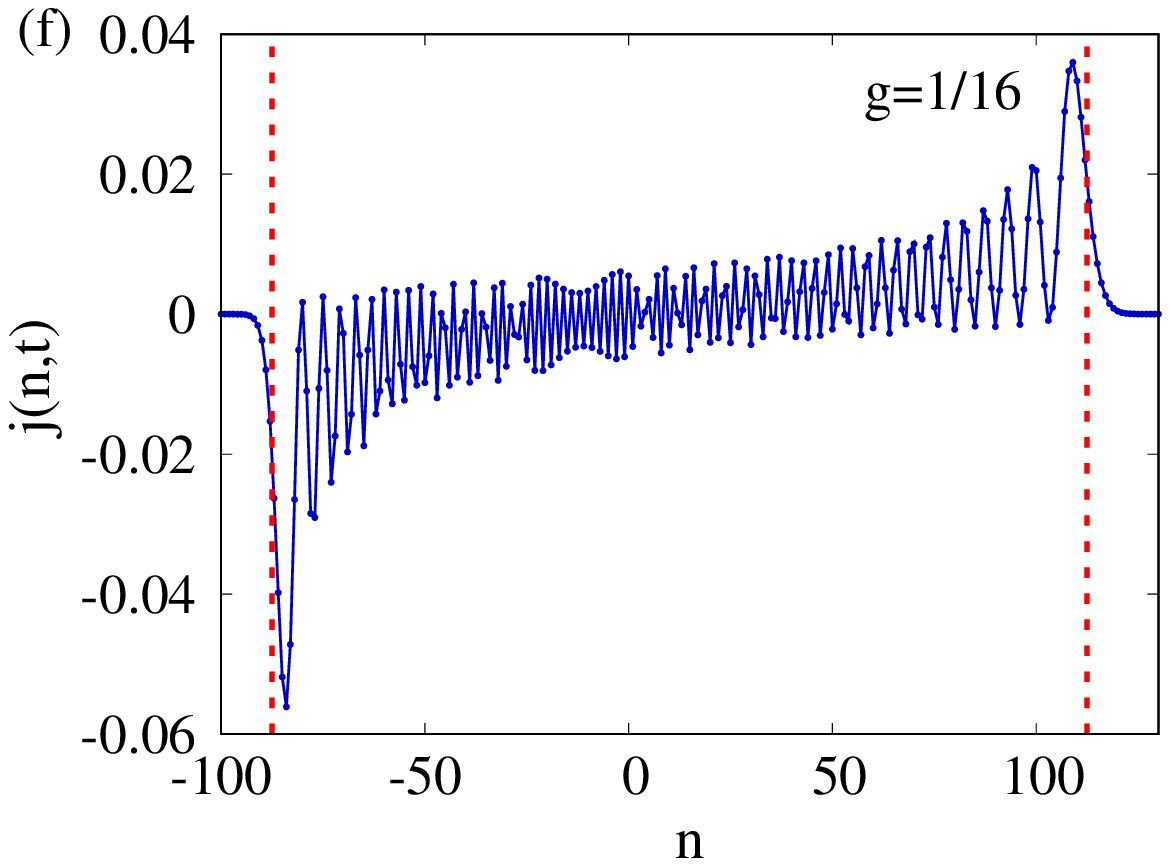}} \quad
  
  \subfigure{\label{fig:disp_0p125}}{\includegraphics[width=5.6cm,height=4.6cm,keepaspectratio]{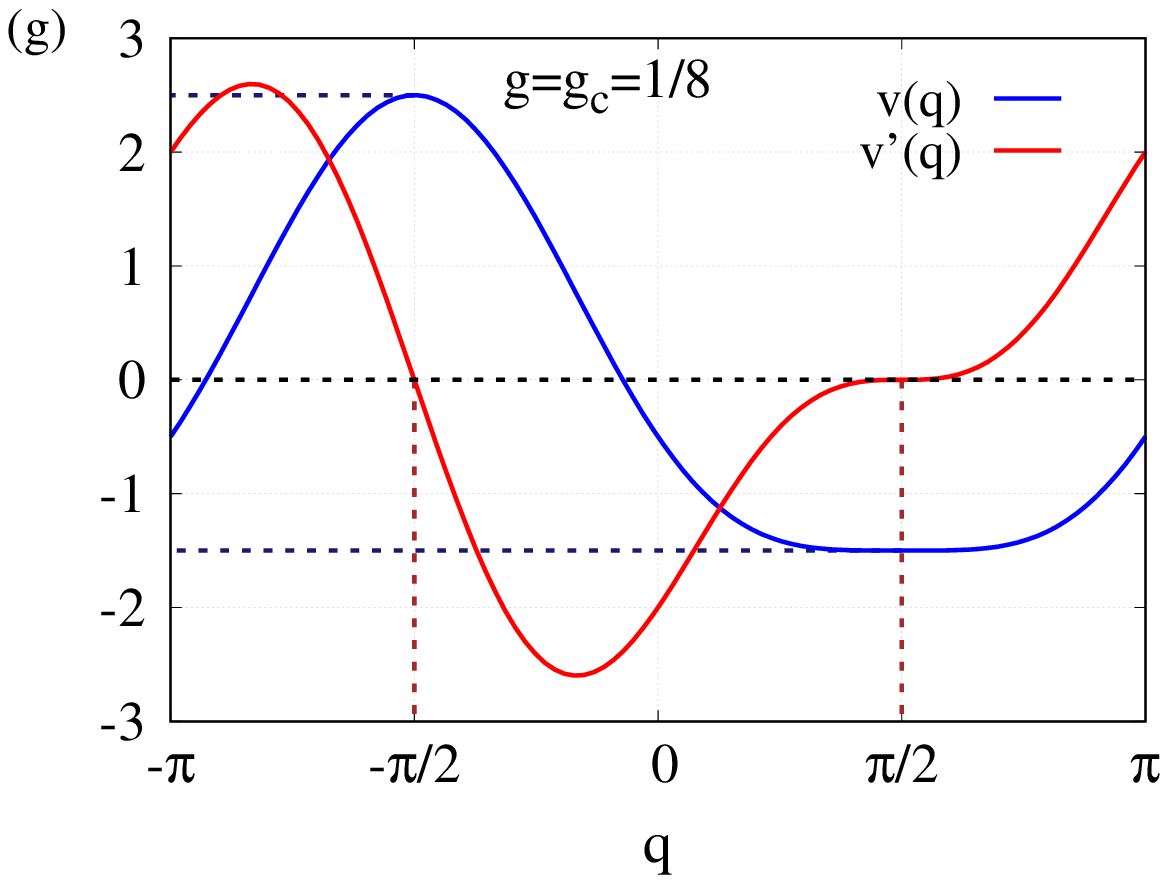}}\quad
 \subfigure{\label{fig:pd_g_1by8}}{\includegraphics[width=5.6cm,height=4.6cm,keepaspectratio]{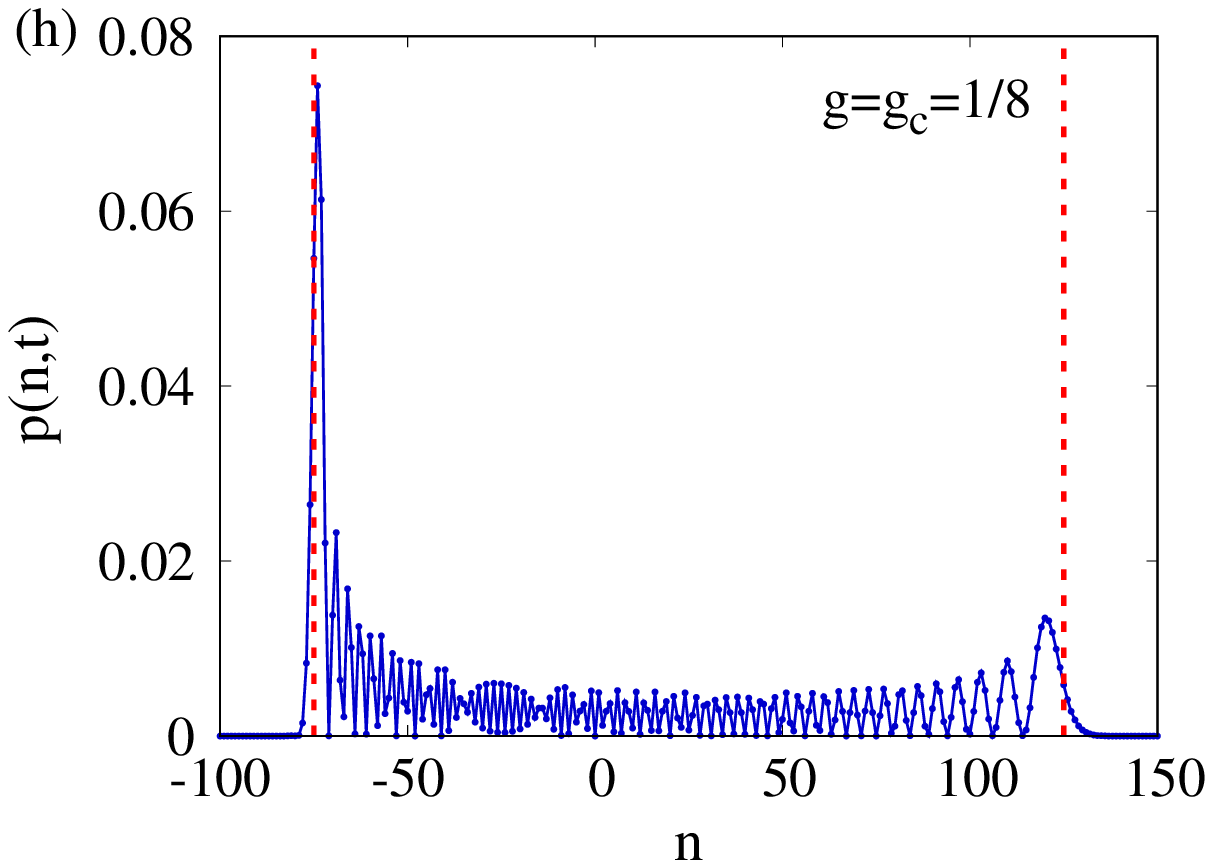}}\qquad
  \subfigure{\label{fig:pc_g_1by8}}{\includegraphics[width=5.6cm,height=4.6cm,keepaspectratio]{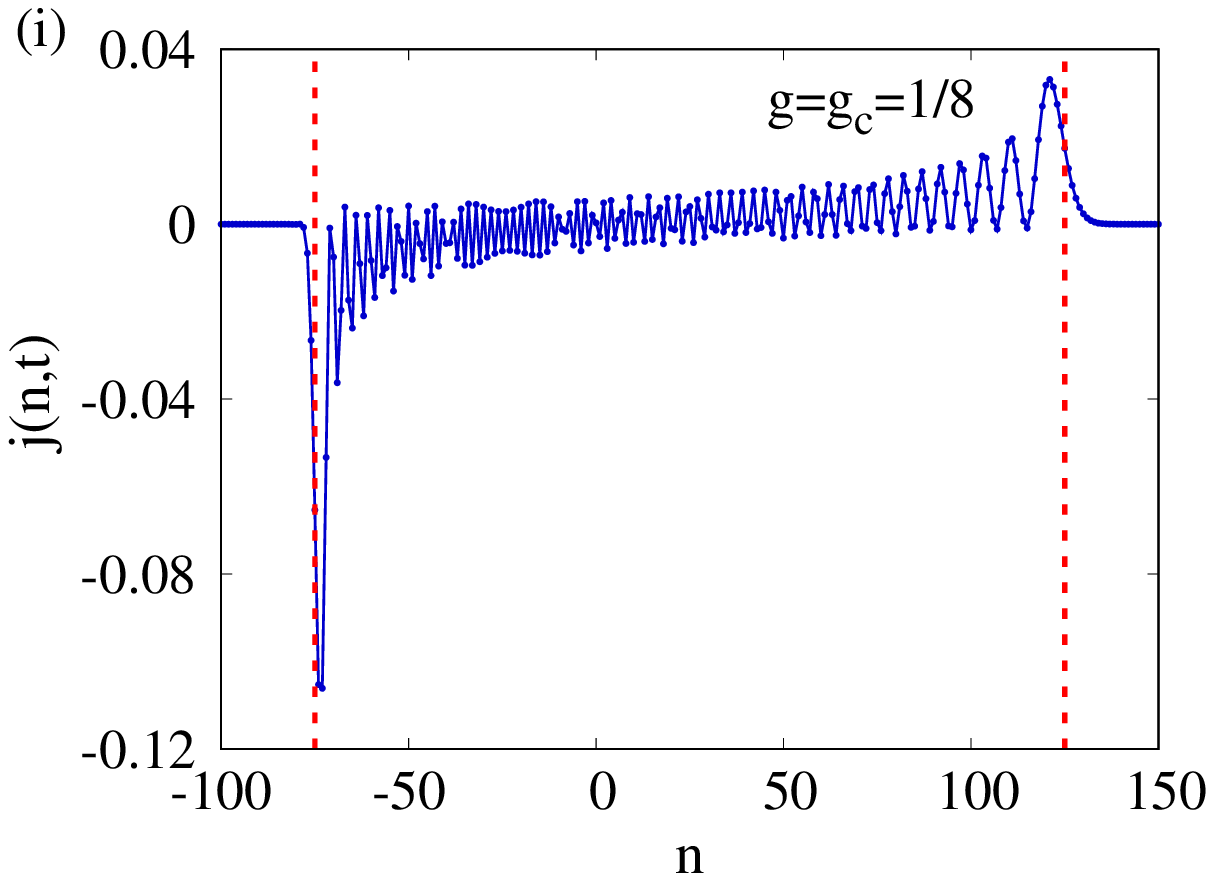}} \quad  
  
   \subfigure{\label{fig:disp_0p25}}{\includegraphics[width=5.6cm,height=4.6cm,keepaspectratio]{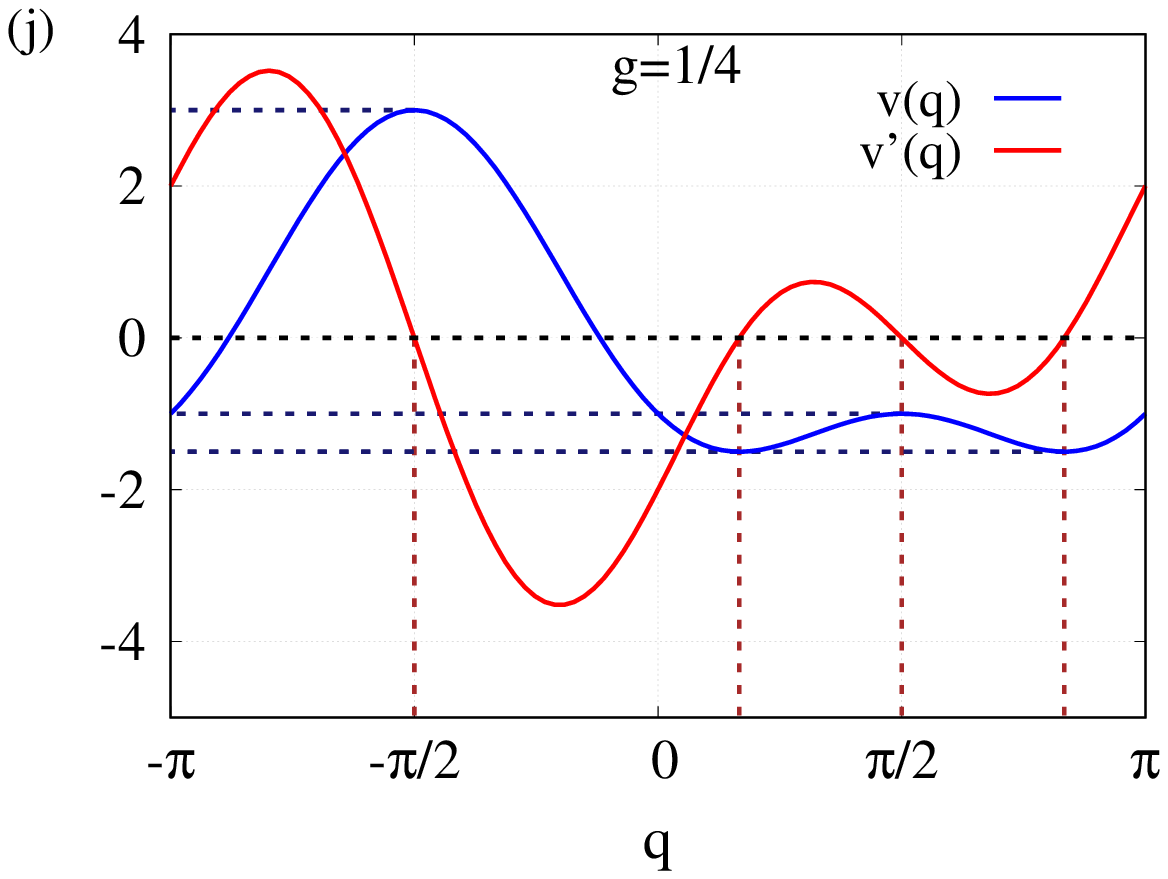}}\quad
 \subfigure{\label{fig:pd_g_1by4}}{\includegraphics[width=5.6cm,height=4.6cm,keepaspectratio]{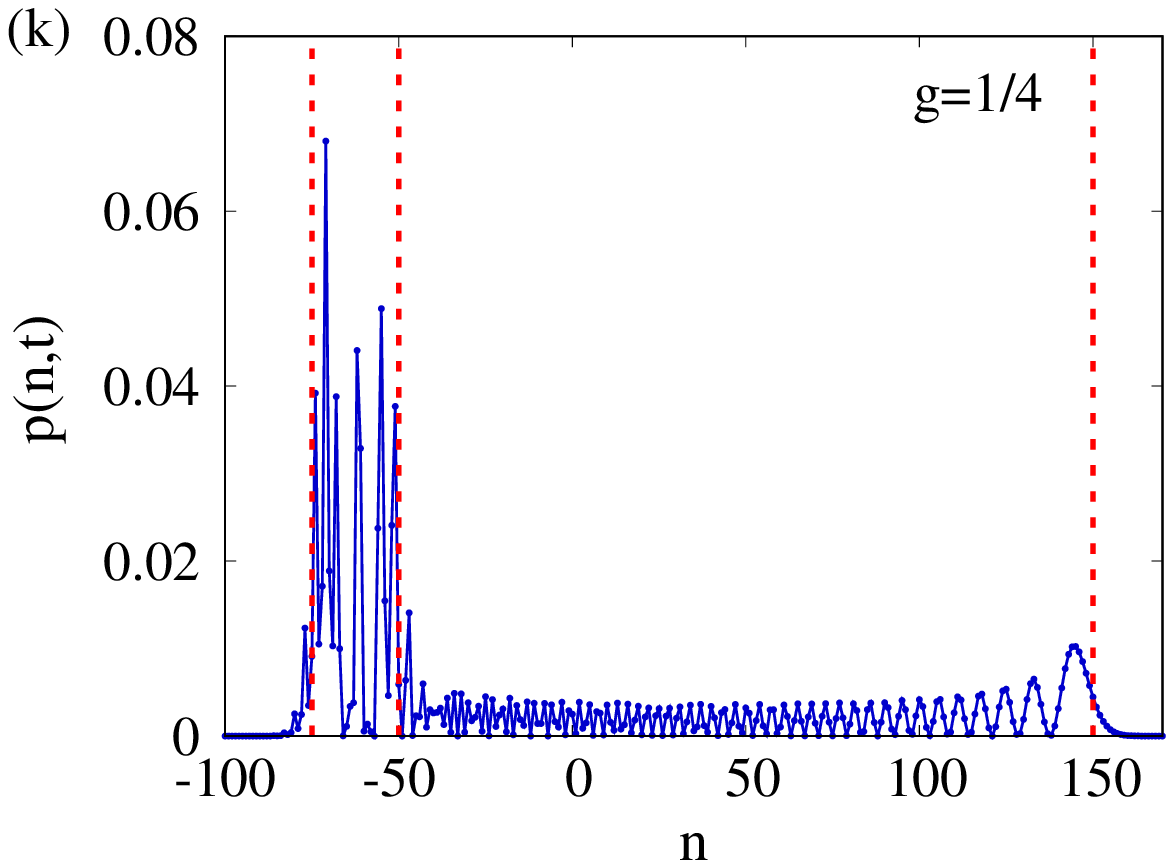}}\qquad
  \subfigure{\label{fig:pc_g_1by4}}{\includegraphics[width=5.6cm,height=4.6cm,keepaspectratio]{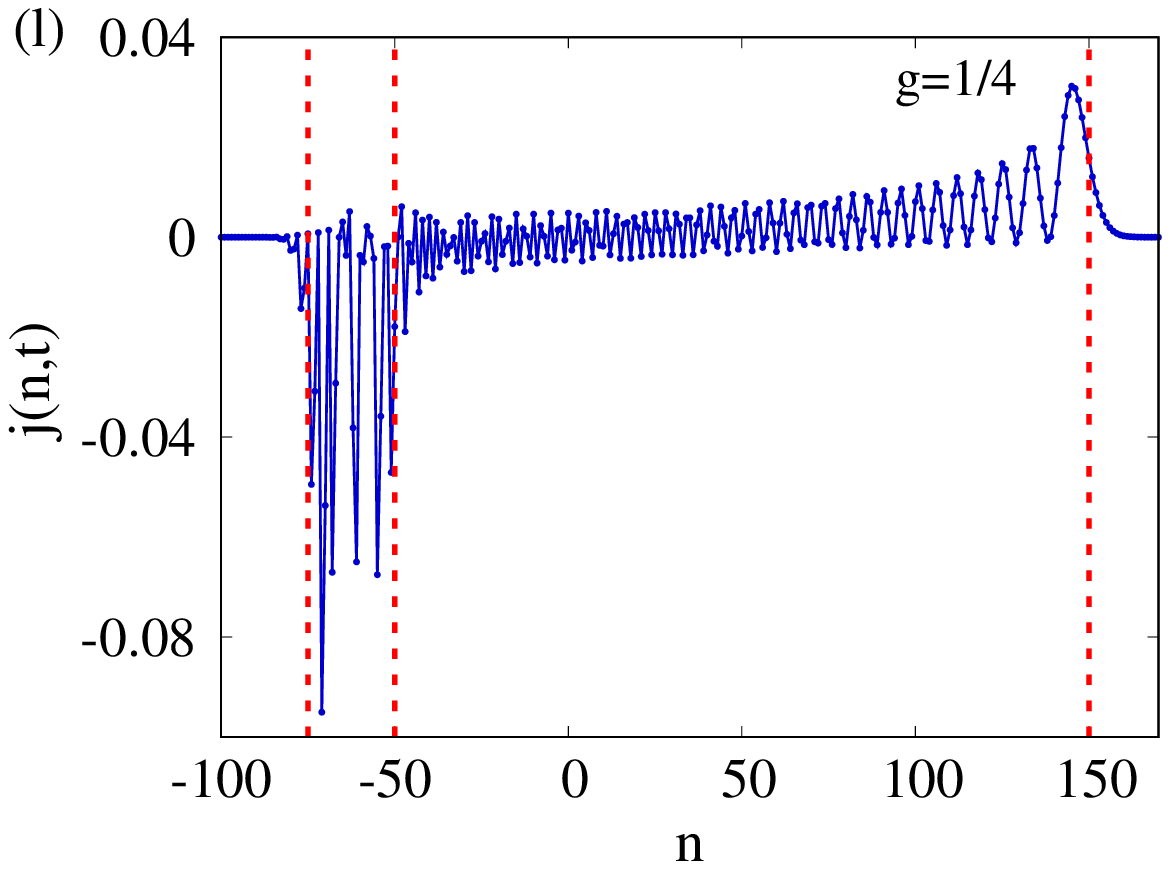}} 
  \caption{The first column shows the dependence of the group velocity $v(q)$ and its first derivative $v'(q)$ on the wave-vector ($q$) for $\phi=\pi/2$ for representative $g$-values. The vertical dotted lines marks the saddle point solutions for the extremal fronts and corresponding horizontal dotted lines show their respective extremal velocities. Panels (a),(d),(g),(j) show that for all the $g$-values we have extremal fronts correponding to $q=\pm \pi/2$. Panel (j) for $g=1/4(>g_c)$ shows the  two additional saddle point solutions which correspond to two-fold degenerate fronts moving with same velocity. The second column of the figure shows the local probability distribution profiles for $\phi=\pi/2$ at time $t=50$ (measured in units of $g_1^{-1}$) for representative $g$-values. Panel (b) shows the distribution for the case with only real NN hopping ($g=0$), the distribution is symmetric about origin and $p(n,t)=p(-n,t)$. Panels (e), (h), (k) shows chiral nature of propagation for $\phi=\pi/2$ with $p(n,t) \neq p(-n,t)$. The third column shows the local current density plots for representative $g$-values for $\phi=\pi/2$ at time $t=50$ (measured in units of $g_1^{-1}$). The current densities show the oscillatory behaviour taking both positive and negative values. Panel (c) for real NN hopping ($g=0$) is symmetric about the origin. Panels (f),(i),(l) shows the asymmetry in the current density for $\phi=\pi/2$. Panels (k), (l) shows the emergence of doubly-degenerate left moving maximal fronts for $g>g_c$ in local probability density and current density profiles. Vertical red lines in second and third column shows the theoretical position of extremal fronts. }\label{fig:local_observables}
\end{figure*}

Thus we find that for purely imaginary NNN hopping, there is a transition at a critical coupling, $g_c =1/8$, from a regime ($g<g_c$)with two first order extremal fronts to a regime ($g>g_c$) with four first order extremal fronts, two of which are degenerate (move with the same velocity).  At the critical coupling $g_c$,  there are  only two extremal fronts; one of which is first order while the other is third order.  
The dependence of the extremal front velocities on the NNN hopping strength $g$ is shown in Fig.~\ref{fig:front_phi} while the $q$ dependence of the group velocity $v(q)=\omega'(q)$ and the derivative of the group velocity $v'(q)=\omega''(q)$ are shown in the first column of Fig. \ref{fig:local_observables} for some representative $g$ values.  The extremal $q$ values and the corresponding extremal velocities can be seen from the plots.  It can be seen from the plots that for all $g$, two extremal solutions occur at $ q^*= \pm \pi/2$.  Further, we can see that these are first order fronts for all $g \neq g_c$ while for $g=g_c$, the front corresponding to $-\pi/2$ is first order while the front with $q=\pi/2$ is third order (see Figs.~\ref{fig:disp_g_0}, ~\ref{fig:disp_g_0p0625}, ~\ref{fig:disp_0p125}).  For $g=1/4 >g_c$, it can be seen from Fig.~\ref{fig:disp_0p25},  that in addition to the extremal fronts at $q^*=\pm \pi/2$,  two additional extremal solutions emerge at $q_{3(4)}^*$'s. These correspond  to fronts with dispersion relation $\omega(q_3^{*})= - \omega(q_4^*)$  and with degenerate velocities $v(q_{3}^*) =v(q_{4}^*)$ and are both first order in nature. 
  
                 From the analysis given above, we expect different long time behaviour for a purely imaginary NNN coupling as compared to that with a real component.  We therefore restrict in the following to the case of a purely imaginary NNN hopping and set $\phi = \pi/2$. We begin by discussing the causal behaviour of the local probability and current densities.   We show in the second and third columns of Fig. {\ref{fig:local_observables}}, the local probability density and current density distributions obtained by a numerical solution of Eq. \ref{eq:fourier_integral} at time $t=50$ for representative $g$-values. 
 
 
     The probability and current densities shown in the second and third columns of Fig.~\ref{fig:local_observables} are non-zero and oscillatory inside an 'allowed' region bounded by extremal fronts moving with maximal velocities.  A 'forbidden region"  with exponentially decaying probabilities lies outside the allowed region. The probability distribution is symmetric about the origin for $g=0$ (Fig.~\ref{fig:pd_g_0}) because we have considered real NN coupling. For $g\neq 0$,  the chiral nature and asymmetry of the distributions about the origin can be seen from the plots (Figs.~\ref{fig:pd_g_1by16}~\ref{fig:pc_g_1by16}, ~\ref{fig:pd_g_1by8}, ~\ref{fig:pc_g_1by8}, ~\ref{fig:pd_g_1by4} and  \ref{fig:pc_g_1by4}).  For $g < g_c$, there are two first order fronts; one moving to the left and the other moving to right; since $v_{rm} = 2 + 4g > |v_{lm}|= |4g -2|$,  the probability distribution drifts to the right as can be seen in Fig. ~\ref{fig:pd_g_1by16}. The width of the allowed region is ($\approx 4t$). The probability of finding the particle increases near the left front as compared to the right front.  For $g >g_c$ (Fig.~\ref{fig:pd_g_1by4}), there is one  right moving front with maximal velocity $ v_{rm} = 2 +4 g$ and two left moving fronts with the same degenerate velocity $v_{lm} = -4g -\frac{1}{8g}$.  There is an additional internal front moving with velocity  $v_{i} = -2 +4 g$. The width of the allowed region increases with $g$ as  $ 8g + \frac{1}{8g} + 2$.  The probabilities are large within the region bounded by the fronts moving with velocities $v_{lm}$ and $v_i$ as compared to that in the region between the fronts located at $v_i t$ and $v_{rm}t$. They also show aperiodic oscillatory behaviour in the region $v_{lm} t <n<v_i t$  in contrast to the periodic oscillations in the region between $v_i t$ and $v_{rm} t$ (Fig. ~\ref{fig:pd_g_1by4}).  At $g=g_c$, there are two extremal fronts as shown in Fig. ~\ref{fig:pd_g_1by8}; the particle has maximum probability to be at the extremal third order  left moving front. The local current density profiles show similar behaviour as can be seen from Fig.~\ref{fig:local_observables}. The densities show oscillatory behaviour inside the allowed region; taking both positive and negative values. The current density distribution is asymmetric about the origin for a non-zero $g$; however, the total probability conservation ensures that the total current sums up to zero.

 \section{\label{sec:Hydrodynamic} Hydrodynamic description at asymptotically long times and distances}
\begin{figure*}[htp]           
  \centering
  \subfigure{\label{fig:velocity_cum_prob_a}}{\includegraphics[width=5.6cm,height=4.6cm,keepaspectratio]{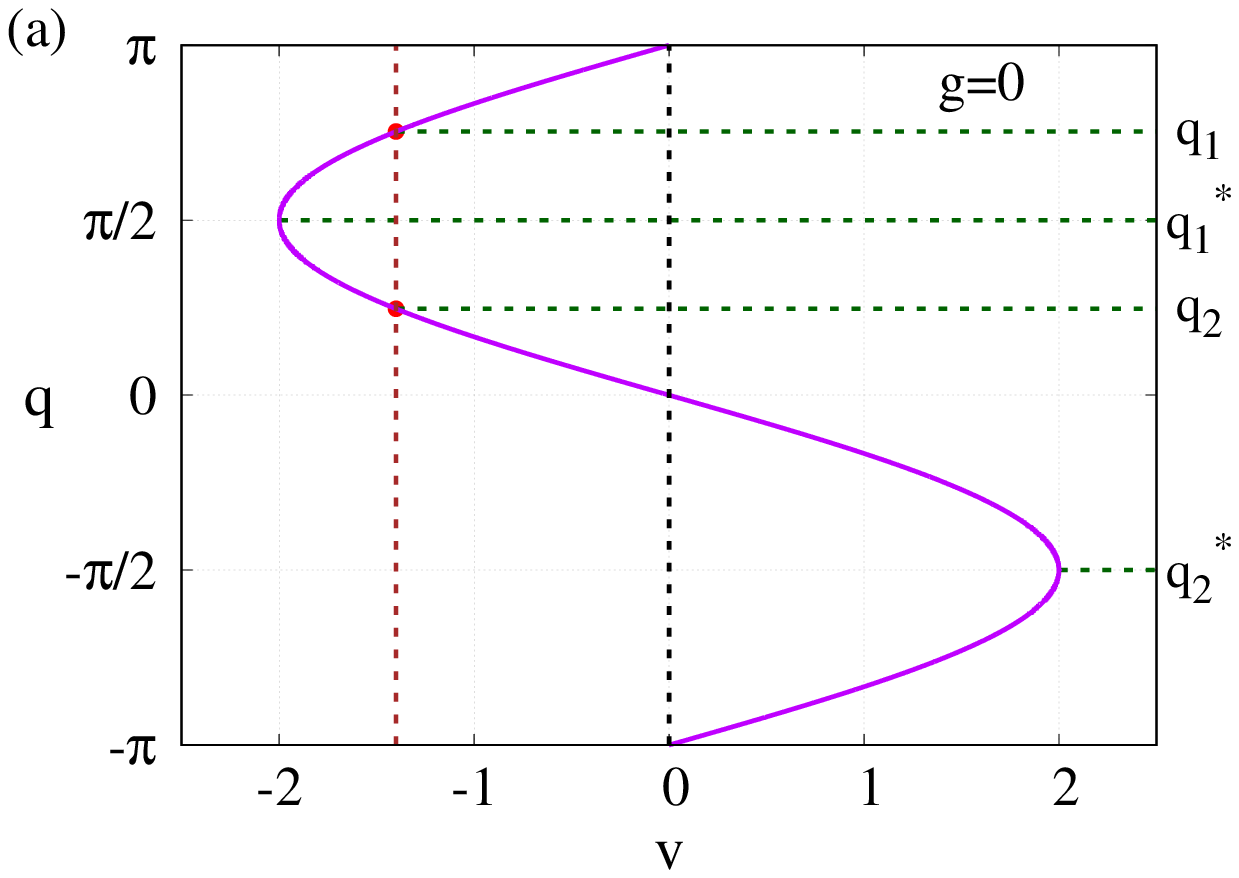}}\qquad 
\subfigure{\label{fig:cum_prob_a}}{\includegraphics[width=5.6cm,height=4.6cm,keepaspectratio]{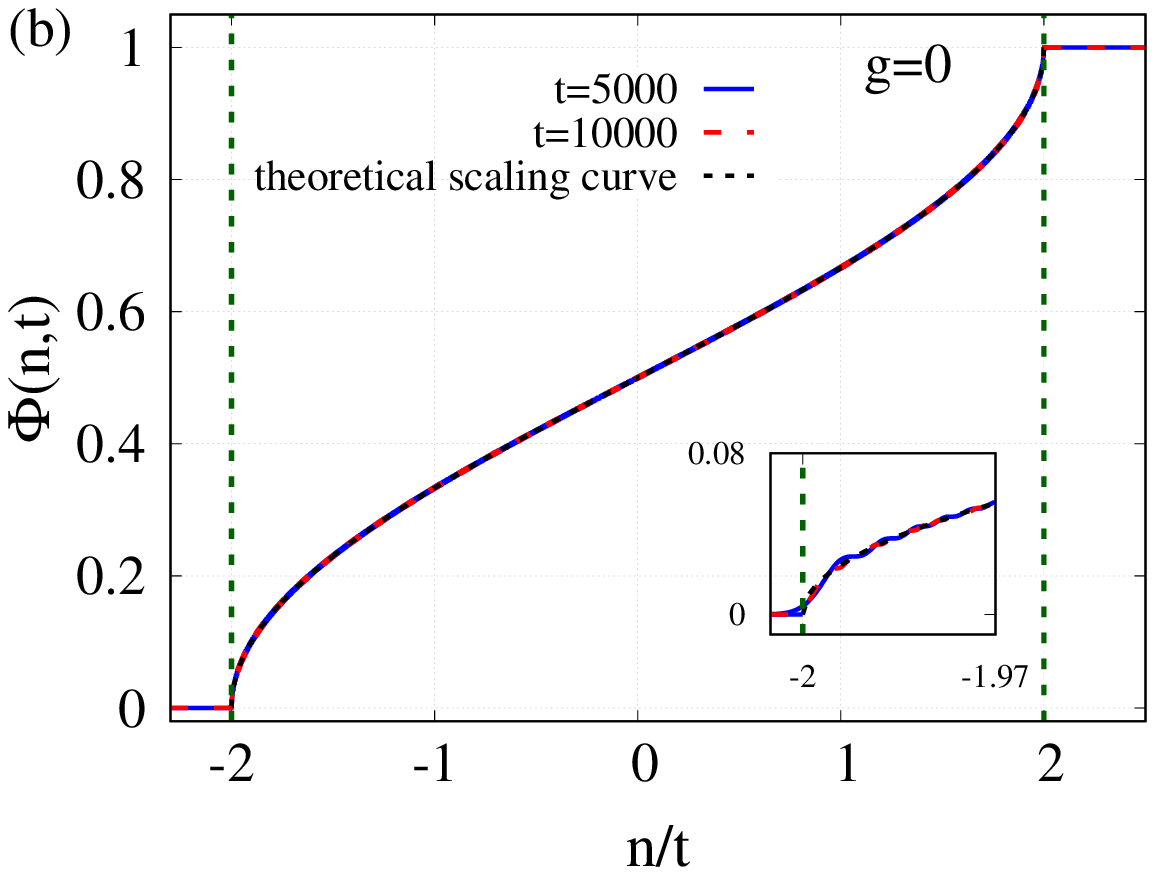}}\quad
\subfigure{\label{fig:cum_cur_a}}{\includegraphics[width=5.6cm,height=4.6cm,keepaspectratio]{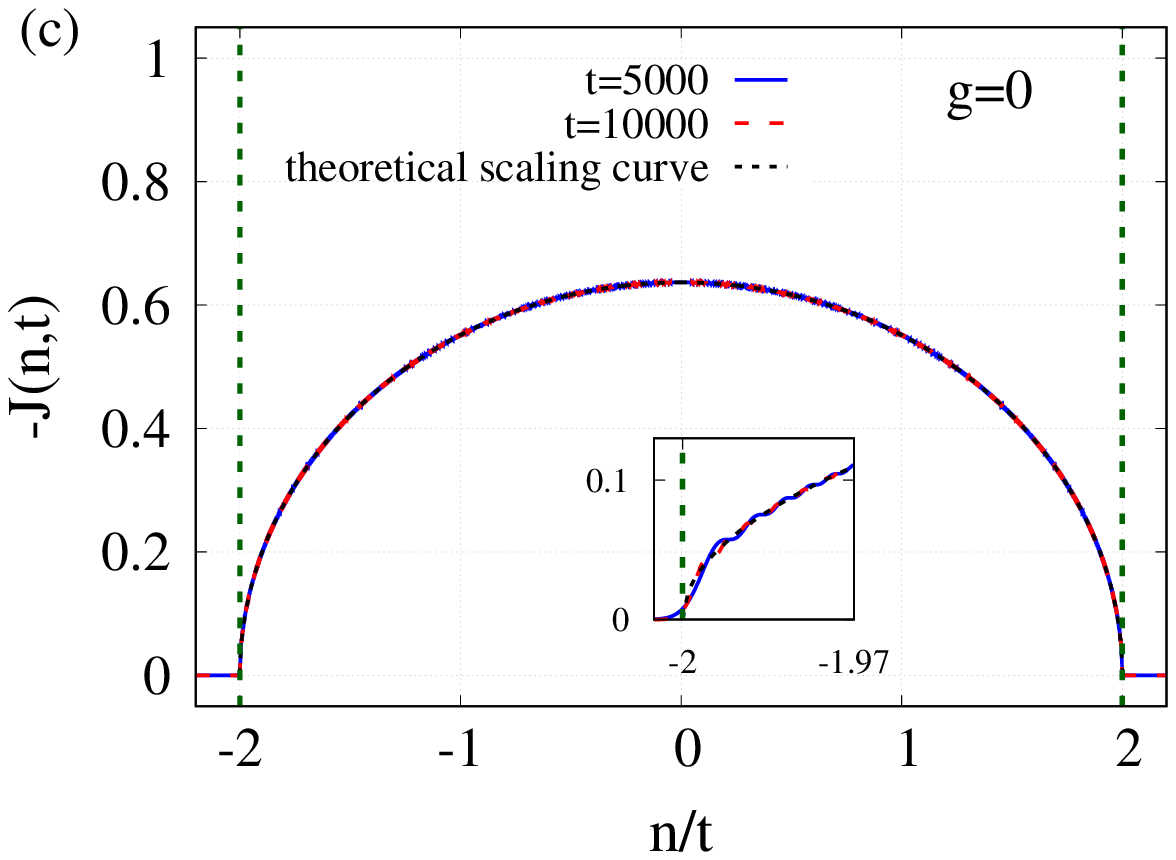}}\qquad

  \subfigure{\label{fig:velocity_cum_prob_b}}{\includegraphics[width=5.6cm,height=4.6cm,keepaspectratio]{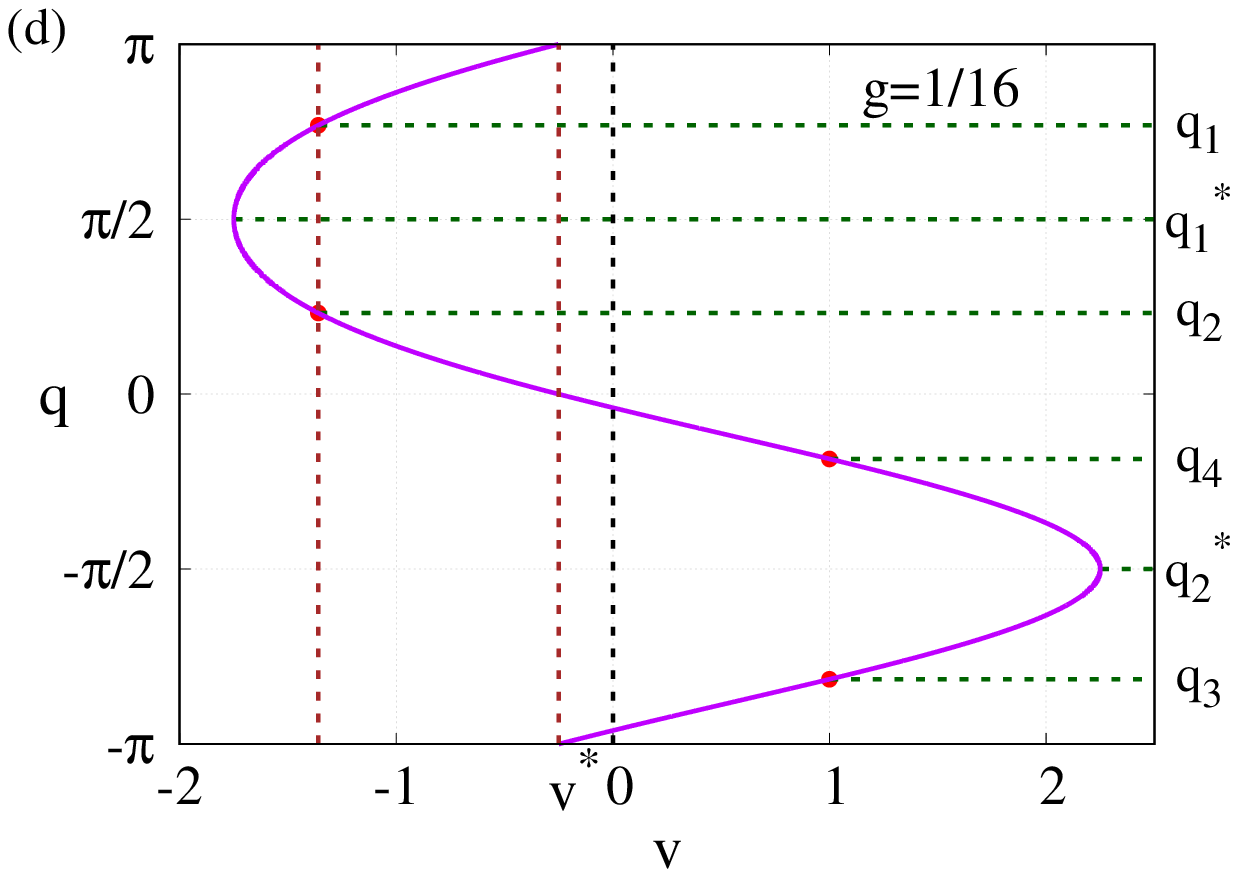}}\quad
  \subfigure{\label{fig:cum_prob_b}}{\includegraphics[width=5.6cm,height=4.6cm,keepaspectratio]{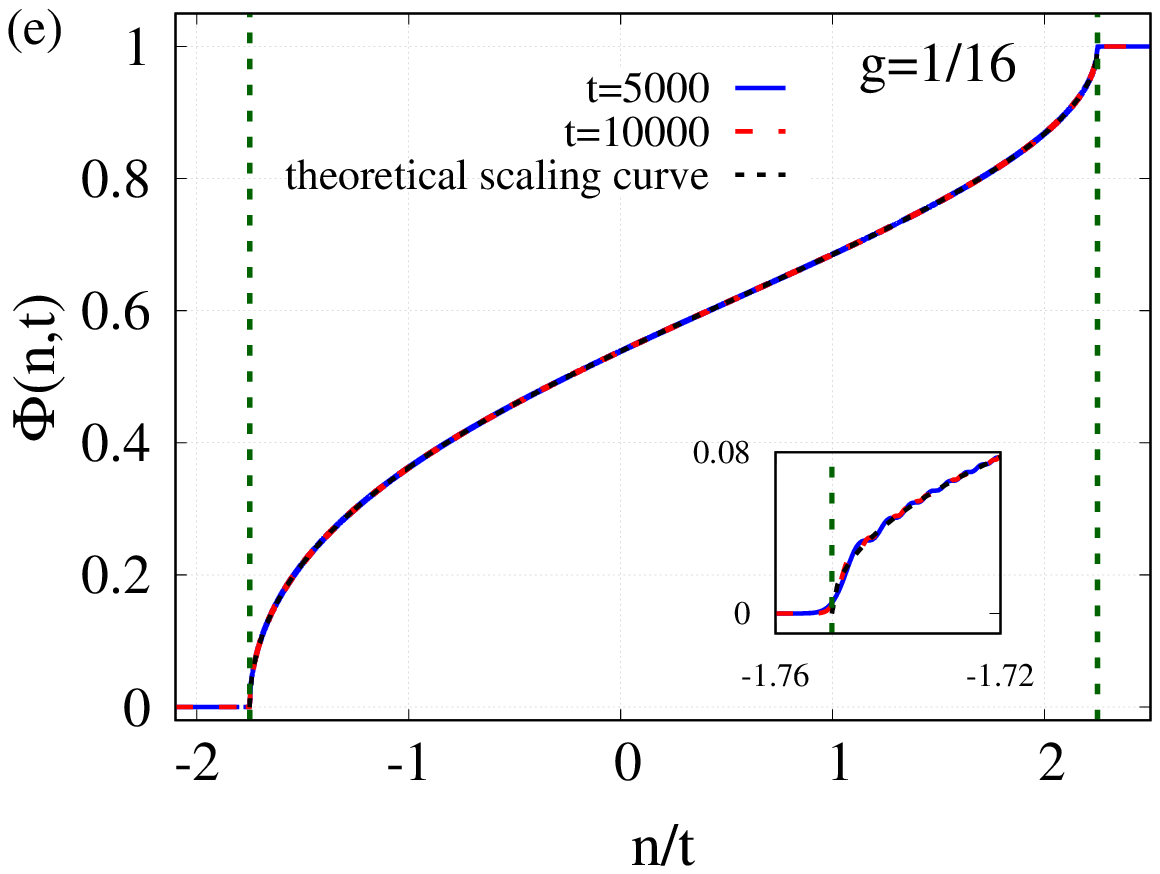}}\qquad
  \subfigure{\label{fig:cum_cur_b}}{\includegraphics[width=5.6cm,height=4.6cm,keepaspectratio]{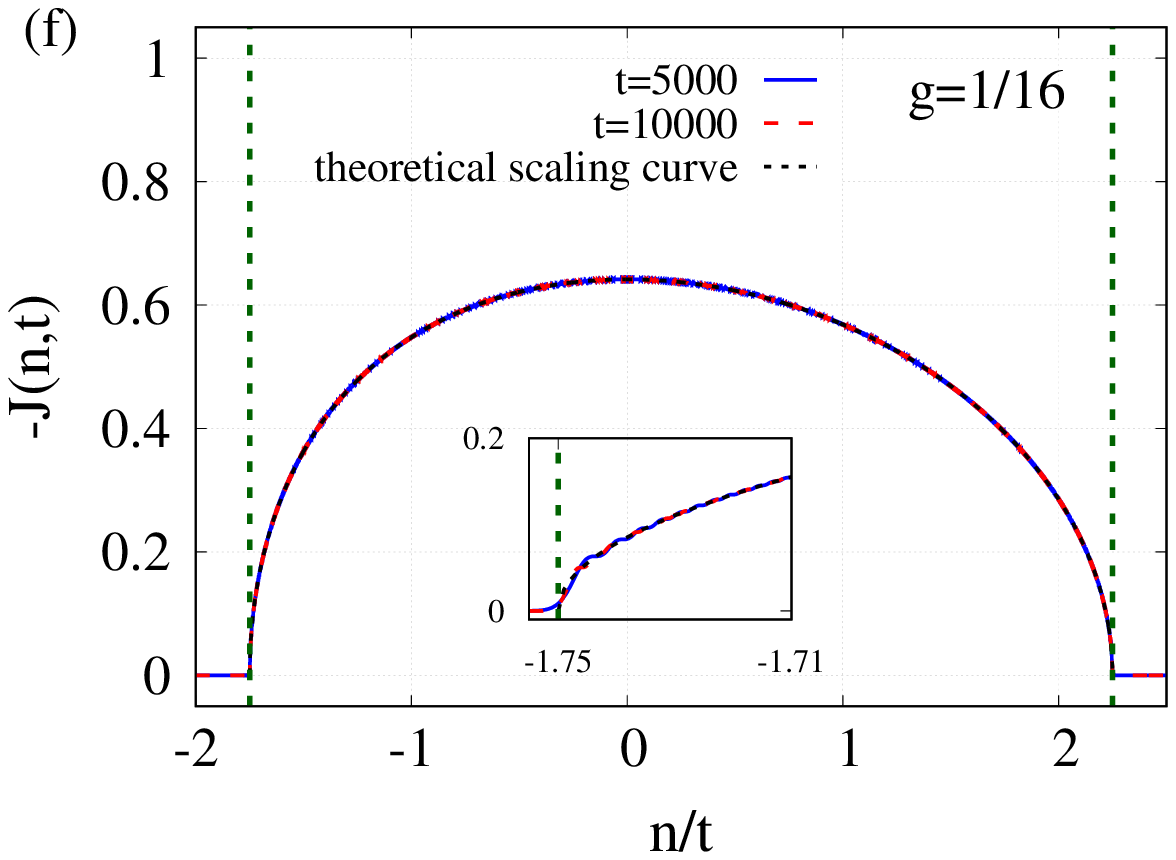}} \quad
  
  \subfigure{\label{fig:velocity_cum_prob_c}}{\includegraphics[width=5.6cm,height=4.6cm,keepaspectratio]{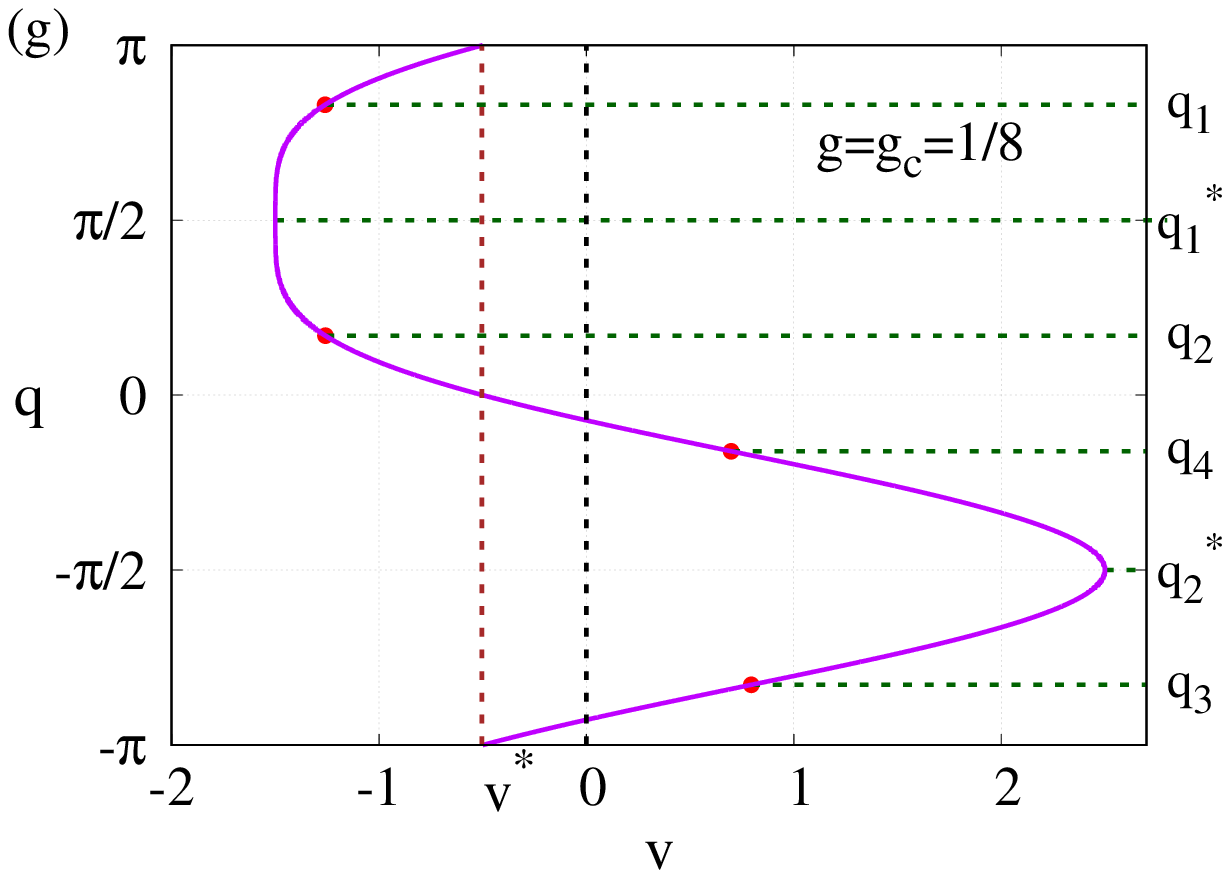}}\quad
 \subfigure{\label{fig:cum_prob_c}}{\includegraphics[width=5.6cm,height=4.6cm,keepaspectratio]{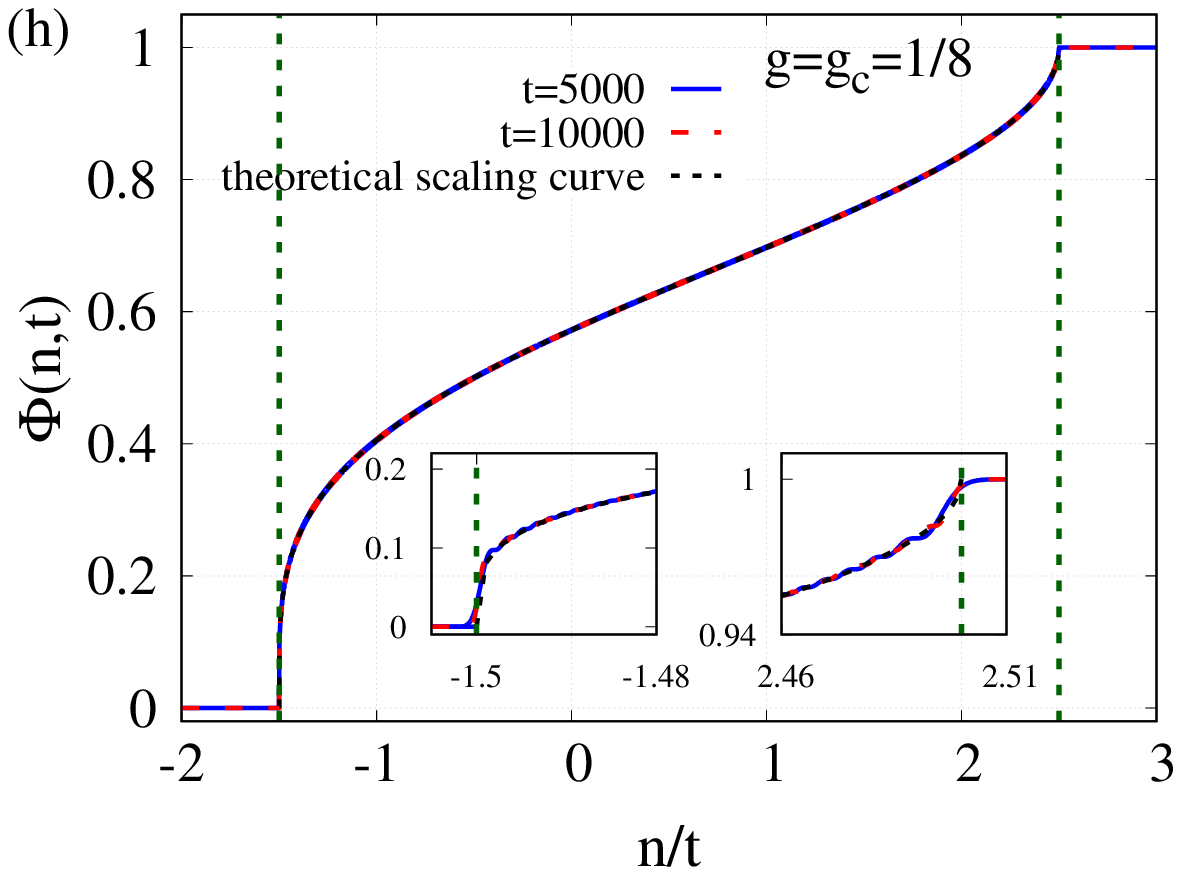}}\qquad
  \subfigure{\label{fig:cum_cur_c}}{\includegraphics[width=5.6cm,height=4.6cm,keepaspectratio]{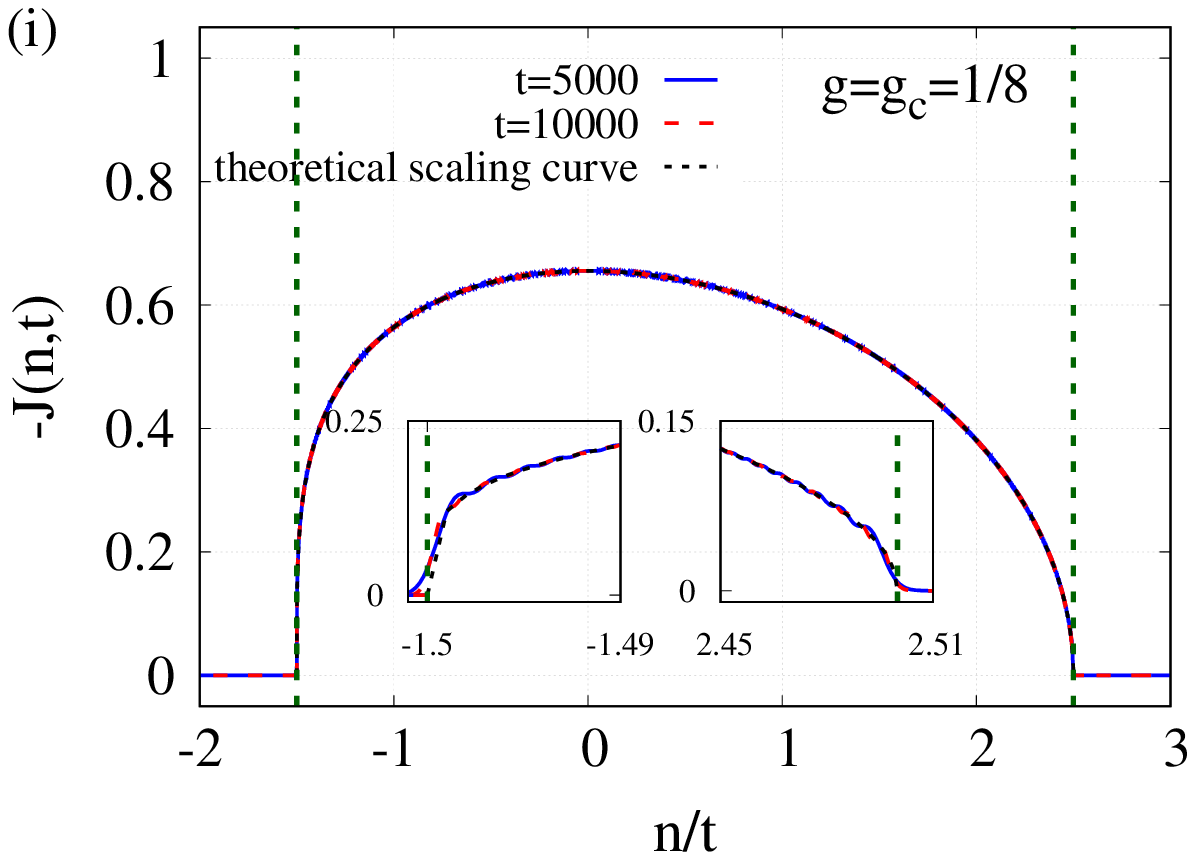}} \quad  
  
   \subfigure{\label{fig:velocity_cum_prob_d}}{\includegraphics[width=5.6cm,height=4.6cm,keepaspectratio]{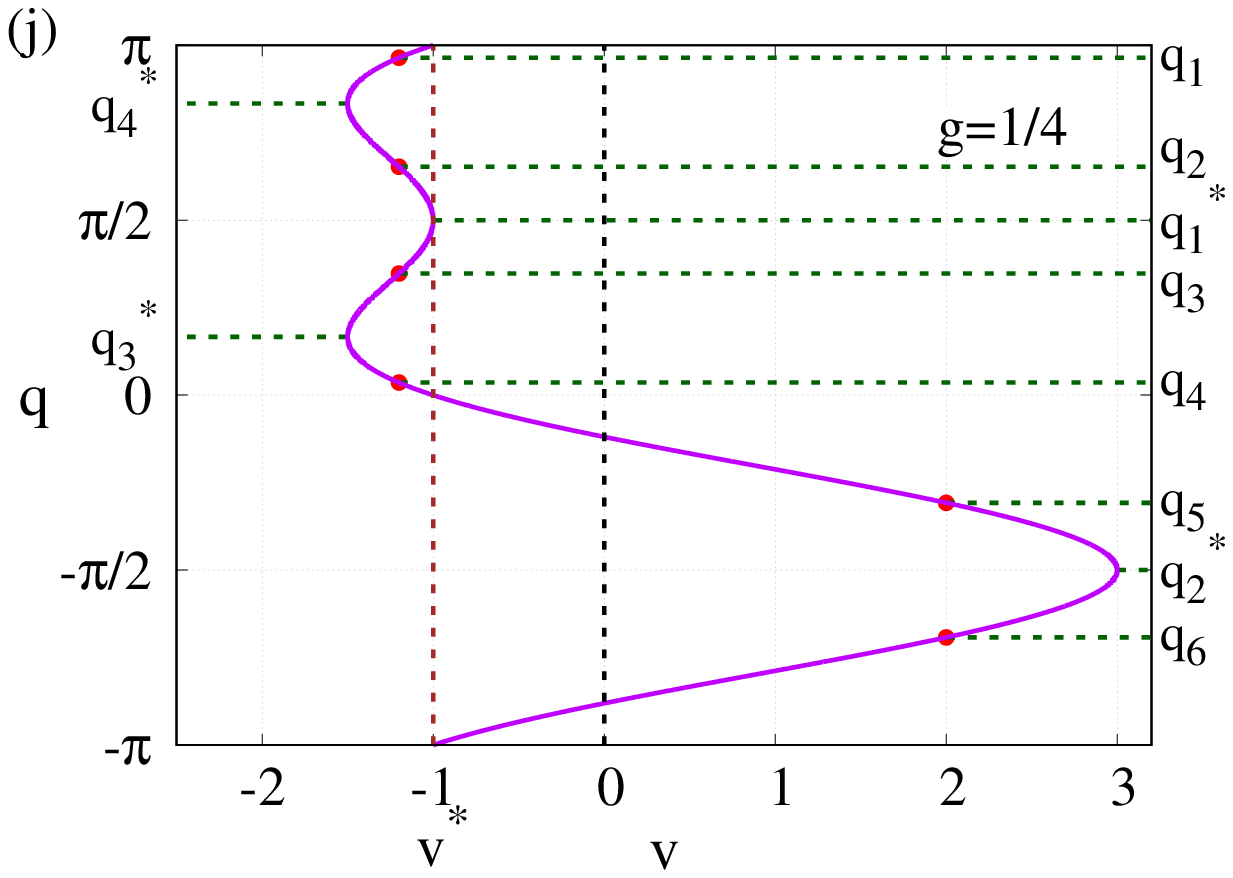}}\quad
 \subfigure{\label{fig:cum_prob_d}}{\includegraphics[width=5.6cm,height=4.6cm,keepaspectratio]{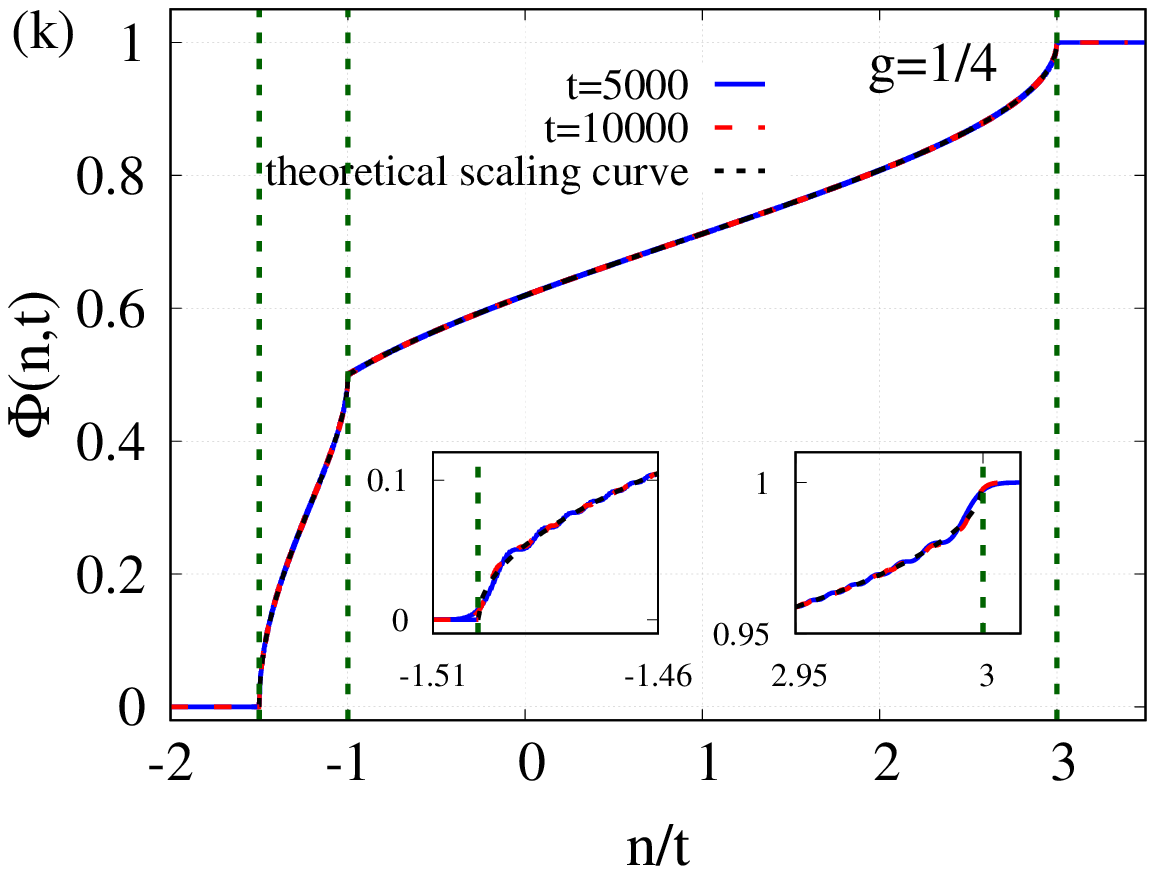}}\qquad
  \subfigure{\label{fig:cum_cur_d}}{\includegraphics[width=5.6cm,height=4.6cm,keepaspectratio]{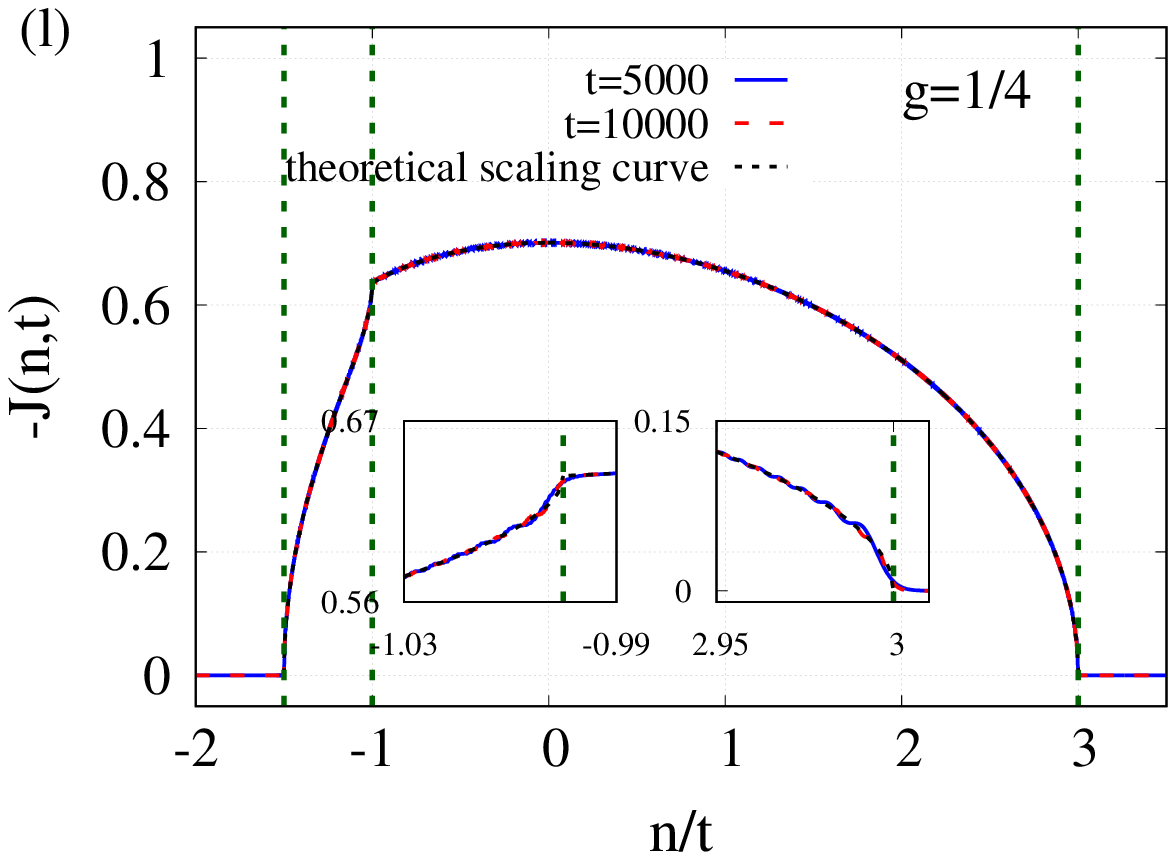}} 
\caption{First column shows the wave vector $q$ (in units $a^{-1}$) dependence on the velocity $v$ (in units of $ag_1$) for representative $g$-values for $\phi=\pi/2$. Second column shows the bulk scaling of the cumulative probability distribution function ($\Phi(n,t)=\Phi(n/t)$) for $\phi=\pi/2$ obtained numerically at times $t=5000, 10000$ (in units of $g_1^{-1}$). Theoretical scaling curve obtained using stationary phase approximation is also shown for comparison. The asymmetry in local distribution for ($g \neq 0$) is reflected in cumulative probabilities ($\Phi(-n,t)\neq 1-\Phi(n,t)$) (panels (e),(h),(k)). Panel (k) shows cumulative probability distribution for $g=1/4(>g_c)$ where the kink structure shows the emergence of additional internal front. Third column shows the bulk scaling for cumulative current density ($J(n,t)=J(n/t)$) for $\phi=\pi/2$ obtained numerically at times $t=5000,10000$ (in units of $g_1^{-1}$). Theoretical scaling curve obtained using stationary phase approximation are plotted for comparison. Panels (f),(i),(l) shows asymmetry in the cumulative current density for $\phi=\pi/2$. For $g=1/4(>g_c)$ (panel (l)), the kink structure shows the emergence of additional internal front. Insets in each plot in second and third column shows the deviation from bulk scaling near the extremal fronts. Vertical lines in second and third column shows the theoretical position of extremal fronts. }\label{fig:global_scaling_observables}
\end{figure*} 


 In this section, we show that at asymptotically long times and distances, the quantum walk can be described by a quasi-stationary state and provide a hydrodynamic description in terms of the local density of quasi-particle excitations.  We show, in particular, that the cumulative probability distribution satisfies an Euler-type hydrodynamic equation representing the large space-time ballistic propagation of the quasi-particle density and current.  Further, we show that the conservation law for the cumulative probability distribution is the lowest in the hierarchy of an infinite set of conservation laws satisfied by the scaled cumulative position moments, characterizing  the long time behaviour of the quantum walk.
 
      We begin by discussing the global scaling forms for the cumulative probability densities and current densities.  The cumulative probability density (CPD) $\Phi(n,t)$  and current density (CCD) $J(n,t)$ are defined as~\citep{Hemlata}:
\begin{equation}\label{eq:cumulative_prob}
\Phi(n,t)=\sum_{-\infty < m \leq n}^{} p(m,t) ; \quad J(n,t)=\sum_{-\infty < m \leq n}^{} j(m,t)
\end{equation}      
The local probability density $p(n,t) $ can be expressed as:
 \begin{equation}\label{eq:prob_density}
p(n,t) =|\psi(n,t)|^2 =\int\limits_{-\pi}^{\pi}\frac{dq}{2\pi}\int\limits_{-\pi}^{\pi}\frac{dp}{2\pi}e^{in(p-q)}e^{i(w(q)-w(p))t}
\end{equation}
Introducing new variables $q=K+Q/2$ and $p=K-Q/2$, we can Taylor expand $\omega(q) -\omega(p)$ around $Q=0$ to leading order at a zero-th order front as:
\begin{equation}
w(q)-w(p)=w(K+Q/2)-w(K-Q/2)= Q v(K)+ O(Q^3)
\end{equation}
and write the probability density (Eq.~\ref{eq:prob_density}) at asymptotically large times and for large $n$ as:
\begin{equation}\label{eq:prob_bulk_scaling}
p(n,t)=\int\limits_{-\pi}^{\pi}\frac{dK}{2\pi}\int\limits_{-\pi}^{\pi}\frac{dQ}{2\pi}e^{i(v(K)t-n)Q} = \int\limits_{-\pi}^{\pi}\frac{dK}{2\pi}\delta(n-v(K)t)
\end{equation}
 Using the 
continuity equation $\frac{\partial}{\partial \,t}p(n,t) + \frac{\partial}{\partial\, n} j(n,t) =0$, the local current density $j(n,t)$ can be expressed as: 
\begin{eqnarray}\label{eq:current_global_scaling}
j(n,t)&&=-\int\limits_{-\infty}^{n}dm \frac{\partial}{\partial t} p(m,t)\nonumber \\
&&=-\int\limits_{-\infty}^{n}dm \frac{\partial}{\partial t}\left(\int\limits_{-\pi}^{\pi}\frac{dq}{2 \pi}\delta(m-v(q)t)\right)\nonumber \\
&&=-\int\limits_{-\infty}^{n}dm \int\limits_{-\pi}^{\pi} \frac{dq}{2 \pi}v(q)\left[-\frac{\partial}{\partial m} \delta(m-v(q) t )\right]\nonumber \\
&&= \int\limits_{-\pi}^{\pi}\frac{dq}{2 \pi}v(q)\delta(n- v(q) t )
\end{eqnarray}

The CPD, $\Phi(n,t)$, and CCD, $J(n,t)$, at an ordinary front at site $n$, can be therefore obtained as: 
\begin{eqnarray}\label{eq: Global_cum_prob}
\Phi(n,t) && =\sum_{-\infty<m \leq n}^{} p(m,t) =\sum_{-\infty<m \leq n}^{} \int\limits_{-\pi}^{\pi}\frac{dq}{2\pi}\delta(m-v(q)t)\nonumber\\
&& = \int\limits_{-\pi}^{\pi}\frac{dq}{2\pi}\rho(n,q,t)
\end{eqnarray}
\begin{eqnarray}\label{eq: Global_cum_current}
J(n,t) &&=\sum_{-\infty<m \leq n}^{} j(m,t) =\sum_{-\infty<m \leq n}^{} \int\limits_{-\pi}^{\pi}\frac{dq}{2\pi} v(q) \delta(m-v(q)t)\nonumber\\
&& = \int\limits_{-\pi}^{\pi}\frac{dq}{2\pi} v(q) \rho(n,q,t)
\end{eqnarray}
where we have defined the local density of excitations, $\rho(n,q,t)$, as:
\begin{equation}
\rho(n,q,t) = \sum_{-\infty <m \leq n}^{} \delta (m-v(q)t )
\end{equation}

       The local density of excitations $\rho(n,q,t)$  can be obtained from the $q-v$ dependencies which we show graphically in the first column of Fig. \ref{fig:global_scaling_observables} for representative values of $g$. This allows us to obtain the  global scaling forms for the cumulative probability and current densities in the saddle point approximation limit ($v(q)=n/t$).
For $g \leq g_c$, the cumulative probability and current density take the global scaling form: 
\begin{equation}\label{eq: cum_prob_scaling_gless}
 \Phi(n,t) = \Phi\left(\frac{n}{t}\right)= \left \{
  \begin{aligned}
    &0;&& n \leq n_{lm}  \\
    &N_1\left(\frac{n}{t}\right) ; && n_{lm}<n<n^{*}\\
        &1-N_2\left(\frac{n}{t}\right) ; && n^*<n<n_{rm}\\
    &1; &&  n \geq n_{rm}
  \end{aligned} \right.
\end{equation}

 \begin{equation}\label{eq:cum_current_scaling_gless}
 J(n,t) = J\left(\frac{n}{t}\right)= \left \{
  \begin{aligned}
    &0;&& n <n_{lm}  \\
    &J_1(n/t) ; &&  n_{lm}<n< n_{rm}\\
    &0; &&  n \geq n_{rm}
  \end{aligned} \right.
\end{equation}

 In the above, we defined $n^* =v^* t$ as the site where $\Phi(n^*,t)=1/2$.
$N_{1(2)}(n/t)$ denotes the density of excitations at particular site $n$ for $n < n^*(>n_{lm})$ defined below. 
Also, as shown in Figs. \ref{fig:velocity_cum_prob_b},\ref{fig:velocity_cum_prob_c}, $q_1\, q_2, \,q_3, q_4$  lie on the respective branches:\\
 $q_1$: \, \, $v_{lm}(q)<v(q)<v^*(q^*),  \quad  \pi/2 \leq  q \leq \pi;$\\  
 $q_2$:  \,\, $ v^*(q^*)>v(q)> v_{lm},   \quad q^*\leq  q \leq  \pi/2;$\\
 $q_3$: \,\,  $ v_{rm}>v(q)> v^*(q^*),   \quad -\pi/2 \geq  q \geq  -\pi,$\\
 $q_4$:\, \, $ v^*(q^*)<v(q)<v_{rm},   \quad -\pi/2 \leq q \leq q^*.$ \\

For $g > g_c$, (say $g=1/4$ as shown in Fig.~\ref{fig:velocity_cum_prob_d}),  the bulk scaling forms for the cumulative probability and current density are obtained as:
\begin{equation}\label{eq: cum_prob_scaling_g_greater}
 \Phi(n,t) = \Phi\left(\frac{n}{t}\right)= \left \{
  \begin{aligned}
    &0;&& n \leq n_{lm}  \\
    &N_1 \left(\frac{n}{t}\right)+N_2 \left(\frac{n}{t}\right) ; && n_{lm}<n<n^{*}\\
        &1-N_3\left(\frac{n}{t}\right) ; && n^*<n<n_{rm}\\
    &1; &&  n \geq n_{rm}
  \end{aligned} \right.
\end{equation}
\begin{equation}\label{eq: cum_current_scaling_g_greater}
 J(n,t) = J\left(\frac{n}{t}\right)= \left \{
  \begin{aligned}
    &0;&& n \leq n_{lm}  \\
    &J_2(n/t); && n_{lm}<n<n^{*}\\
        & J_3(n/t); && n^*<n<n_{rm}\\
    &0; &&  n \geq n_{rm}
  \end{aligned} \right.
\end{equation}
where, as shown in Fig. ~\ref{fig:velocity_cum_prob_d},
\begin{eqnarray}
N_1 \left(\frac{n}{t}\right)&& =\frac{q_1-q_2}{2 \pi};\quad N_2 \left(\frac{n}{t}\right)=\frac{|q_3-q_4|}{2 \pi};\nonumber\\
N_3 \left(\frac{n}{t}\right)&& =\frac{|q_5-q_6|}{2 \pi}
\end{eqnarray}
and
\begin{eqnarray}
J_1(n/t)&& = \frac{1}{2 \pi} [\omega(q_1)-  \omega(q_2)] \\
J_2(n/t)&& =\frac{1}{2\pi}[\omega(q_1) - \omega(q_2)+\omega(q_3) -\omega(q_4)] 
\end{eqnarray}
and 
\begin{equation}
J_3(n/t)=\frac{1}{2\pi}[\omega(q_6)-\omega(q_5)]
\end{equation}
where $q_1, q_2, q_3, q_4, q_5,  \, \,q_6$ are  defined as points lying on the branches given below:\\
$q_1$:  $v_{lm}<v<v^*, \,\,q^{*}_4\leq q \leq \pi$ \\
$q_2$:  $v_{lm}<v<v^{*}, \,\,\pi/2 \leq q \leq q^{*}_4$ \\
$q_3$:  $v_{lm}<v<v^{*},\,\,   q^{*}_3\leq q \leq  \pi/2$\\
$q_4$:  $v_{lm}<v<v^{*}, \,\,  q^{*} \leq q \leq  q^{*}_3$\\
$q_5$:  $v^{*}<v<v_{rm},\,\,   -\pi/2 \leq q \leq  q^{*}$\\
$q_6$:  $v^{*}<v<v_{rm}, \,\,  -\pi \leq q \leq  -\pi/2$\\

In  second and third columns of Fig. {\ref{fig:global_scaling_observables}}, we plot the cumulative probability and current distributions obtained by exact numerical calculations and the scaling forms obtained above (Eqs.~\ref{eq: cum_prob_scaling_gless}, ~\ref{eq: cum_prob_scaling_g_greater}, ~\ref{eq:cum_current_scaling_gless} and ~\ref{eq: cum_current_scaling_g_greater}). It can be seen from the plots that there is good agreement between the analytic and numerical results.  The CPD and CCD are both  flat outside the causal cone structure showing that correlations exist only inside the 'allowed region'.  For $g=0$, the cumulative distributions are symmetric about the origin with  value $1/2$ at origin, while for non-zero $g$-values, the asymmetry seen in local probability and current distributions is also reflected in cumulative probability and current distribution profiles. Since the local probability densities are larger near the left extremal front, the site corresponding to cumulative probability value $1/2$ shifts to left.  For $g > g_c$,  an internal kink occurs in the CPD and CCD profiles  due to the presence of the additional extremal front corresponding to the extremal velocity $v_i$ (Figs. \ref{fig:cum_prob_d}, \ref{fig:cum_cur_d}). Unlike the case  for $\phi=0$,  here, because of the degeneracy of the extremal fronts as discussed in Sec.~\ref{sec:model},  we observe only one additional kink.  Also, unlike the case with a real component of NNN hopping \citep{Hemlata}, we do not observe any internal kink at $g_c$ because there are only two extremal fronts.  The inset in the figures show that the global scaling is violated near the extremal fronts. We will describe the  scaling behaviour at sites very close to the extremal fronts in the next section.

     Using the equation of motion for the  cumulative probability density :
\begin{equation}
\frac{\partial}{\partial t}\sum_{m \leq n}^{}p(m,t)= i [H,\sum_{m \leq n}^{}p(m,t)],
\end{equation}     
we can obtain at asymptotically long times and distances (and going to the continuum limit of the spatial lattice), the conservation law:

\begin{equation}\label{eq: Euler_eqn_M0}
\frac{ \partial}{\partial t} \Phi(n,t) +\frac{\partial}{\partial n} J(n,t)=0
\end{equation}

Substituting the global scaling equations Eq. \ref{eq: Global_cum_prob} and Eq. \ref{eq: Global_cum_current} into Eq. \ref{eq: Euler_eqn_M0} gives the local first order differential equation valid in the ballistic region:
\begin{equation}\label{eq: hydrodynamics_bulk_g_0}
\frac{\partial}{\partial t}{\rho(q,n,t)}+\frac{\partial}{\partial n} (v(q) \rho(q,n,t)) =0
\end{equation}
This is an Euler-type hydrodynamic equation representing the large space-time quantum ballistic propagation of the quasi-particle density and current.
  In fact, one can show that the conservation law for the CPD (Eq.~\ref{eq: Euler_eqn_M0}) is the lowest in an infinite hierarchy of conservation laws satisfied by  the scaled cumulative position moment distributions. The $k-th$, $k=0,1,2, \cdots$  position moment $\mu_k$ of the distribution is given as:
 \begin{eqnarray}
 \mu_k & = &  \sum_n \psi^*(n,t) n^k \psi(n,t) = \int\limits_{-\pi}^{\pi} \frac{dq}{2 \pi} \psi^*(q,t) (i \frac{d}{dq})^k \psi(q,t)\nonumber \\
 & = &  \int\limits_{-\pi} ^{\pi} \frac{dq}{2 \pi}\hat\psi^{*}(q,0)e^{it \omega(q)}(i^k \frac{d^k}{dq^{k}})e^{-it\omega(q)}\hat\psi(q,0)\nonumber \\
 &= & \int\limits_{-\pi} ^{\pi} \frac{dq}{2 \pi} e^{it \omega(q)}(i^k \frac{d^k}{dq^{k}})e^{-it\omega(q)}
 \label{eq:moment-def}
 \end{eqnarray}
  In the second equality of first line in the above equation, we have expressed the wave function in the momentum representation and in the last line of the above equation, we have used the fact that the state is initially localized at the origin. It is easy to see from Eq.~\ref{eq:moment-def} and the dispersion relation (Eq.~\ref{eq:dispersion}),  that the total  first moment $\mu_1$ vanishes for any $g$ and $\phi$.
The second, third and fourth moments can be computed to be:
\begin{eqnarray}\label{eq:moments_1_2}
\begin{aligned}
\mu_2 & = 2(1+4 g^{2})t^{2}; \\
\mu_3& =12gt^{3} \sin{\phi}\\
\mu_4&= 6 (1+16 g^2+16 g^4)t^4+2(1+16 g^2)t^2  \\
\end{aligned}
\end{eqnarray}
From Eq.~\ref{eq:moment-def}, we can also see that:
\begin{equation}
\frac{\mu_k}{t^k} = C_k + O\left(\frac{1}{t^2}\right)
\end{equation}
where $C_k$ is a constant independent of $t$. Therefore, at asymptotically long times, $\mu_k/t^k \sim C_k$.

\begin{figure}[htp!]
  \subfigure{\includegraphics[width=7cm,height=5cm,keepaspectratio]{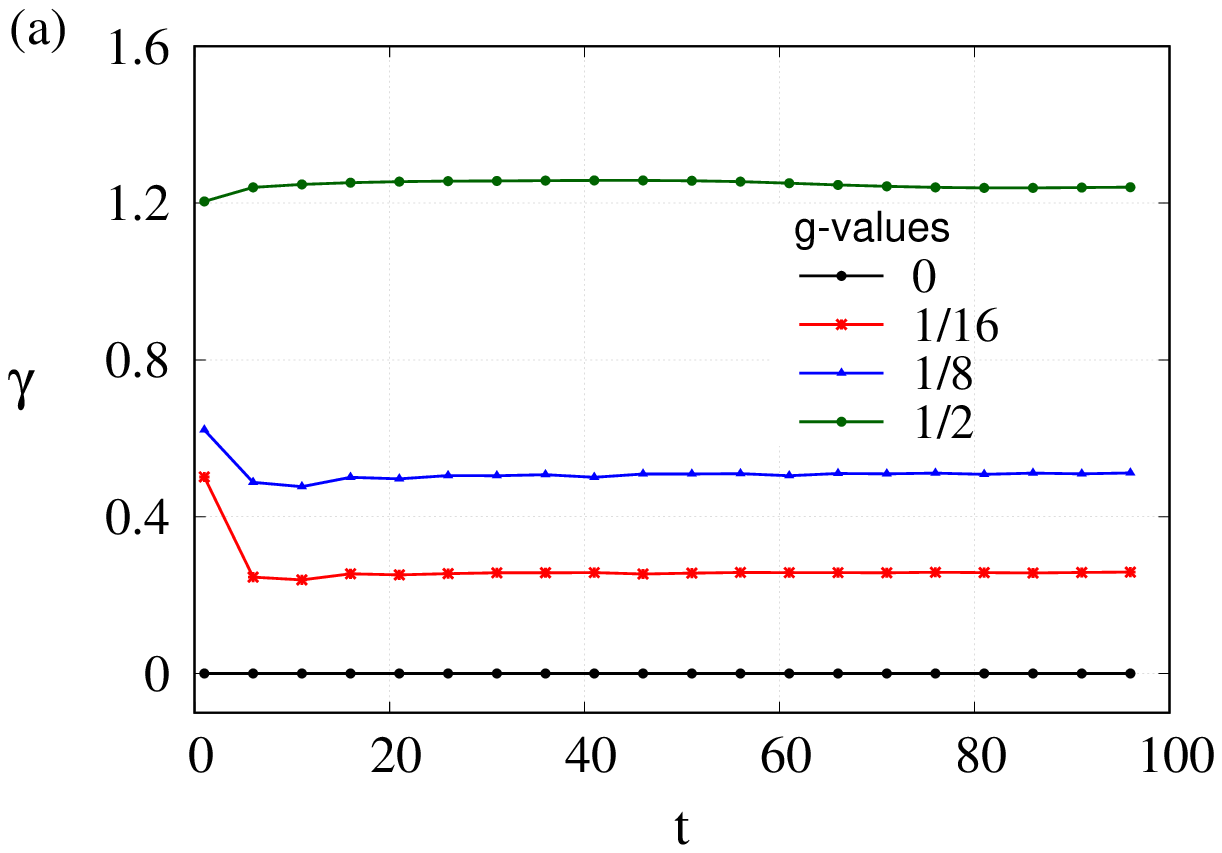}}
  
  \subfigure{\includegraphics[width=7cm,height=5cm,keepaspectratio]{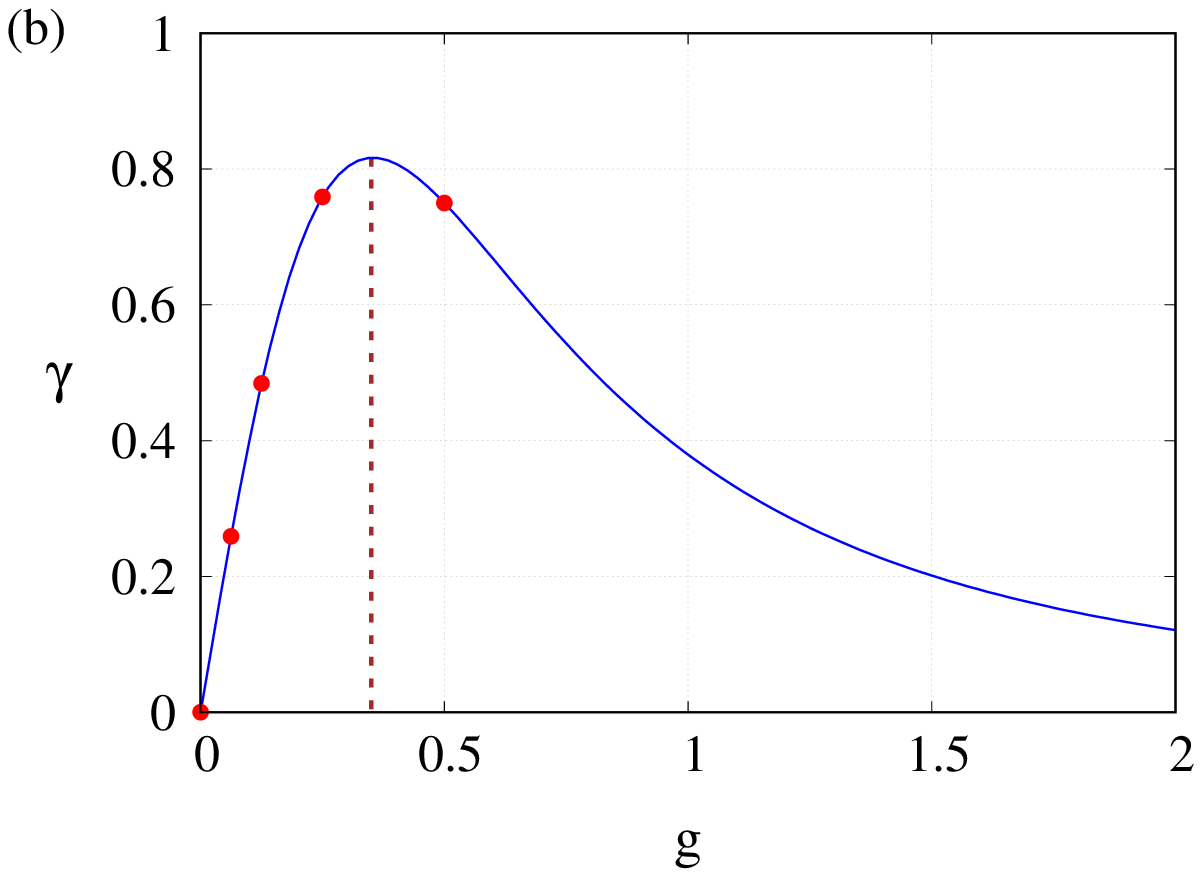}}
   \caption{The plot showing exact numerical results for the nature of skewness for $\phi=\pi/2$. (a) as a function of time for different $g$-values. Skewness remains constant over time for all the $g$-values as expected from Eq. \ref{eq:skew_eqn}. (b) as a function of $g$. The continuous curve corresponds to theoretical curve (Eq. \ref{eq:skew_eqn}) and points on it represent the numerically obtained values for discrete $g$-values at time $t=10000$ (in units of $g_1^{-1}$). Positive value of $\gamma$ indicates distribution drift to right showing maximum at $g=0.35$.}
\label{fig:skewness} 
\end{figure}

 A natural measure to characterize the asymmetry of a distribution is the skewness which is defined as: \citep{SciPost_Phys_6}
\begin{equation}\label{eq:skew_eqn}
\gamma=\frac{ \mu_3}{\mu_2^{3/2}} = \frac{3\sqrt{2} \,g \, \sin \phi}{(1+4 g^{2})^{3/2}}
\end{equation}
From Eqns.~\ref{eq:moments_1_2} and ~\ref{eq:skew_eqn}, we can see that the skewness $\gamma$ vanishes for any real NNN hopping ($\phi=0$),  indicating the symmetry of the distribution about the origin. It can also be observed that the skewness  takes its maximum value at $\phi =\pi/2$ , i.e., for a purely imaginary NNN hopping.  In general, a positive value of $\gamma$ shows drift of the distribution to the right of the origin whereas a negative value shows  a left drift in the distribution.
We show the time dependence and $g$ dependence of $\gamma$ in Fig.~\ref{fig:skewness}.  From Eq. \ref{eq:skew_eqn}, we can see that $\gamma$ is time independent for any $g$-value which is also observed from numerical results shown in Fig. \ref{fig:skewness}(a).  The theoretically predicted and numerically calculated $g$-dependence of $\gamma$ is shown  Fig.~\ref{fig:skewness}(b). From Eq.~\ref{eq:skew_eqn}, we can see that it has a linear $g$ dependence for  small $g, g<<g_c$ while for large $g$-values, it  exhibits a  power law decay $ \gamma \approx 1 /2 g^2$, which again agrees well with our numerical results. 
 

\begin{figure*}[htp]           
  \centering
  \subfigure{\label{fig:cum_first_moment_phi_0}}{\includegraphics[width=15cm,height=5cm,keepaspectratio]{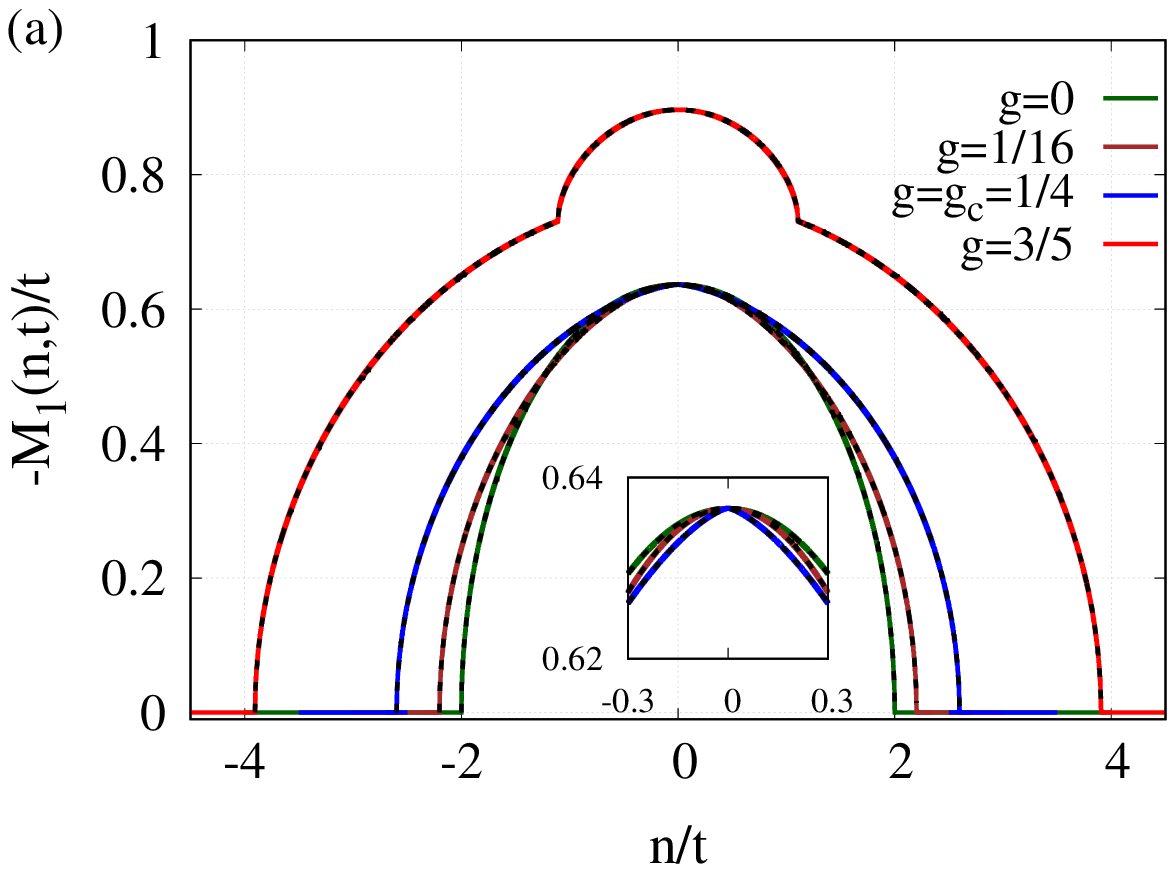}}\quad
\subfigure{\label{fig:cum_first_moment_phi_pi_by_2}}{\includegraphics[width=15cm,height=5cm,keepaspectratio]{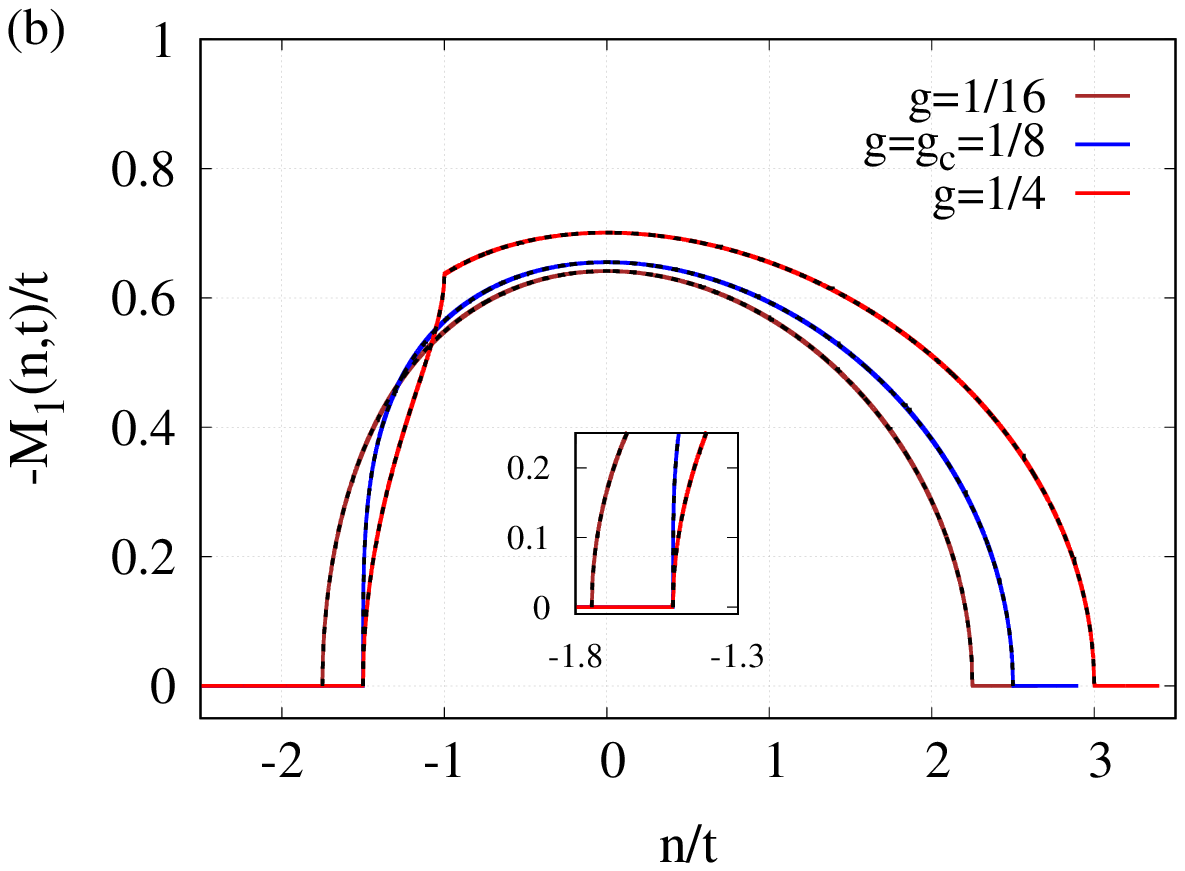}}\\

\subfigure{\label{fig:cum_second_moment_phi_0}}{\includegraphics[width=15cm,height=5cm,keepaspectratio]{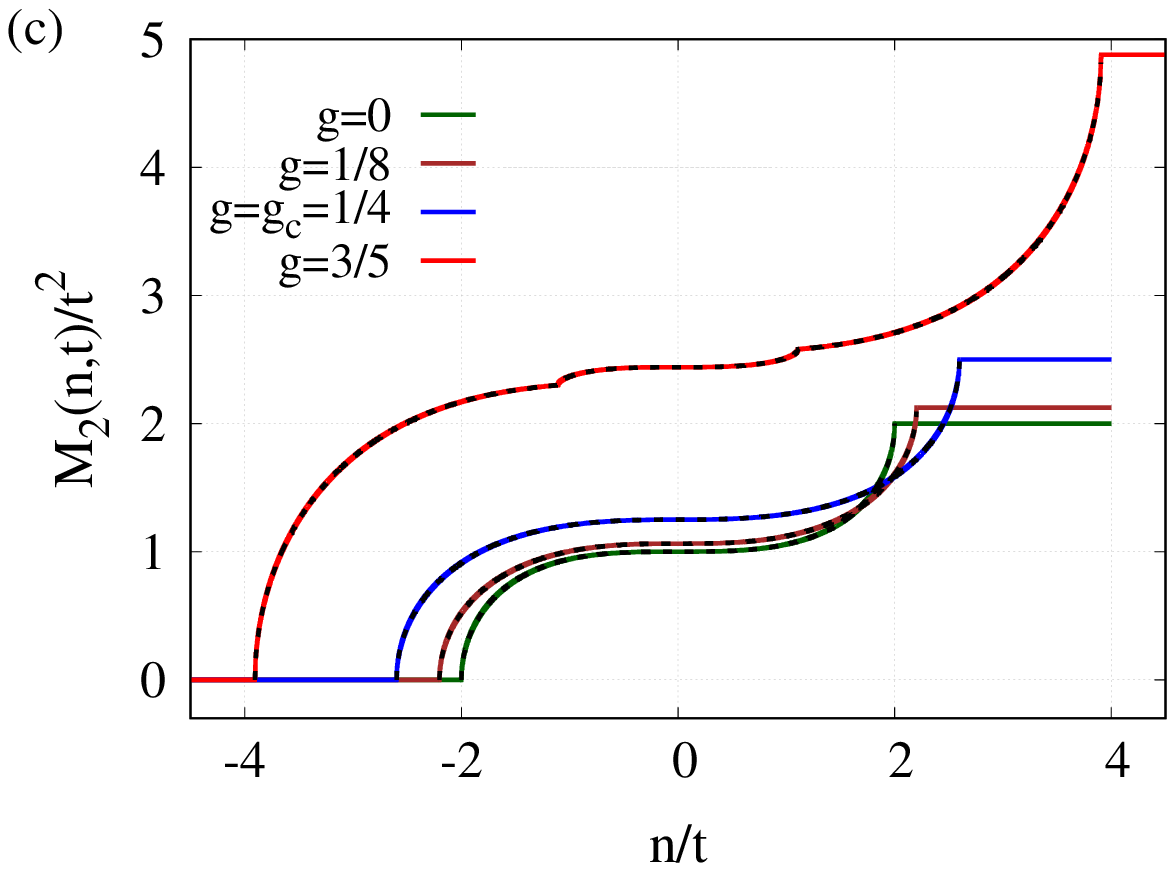}}\qquad 
\subfigure{\label{fig:cum_second_moment_phi_pi_by_2}}{\includegraphics[width=15cm,height=5cm,keepaspectratio]{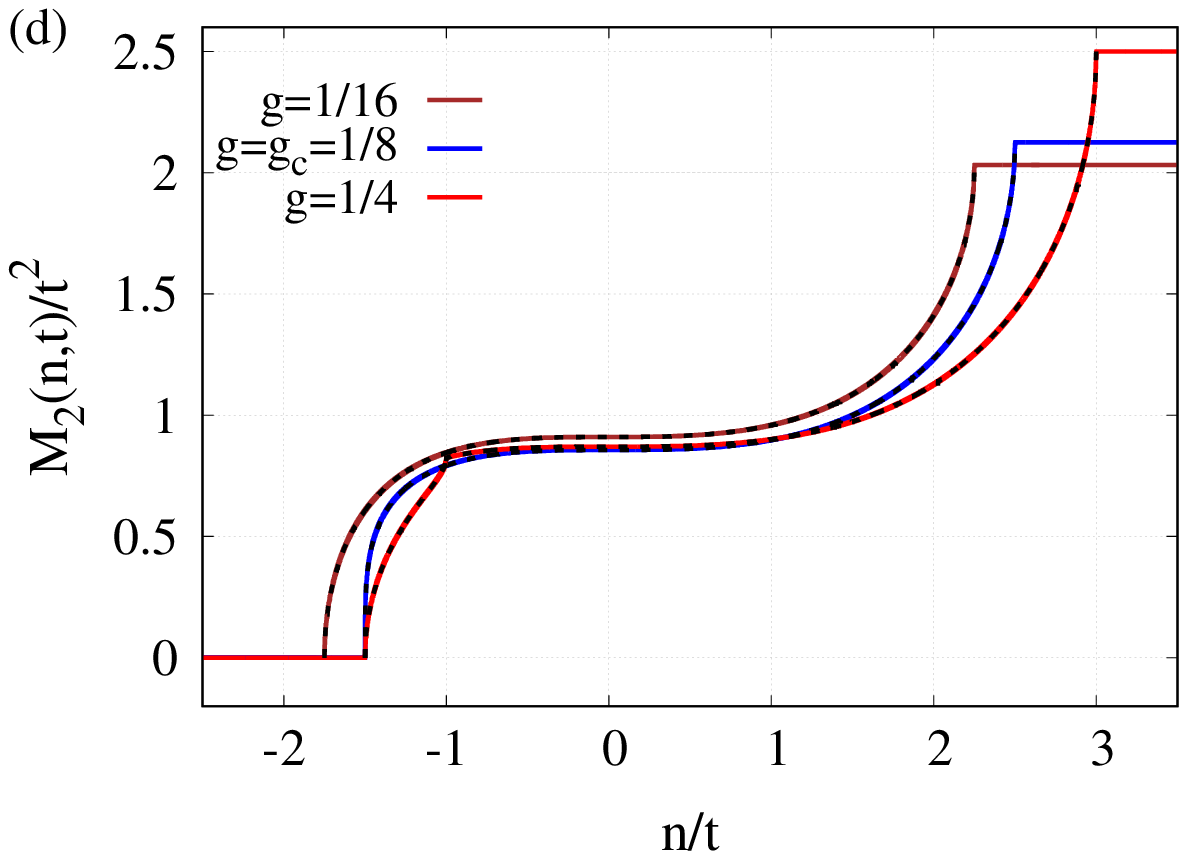}}

\subfigure{\label{fig:cum_third_moment_phi_0}}{\includegraphics[width=15cm,height=5cm,keepaspectratio]{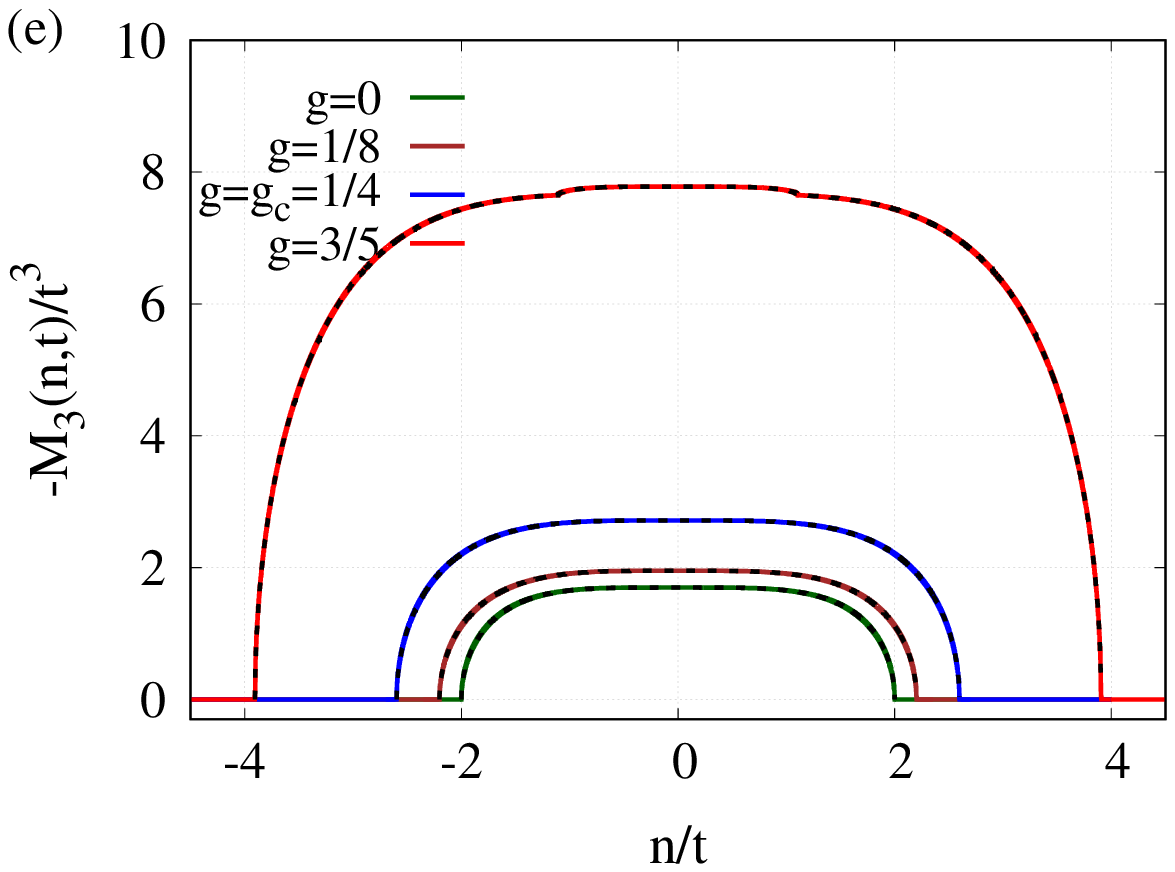}}\qquad 
\subfigure{\label{fig:cum_third_moment_phi_pi_by_2}}{\includegraphics[width=15cm,height=5cm,keepaspectratio]{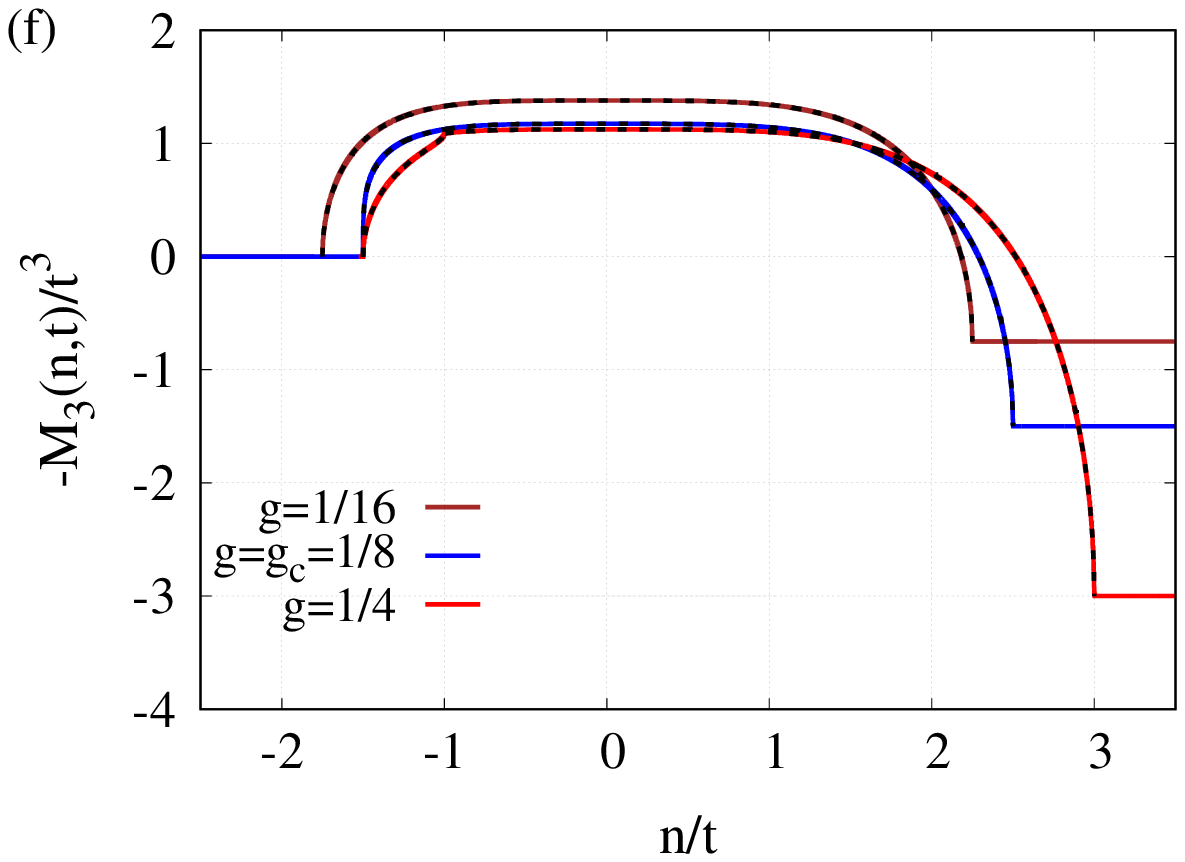}}
\caption{ Cumulative scaled position moments for $\phi=0$ (first column) and $\phi=\pi/2$ (second column) for representative $g$-values obtained numerically at $t=10000$ (measured in units of $g_1^{-1}$). The analytical solutions obtained using saddle point approximation are plotted (black dotted lines) for comparison with numerics which shows reasonable agreement in both. Panel (a) and (b) shows the first scaled cumulative position moments for $\phi=0$ and $\phi=\pi/2$ for different $g$-values. For $\phi=0$, profiles are symmetric about the origin whereas we observe asymmetry for $\phi=\pi/2$. The emergence of additional extremal fronts can be seen from the kink structure observed for $g \geq g_c$ for $\phi=0$ and for $g>g_c$ for $\phi=\pi/2$. The insets in the plots shows the change in the curvature near the second order (panel (a)) and third order (panel (b)) fronts at the respective critical NNN hopping strengths $g_c$. In both the cases total cumulative first moment  i.e position expectation value is zero. Panel (c) and (d) shows the second scaled cumulative position moments for $\phi=0$ and $\phi=\pi/2$ for different $g$-values. We see the asymmetry for imaginary NNN hopping strength (panel (d)) as compared to the real NNN hopping strength (panel (c)). The emergence of additional extremal fronts can be seen from the kink structure observed for $g > g_c$ in both cases. The second cumulative moment for both the cases increases quadratically with $g$ (Eq. \ref{eq:moments_1_2}) and saturates at the same value for respective $g$-values for both $\phi=0$ and $\phi=\pi/2$ cases. Panel (e) and (f) shows the third scaled cumulative position moments for $\phi=0$ and $\phi=\pi/2$ for different $g$-values. Since the third position moment is zero for $\phi=0$ we see symmetric plots which sum up to zero whereas for $\phi=\pi/2$ case we see asymmetry with non-zero third position moment (Eq. \ref{eq:moments_1_2}). The additional extremal fronts for $g>g_c$ for both the cases can be seen from the kink structures emerging when $g>g_c$.}\label{fig:cum_moment} 
\end{figure*} 

    The corresponding cumulative moments $M_k(n,t)$ are defined as 
   \begin{equation}
M_k(n,t)=\sum_{-\infty < m \leq n}^{}m^k p(m,t)
\end{equation} 
 The cumulative first position moment is then given as
\begin{equation}
M_1(n,t)=\sum_{-\infty <m \leq n}^{}m  p(m,t)
\end{equation} 
At asymptotically long times,  using Eq.~\ref{eq:prob_bulk_scaling}  for the bulk scaling form for the probability,  we can obtain a global scaling for $M_1(n,t)$ as
\begin{eqnarray}\label{m1_longtime}
M_1(n,t)=M_1(n/t) &=&\sum_{-\infty < m \leq n}^{}m \left(\int\limits_{-\pi}^{\pi}\frac{dq}{2 \pi}\delta(m-v(q)t)\right)\nonumber \\
&=&\int\limits_{-\pi}^{\pi}\frac{dq}{2 \pi}\sum_{-\infty < m \leq n}^{}m \,\delta(m-v(q)t)\nonumber \\
&=& \int\limits_{-\pi}^{\pi}\frac{dq}{2 \pi}\sum_{-\infty < m \leq n}^{}v(q) t  \,\delta(m-v(q)t)\nonumber \\
&=& \int\limits_{-\pi}^{\pi}\frac{dq}{2 \pi} v(q) t  \,\rho(n, q,t) \nonumber \\
&=& t J(n,t)
\end{eqnarray}
Similarly, we find that the $k-th$ cumulative  moment satisfies the global scaling relation:
\begin{equation}\label{eq:Euler_eqn_Mk}
M_k(n,t)=M_{k}\left(\frac{n}{t}\right)=t^k\int\limits_{-\pi}^{\pi}\frac{dq}{2\pi}v^k(q)\rho(n,q,t)\\
\end{equation}
Thus, we obtain at asymptotically long times, global scaling forms for  the cumulative position moments in terms of the local density of excitations.  Defining ${\widetilde M}_k = \frac{M_k}{t^k}$,  we can then see by  using  Eq.~\ref{eq: hydrodynamics_bulk_g_0}, that at large times and distances, the  cumulative moments ${\widetilde M}_k$ satisfy continuity equations of the form:
\begin{equation}
\frac{\partial}{\partial t}{\widetilde M}_k+\frac{\partial }{\partial n}{\widetilde M}_{k+1}=0; \qquad k=0,1,2, 3 \cdots
\end{equation} 
which constitute an infinite set of conservation laws. The conservation law for the CPD (Eq.~\ref{eq: Euler_eqn_M0}) is the lowest ($k=0$) in the set.
We show in  Fig. \ref{fig:cum_moment}, the  plots for the cumulative first, second and third  position moment  obtained from the theoretical (Eq.~\ref{eq:Euler_eqn_Mk}) and exact numerical calculations for real NNN hopping ($\phi=0$) and imaginary NNN hopping, $\phi=\pi/2$.  It can be seen from the plots that the numerical results match well with the theoretical computations and show the global scaling behaviour of the moments. The plots are symmetric about the origin for $\phi =0$ whereas they are asymmetric for $\phi=\pi/2$.  We also observe that beyond the critical value of the NNN coupling strength, a kink structure emerges within the allowed region characterizing the change in number of extremal fronts. The first moment saturates to a zero value for all $g$ values and any phase since the total current is zero. The total cumulative second moment saturates at the same value for respective $g$-values for both $\phi=0$ and $\phi=\pi/2$ which agrees with Eq.~\ref{eq:moments_1_2} that the total second position moment ($\mu_2=2(1+4g^2)t^2$) is independent of $\phi$.  The third moment saturates to a zero value for $\phi=0$ whereas for $\phi=\pi/2$,  it increases with $g$.  This again agrees with the result of  Eq.~\ref{eq:moments_1_2} for the third moment.

\section{\label{sec:local}{Anamolous scaling and staircase structure for cumulative probability near extremal fronts}}

  We observe that extremal front edges in the CPD and CCD deviate from global scaling (shown in the inset of Figs. \ref{fig:global_scaling_observables}).  In this section, we analyze in a manner similar to that in Ref.~\citep{Hemlata} and show that  at asymptotically long times and distances, the deviations of the CPD and CCD  exhibit anomalous sub-diffusive scaling behaviour at the extremal front edges. We also compare the analytic results with exact numerical results. 
 The deviation of cumulative probability and the current density from the value at an extremal front at site $n_e$  is defined as~\citep{Hemlata}:
\begin{equation}
\delta \Phi(n,t)\equiv \Phi(n_e,t)-\Phi(n,t);\quad \delta J(n,t) \equiv J(n_e,t)-J(n,t)
\end{equation} 

\begin{figure*}[htp]           
  \centering
  \subfigure{\label{fig:stair_cum_prob_phi_a}}{\includegraphics[width=9cm,height=4.3cm,keepaspectratio]{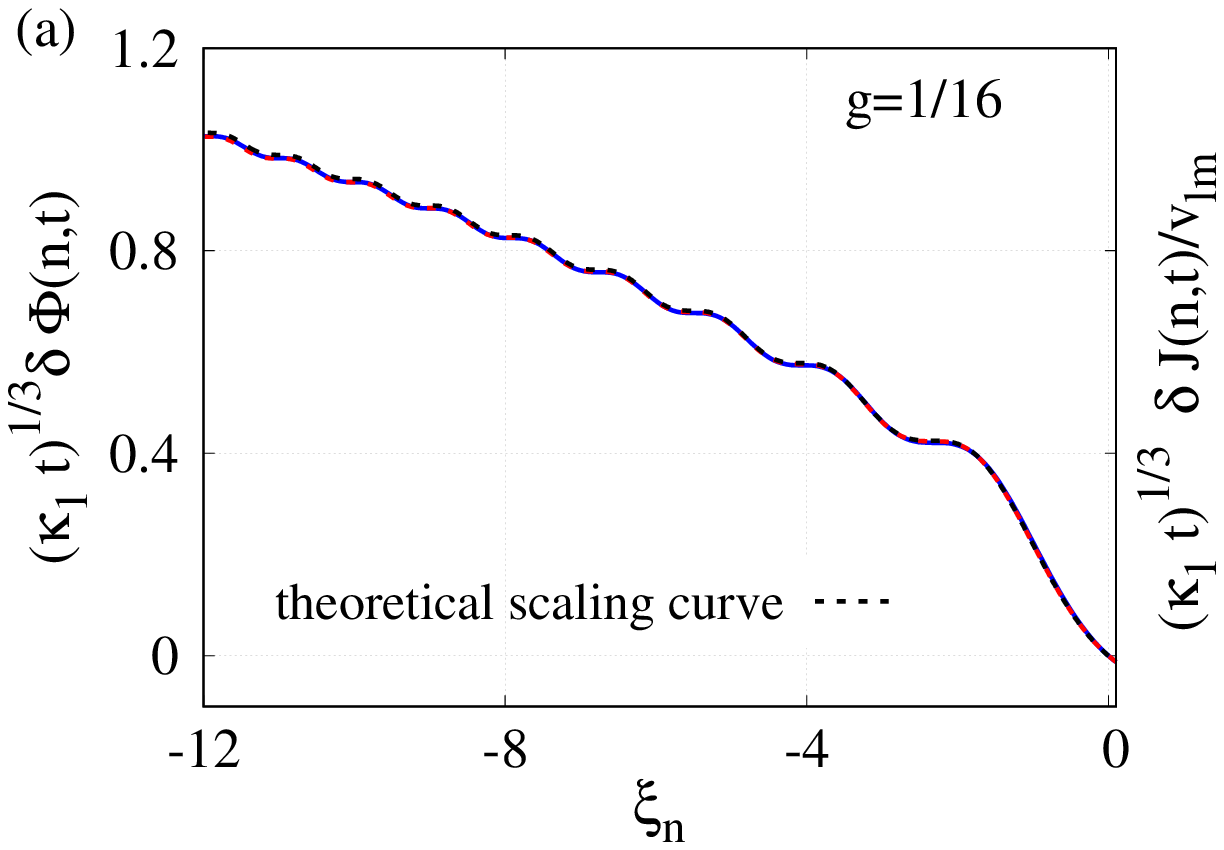}}\qquad 
\subfigure{}{}\quad

\subfigure{\label{fig:stair_cum_prob_phi_b}}{\includegraphics[width=9cm,height=4.3cm,keepaspectratio]{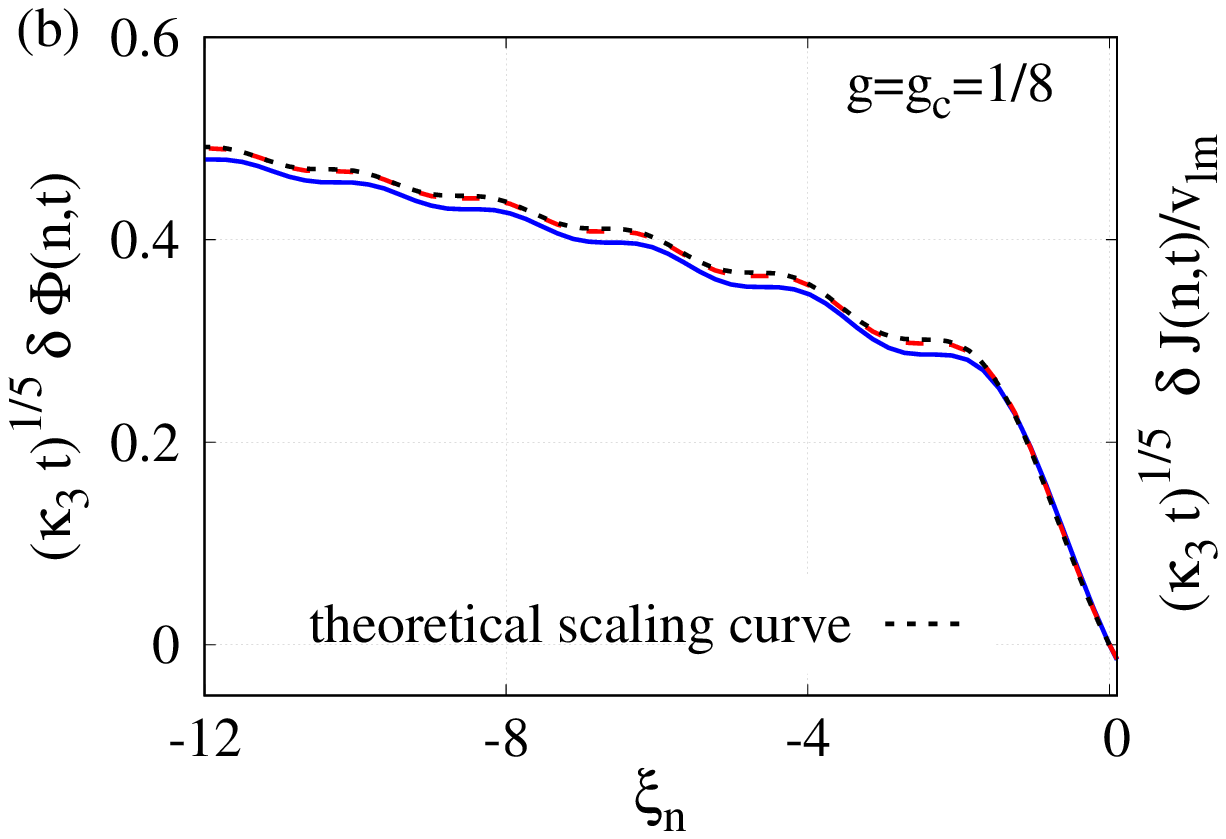}}\qquad
  \subfigure{\label{fig:stair_cum_prob_phi_c}}{\includegraphics[width=9cm,height=4.3cm,keepaspectratio]{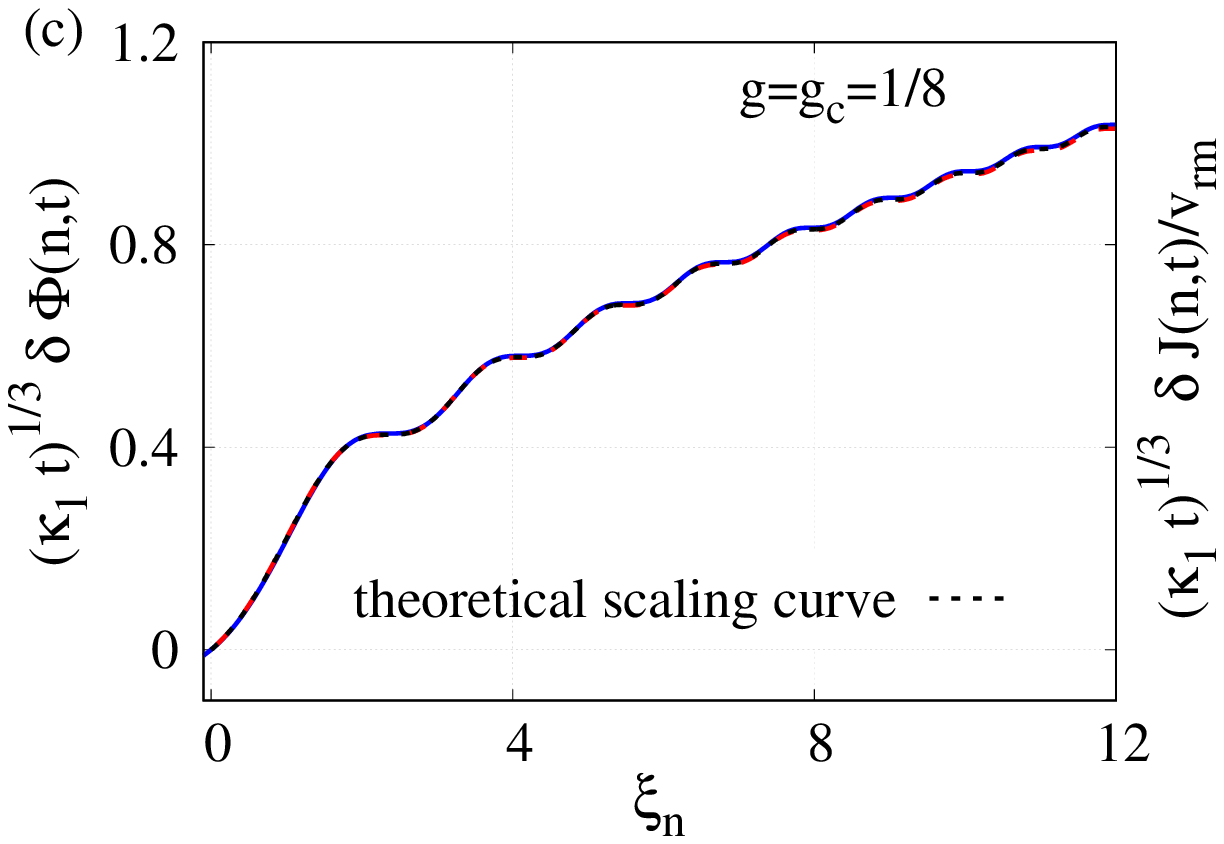}}\quad

  \subfigure{\label{fig:stair_cum_prob_phi_d}}{\includegraphics[width=9cm,height=4.3cm,keepaspectratio]{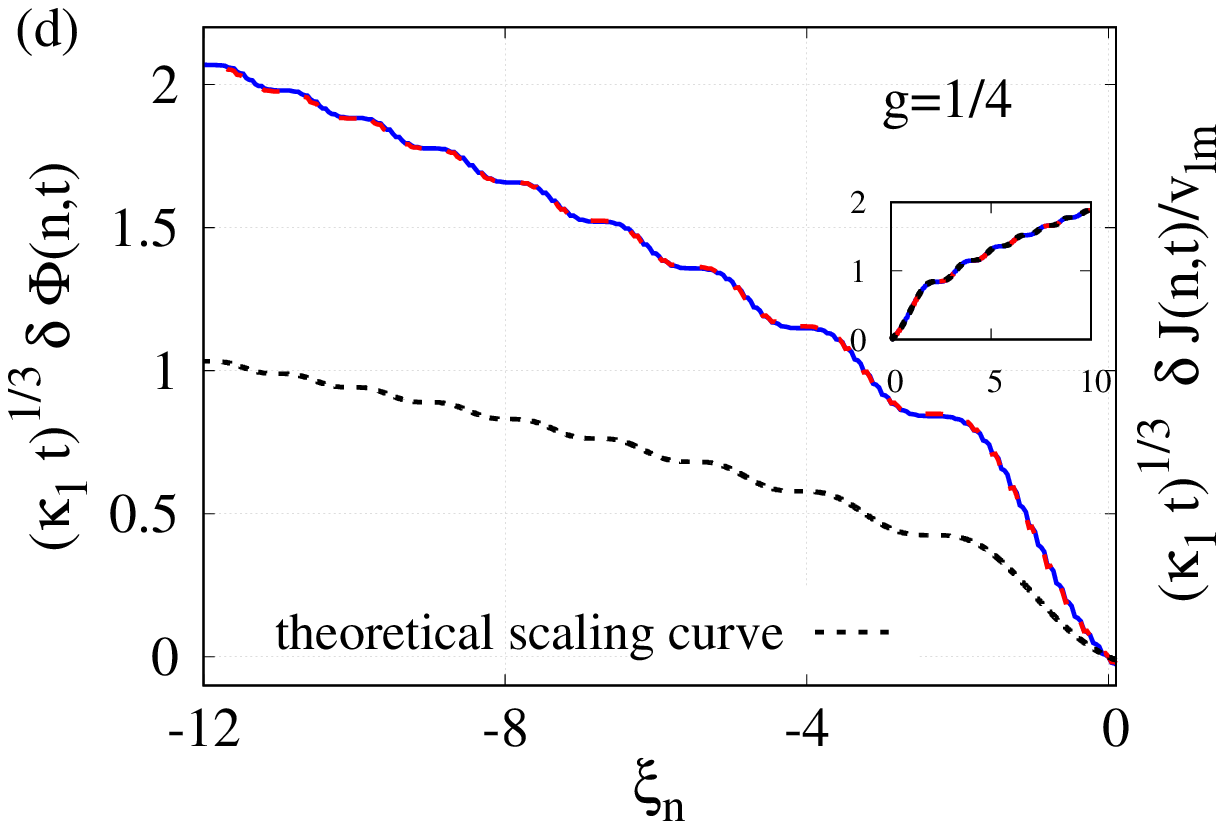}}\qquad
  \subfigure{\label{fig:stair_cum_prob_phi_e}}{\includegraphics[width=9cm,height=4.3cm,keepaspectratio]{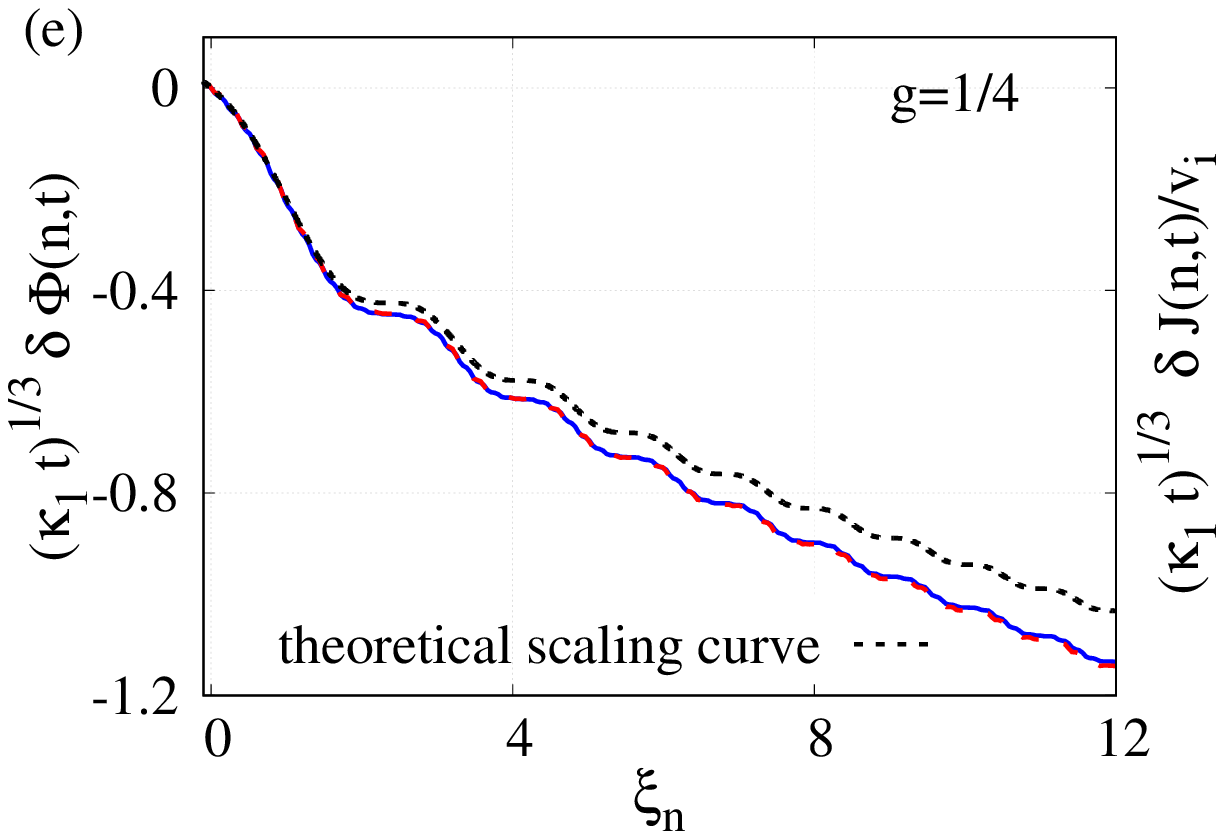}}\quad

\caption{ Local extremal front scaling for cumulative probability and cumulative current density profiles obtained from exact numerical calculations for $\phi=\pi/2$ at time $t=10000$ (in units of $g_1^{-1}$). The internal staircase structure emerging at the fronts is shown for different $g$-values and is compared with the analytical result obtained using saddle-point method. Panel (a) shows the internal staircase structure at the left extremal front for $g<g_c$ described by Airy scaling where distances from the front scale as $(n_{lm}-n)/n_{lm}^{1/3}$, cumulative probabilty scale as $n_{lm}^{1/3}\delta \Phi$ and cumulative current scale as $n_{lm}^{1/3}\delta J=v_{lm}n_{lm}^{1/3}\delta \Phi$. For $g=g_c$ we observe similar kind of Airy scaling near right extremal front as shown in Panel (c). Panel (b) shows the staircase structure near the third order extremal front (left) for critical NNN hopping strength $g=g_c$ where distances from the front scale as $(n_{lm}-n)/n_{lm}^{1/5}$, cumulative probabilty scale as $n_{lm}^{1/5}\delta \Phi$ and cumulative current scale as $n_{lm}^{1/5}\delta J=v_{lm}n_{lm}^{1/5}\delta \Phi$. Panel (d) shows local scaling for two degenerate left moving fronts observed for $g>g_c$. Both the fronts shows Airy scaling where distances from the front scale as $(n_{lm}-n)/n_{lm}^{1/3}$, cumulative probabilty scale as $n_{lm}^{1/3}\delta \Phi$ and cumulative current scale as $n_{lm}^{1/3}\delta J=v_{lm}n_{lm}^{1/3}\delta \Phi$. Inset shows the degeneracy of fronts where we plot twice the analytical curve which is in agreement with numerically obtained results. Heights of the steps obtained here are twice compared to the case where we have single first order front. Panel (e) shows local scaling near the internal front for $g>g_c$ where distances from the front scale as $(n_{i}-n)/n_{i}^{1/3}$, cumulative probabilty scale as $n_{i}^{1/3}\delta \Phi$ and cumulative current scale as $n_{i}^{1/3}\delta J=v_{i}n_{i}^{1/3}\delta \Phi$.}\label{fig:stair_cum_prob_phi_pi_by_2}
\end{figure*} 


         We demonstrate the local scaling behaviour of the probability and current densities by evaluating the wave function at a site very close to the extremal front  by performing the Fourier integral in Eq.~\ref{eq:fourier_integral} using the stationary phase approximation. 
Near a $k$-th order extremal front, we can expand $\omega(q)$ to leading order as :
\begin{equation}
\omega(q) \approx  \omega(q_e) + (q-q_e) v_e + \frac{(q-q_e)^{k+2}}{(k+2)!}  \omega ^{(k+2)} (q_e)
\end{equation}         
Then, we can expand $ \varphi(n,q) $  in Eq.~\ref{eq:saddle-point} for $n$ near $n_e =v_e t$  as:
\begin{equation}
\varphi (n,q)  \approx   (\frac{n}{t} q_e -\omega(q_e) )  + (\frac{n}{t}- v_e )(q-q_e) - \frac{(q-q_e)^{k+2}}{(k+2)!} \omega ^{(k+2)}(q_e) 
\end{equation} 
The wave-function at  the site $n$ can be then written as:
\begin{equation}\label{eq:extremal_front_solution}
\psi(n,t) = \int\limits_{-\pi}^{\pi} \frac{dq}{2 \pi} e^{i \varphi(n,q) t }
\approx e^{i (n q_e - \omega (q_e) t)} {\tilde{A}}(n,t)
\end{equation}
where ${\tilde A}(n,t)$ is:
\begin{equation}\label{eq:tildepsi}
\tilde{A}(n,t)=\int\limits_{-\pi}^{\pi}\frac{dq}{2\pi} e^{i [ (n-v_e t)(q-q_e)-\frac{(q-q_e)^{k+2}}{(k+2)!} \omega^{(k+2)}(q_e)t]}
\end{equation}
We observe from Eq.~\ref{eq:tildepsi} that ${\tilde A}(n,t)$ satisfies the partial differential equation:
\begin{equation}\label{eq: gen_differential_equation}
\frac{\partial {\tilde A}(n,t)}{\partial t}+v_e\frac{\partial {\tilde A}(n,t)}{\partial n}=(-i)^{k+3}\frac{\omega ^{(k+2)}(q_e)}{(k+2)!}\frac{\partial^{k+2}{\tilde A}(n,t)}{\partial n^{k+2}}
\end{equation}
This can be converted into a scaling form by introducing the local scaling variable:
\begin{equation}\label{eq:xi-def}
\xi_n = - \frac{n-v_e t }{(\kappa_k t)^{1/(k+2)}}; \,\, \mbox{here}\,\, \kappa_k = \frac{ \omega ^{(k+2)}(q_e)}{ (k+1)!}
\end{equation}
and defining  
\begin{equation}
{\tilde A}(n,t) = \frac{1}{(\kappa_k t)^{1/(k+2)}} A_k(\xi_ n),
\end{equation}
the partial differential equation satisfied by $\tilde A(n,t)$ (Eq.~\ref{eq: gen_differential_equation}) is converted into the ordinary differential equation satisfied by  $A_k(\xi_n)$:
\begin{equation}\label{eq:gen_airy}
A^{(k+1)}_k (\xi) = (-1)^{k} i^{(k+1)} \xi A_k(\xi)
\end{equation}
The solution of the above equation can be obtained as: 
\begin{equation}\label{eq:gen_A_k}
A_k(\xi) =  \int\limits_{-\infty}^{\infty} \frac{d \eta}{ 2 \pi} e^{ -i \, \xi \eta }\,\, e^{- i \frac{\eta ^{k+2}}{k+2}}
\end{equation}
Note that that $A_1(z_n)$ is the Airy function $Ai(z_n)$.
The probability density $p(m,t)$ and current density $j(m,t)$ (Eq.~\ref{eq:prob_current}) at a site $m$ close to the extremal front takes the form: 
\begin{equation}
p(m,t) = \frac{1}{(\kappa_k t)^{2/k+2}} |A_k(\xi_m)|^2; \qquad  j(m,t)  = v(q_e) p(m,t)
\end{equation}
Hence, the deviation $\delta \Phi(n,t)$ and $\delta J(n,t)$ take the local scaling form: 
\begin{equation}\label{eq: gen_delta_phi}
(\kappa_k t)^{1/k+2} \delta \Phi (\xi_n)=  \int_{0}^{\xi_n} d\xi A_k^2 (\xi);\quad \delta J(n,t) = v_e \delta \Phi (n,t)
\end{equation}


\begin{table}[]
\begin{tabular}{|l|l|l|l|l|l|l|l|}
\hline
\multicolumn{8}{|l|}{~~~~~~~~~~~~~~~Area under the steps (t=10000)}                                                                                                     \\ \hline
                       &                                                                        & Steps & 1      & 2      & 3      & 4      & 5      \\ \hline
\multirow{2}{*}{g=1/16} & \multirow{2}{*}{\begin{tabular}[c]{@{}l@{}}Left\\ 1o\end{tabular}} & CPD & 0.9383 & 0.8997 & 0.9103 & 0.9165 & 0.9138 \\ \cline{3-8} 
                       &                                                                        & CCD  & 0.9450 & 0.9064 & 0.9086 & 0.9186 & 0.9239 \\ \hline
\multicolumn{8}{|l|}{}                                                                                                                               \\ \hline
\multirow{4}{*}{g=1/8} & \multirow{2}{*}{\begin{tabular}[c]{@{}l@{}}Left\\ 3o\end{tabular}}     & CPD  & 0.8218 & 0.7543 & 0.7704 & 0.7914 & 0.7905 \\ \cline{3-8} 
                       &                                                                        & CCD  & 0.8540 & 0.7755 & 0.7781 & 0.8043 & 0.8183 \\ \cline{2-8} 
                       & \multirow{2}{*}{\begin{tabular}[c]{@{}l@{}}Right\\ 1o\end{tabular}}    & CPD  & 0.9519 & 0.9140 & 0.9187 & 0.9249 & 0.9302 \\ \cline{3-8} 
                       &                                                                        & CCD  & 0.9450 & 0.9064 & 0.9130 & 0.9192 & 0.9238 \\ \hline
\multicolumn{8}{|l|}{}                                                                                                                               \\ \hline
\multirow{4}{*}{g=1/4} & \multirow{2}{*}{\begin{tabular}[c]{@{}l@{}}Left\\ 1o\end{tabular}}     & CPD  & 0.9268 & 0.9346 & 0.9140 & 0.8672 & 0.8930 \\ \cline{3-8} 
                       &                                                                        & CCD  & 0.9713 & 0.9069 & 0.9045 & 0.8800 & 1.1974 \\ \cline{2-8} 
                       & \multirow{2}{*}{\begin{tabular}[c]{@{}l@{}}Internal\\ 1o\end{tabular}} & CPD  & 1.0209 & 0.9168 & 1.0234 & 0.9860 & 1.0063 \\ \cline{3-8} 
                       &                                                                        & CCD  & 0.9746 & 0.9496 & 1.0327 & 0.9590 & 1.0974 \\ \hline
\end{tabular}
\caption{\label{tab:area} Areas (measured in units of a) under the steps obtained for cumulative probability distribution profile (CPD) and cumulative current density (CCD) from numerical calculations done at $t=10000$ (measured in units of $g_1^{-1}$) for some representative $g$-values. 1o: first order front; 3o: third order front. Here the areas for CCD are divided by extremal velocity to see the agreement with areas obtained for CPD. For $g=1/4$, areas under the steps near two-fold degenerate left first order (1o) front of CCD are divided by 2 to see the agreement with areas for steps obtained for single 1o front. The areas under the steps remain constant for particular $g$-value at respective extremal front. The heights of the step correspond to the value of scaled $\delta \Phi$ plotted in Fig. \ref{fig:stair_cum_prob_phi_pi_by_2} corresponding to the zeros of it's first derivative and the width of each step is obtained from the difference between the two consecutive non-stationary inflection points of scaled $\delta \Phi$.}
\end{table}


   Close to a first order extremal front, the amplitude function ${\tilde A}(n,t)$ satisfies the linearized K-dV equation \citep{haberman}: 
\begin{equation}
\frac{\partial \tilde{A}}{\partial t}+v_e\frac{\partial \tilde{A}}{\partial n}=\frac{\omega^{(3)}(q_e)}{3!}\frac{\partial^{3}\tilde{A}}{\partial n^{3}}.
\end{equation}   
It can be written in terms of local scaling variable $\xi$ and the Airy function; $A_1(\xi) =Ai(\xi)$ as (Eq.~\ref{eq:gen_A_k}):
\begin{equation}
\tilde{A}(n,t) =\frac{1}{(\kappa_1 t)^{1/3}}Ai(\xi_n); \quad\xi_n = -\frac{n-v_et}{\kappa_1^{1/3} t^{1/3}}
\end{equation}
Using  Eq. \ref {eq: gen_delta_phi}, we can obtain the deviation of the CPD and CCD , $\delta \Phi $ and $\delta J$  at first order extremal front in the transition region $n \approx n_e=v_e t, n<v_e t$  as~\citep{Hemlata}:
\begin{equation}\label{eq:g_extremal_front_scaling}
(\kappa_1 t)^{1/3}\delta \Phi (\xi_n) = \int\limits _0^{\xi_n}d \xi  Ai^2 (\xi) ;\quad
\delta J(\xi_n) =  v_e \delta \Phi(\xi_n) 
\end{equation}
Thus, near any first order front, the deviation in the CPD and CCD exhibit Airy scaling. A local staircase structure is also observed near the edges due to the existence of real zeros of the Airy function and its derivatives~\citep{Hemlata}.  The heights and widths of the steps can be obtained from zeros of the first and second derivatives of Airy functions; an analysis of the asymptotic locations of the zeros shows that the area under the steps remains a constant ~\citep{sasvari_pre69,Hemlata}.   For the model at hand, for  $g<g_c$, there are two first order maximal fronts while for $g>g_c$, there are four first order fronts, of which the two left moving fronts move with the same velocity.  We demonstrate, in Fig. \ref{fig:stair_cum_prob_phi_pi_by_2}(a), (d) the local scaling behaviour of the CPD and CCD  near the maximal left moving first order front  as obtained from the theoretical saddle point analysis as well as from exact numerical computations  for representative $g$ values, ($g<g_c$ and $g>g_c$). The plots show reasonable agreement between that predicted by theoretical analysis and that obtained from numerics.  The Airy scaling and local staircase structure can be observed at the edges of the fronts  from these plots.  The multiplicity of the fronts for $g>g_c$ can  also be seen from the inset of Fig.~\ref{fig:stair_cum_prob_phi_c}.  The numerical computations also show that the areas under the steps remains a constant as can be seen from Table ~\ref{tab:area}. Further, we find that for $g >g_c$, 
near the two-fold degenerate left moving front,  we need to divide the area under each step by $ \approx 2$ in order to obtain agreement between numerical and analytical results (for example, for the areas shown in Table ~\ref{tab:area} for $g=1/4$). We do not show this but similar local scaling is observed near the right moving first order fronts as well.

At $g=g_c$,  there is a first order right moving extremal front and a third order left moving extremal front. The local scaling  near the first order right moving front is of the Airy type discussed in Eq.~\ref{eq:g_extremal_front_scaling} as shown in Fig.~\ref{fig:stair_cum_prob_phi_b}.  We also observe a local staircase structure in the distribution; the corresponding area under the steps remains a constant as shown in Table~\ref{tab:area}.  Near the left moving third order extremal front, the amplitude function satisfies the linearized fifth order equation of the K-dV  type (Eq.~\ref{eq: gen_differential_equation}):
\begin{equation}
\frac{\partial}{\partial t} \tilde{A}(n,t)+v(q_{lm})\frac{\partial}{\partial n} \tilde{A}(n,t)= - \frac{\omega^{(5)}(q_{lm})}{5!}\frac{\partial^{5}}{\partial n^ 5}\tilde{A}(n,t)
\end{equation}
Using Eq. \ref{eq: gen_delta_phi}, the deviation of the CPD and CCD,   $\delta \Phi$  and $\delta J$ in the transition region $n \approx n_{lm}, |n|<|n_{lm}|$ can be obtained in terms of $\xi$ (Eq.~\ref{eq:xi-def}) as,  
\begin{equation}
( \kappa_3 t)^{1/5} \delta \Phi(\xi_n) =  \int\limits _0^{\xi_n} d\xi \,\, A_3^2 (\xi_m); \quad
\delta J(\xi_n)  = v_{lm} \delta \Phi (\xi_n)
\end{equation}

Thus, the deviation of the CPD and CCD near the third order front show a $t^{1/5}$ scaling behaviour. The above analytic scaling analysis  obtained within the stationary phase approximation agrees very well with the exact numerical results as shown in Fig.~\ref{fig:stair_cum_prob_phi_pi_by_2}(b).  We also observe a local staircase structure in the distribution. We have extracted the heights and widths of the steps from the numerical computations and find that the area under the steps remains constant as shown in Table~\ref{tab:area}.  We observe that the  staircase structure near the third order front suggests the existence of real zeros of the function $A_3(\xi)$ and its first derivative. We conjecture that the function $A_k(\xi)$ and its derivative has real zeros when $k$ is odd. Therefore, one should expect a local staircase structure for any odd order extremal front. On the other hand, for a front of even order, it can be seen from Eq.~\ref{eq: gen_differential_equation},  since the  higher order dispersion term is imaginary in nature, one does not get solutions with real zeros. Therefore, one does not expect any local staircase structure near an even order extremal front. This has already been demonstrated for real NNN hopping ~\citep{Hemlata}, where we saw that at $g_c$, there is no staircase structure near the second order internal front.

We do not show the numerical plots for these, but we expect from the saddle point analysis, the local anomalous scaling behaviour of the cumulative position moments at site close to a $k$-th order extremal front to be of the form: $\delta \widetilde{M}_k(\xi_n,t) = v_e^{k} \delta \Phi(\xi_n,t)$.

\section{\label{sec:conclusions}{Conclusions}}

We have studied the long-time dynamics of a quantum walk  of a single particle, initially localized at the origin,  on a one dimensional spatial lattice with complex nearest neighbour and next-nearest neighbour hopping.  At long times, wave fronts propagate ballistically; there exist Lieb-Robinson bounds on the speed at which the wave fronts can propagate which gives rise to a causal structure for the propagation which depends on both the magnitude and phase of the complex NNN hopping amplitude.  Complex next-nearest neighbour coupling leads to broken time reversal symmetry  and consequent  inversion symmetry breaking. This gives rise to chiral propagation of the wave-fronts and hence asymmetric L-R bounds for the maximal velocities or, in other words,  the maximal left moving and right moving velocities are different. Hence, the causal cone structure, probability density and current density distributions are asymmetric about the origin. The asymmetry, which can be measured using the skewness depends on the phase and magnitude of the NNN hopping amplitude. It vanishes for a real NNN hopping and reaches its maximal value  for the completely imaginary NNN hopping.  At a certain critical strength of NNN  hopping,  the value of which depends on both the magnitude and phase of the NNN hopping,  there is a Lifshitz transition from a regime with one causal cone to two causal cones.  In case of real NNN hopping, there is a Lifshitz phase transition from the phase with one causal cone to two nested causal cones symmetrically placed about origin \citep{Hemlata}.  In the presence of complex NNN hopping, the causal cones are no longer symmetric about the origin due to the breaking of inversion symmetry; however, the behaviour of the extremal fronts remains similar for any phase  $ 0 \leq \phi <\pi/2$;  for $g<g_c$, there are two first order extremal fronts while for $g>g_c$, there are four first order fronts and exactly at the critical coupling $g=g_c$, there are three extremal fronts: the two fronts with maximal left and right moving are first order while the internal extremal front is second order in nature.  In contrast to this, there is a novel  behaviour of the causal cone structure and behaviour of the extremal fronts for a purely imaginary NNN hopping:  in the regime with two causal cones ($g>g_c$),  there are three  maximal fronts two of which move with the same maximal velocity. This gives rise to partially overlapping causal cones (even a small real component of NNN hopping breaks this degeneracy and one obtains asymmetric but completely nested cones). We also find very different behaviour at the critical coupling for purely imaginary NNN hopping as compared to that for a purely real NNN hopping.  Exactly, at the  critical coupling, for purely imaginary NNN hopping, the phase is characterised by a single causal cone with the two maximal fronts being of different order; one is a first order front while the other is a third order front.

         Further, we showed, using the stationary phase approximation, that at asymptotically long  times and distances,  the quantum walk can be described in the bulk, by a quasi-stationary state and provided a hydrodynamic description in terms of the local density of quasi-particle excitations. The quasi-stationary state is characterized by a hierarchy of an infinite set of conservation laws satisfied by scaled cumulative position moments. The analytic results are shown to be in good agreement with exact numerical computations albeit with deviations from the global scaling near the extremal front edges.  We also studied the scaling behaviour near the extremal front edges and showed that the local scaling behaviour can be described by higher order hydrodynamic equations. In the regimes $g<g_c$ and $g>g_c$, all extremal fronts are first order in nature and show a sub-diffusive $t^{1/3}$  Airy scaling and staircase structure, the two-fold multiplicity of the maximal front (for $g>g_c$ ) is reflected in the area under the steps of the staircase.  Exactly at the critical coupling $g=g_c$,  the different orders of the two extremal fronts leads to different sub-diffusive $ t^{1/3}$ and $t^{1/5}$ scaling near the two front edges.  A local staircase structure and quantization is observed near both edges. This is in contrast to the case with real NNN hopping where at the critical coupling, there are three extremal fronts, two of which are first order and the other which is an internal front is a second order front.  At the second order front there is no local staircase structure~\citep{Hemlata}. Thus the nature of the Lifshitz transition is different in the two cases. We hope that it will be possible to test these results experimentally. While in the process of writing up this work, we came across very recent work~\citep{novo2020floquet} which suggested protocols through Floquet engineering for experimental realizations of quantum walks with complex hopping amplitudes and in particular imaginary next-nearest neighbour hopping amplitudes. 
           
          Also, in our earlier work~\citep{Hemlata},  we connected the long time dynamics of a single particle (initially localized at the origin) quantum walk problem with the long time dynamics of domain wall propagation in spin chains ~\citep{antal_pre59, sasvari_pre69}. We suggest that the present study can be connected to the time evolution of a domain wall in a spin chain model with complex NNN spin hopping \citep{suzuki,pradeep2}. Quantum quenches in spin chain systems and more generally, non-equilibrium dynamics of integrable systems have been studied within a generalized hydrodynamical framework \citep{Fagotti_prl, Doyon_PhysRevX.6.041065,Fagotti, DOYON_2018, Agrawal_PhysRevB.99.174203}.  
            
\acknowledgments{ We thank L. Novo and S. Ribeiro for useful discussions and for pointing out some references on quantum walks. P.D thanks SERB(CRG/2019/003757), DST India for financial support through research grant. H.B thanks UGC for providing fellowship. } 

\bibliography{qw_chiral}

\begin{thebibliography}{46}%
\makeatletter
\providecommand \@ifxundefined [1]{%
 \@ifx{#1\undefined}
}%
\providecommand \@ifnum [1]{%
 \ifnum #1\expandafter \@firstoftwo
 \else \expandafter \@secondoftwo
 \fi
}%
\providecommand \@ifx [1]{%
 \ifx #1\expandafter \@firstoftwo
 \else \expandafter \@secondoftwo
 \fi
}%
\providecommand \natexlab [1]{#1}%
\providecommand \enquote  [1]{``#1''}%
\providecommand \bibnamefont  [1]{#1}%
\providecommand \bibfnamefont [1]{#1}%
\providecommand \citenamefont [1]{#1}%
\providecommand \href@noop [0]{\@secondoftwo}%
\providecommand \href [0]{\begingroup \@sanitize@url \@href}%
\providecommand \@href[1]{\@@startlink{#1}\@@href}%
\providecommand \@@href[1]{\endgroup#1\@@endlink}%
\providecommand \@sanitize@url [0]{\catcode `\\12\catcode `\$12\catcode
  `\&12\catcode `\#12\catcode `\^12\catcode `\_12\catcode `\%12\relax}%
\providecommand \@@startlink[1]{}%
\providecommand \@@endlink[0]{}%
\providecommand \url  [0]{\begingroup\@sanitize@url \@url }%
\providecommand \@url [1]{\endgroup\@href {#1}{\urlprefix }}%
\providecommand \urlprefix  [0]{URL }%
\providecommand \Eprint [0]{\href }%
\providecommand \doibase [0]{http://dx.doi.org/}%
\providecommand \selectlanguage [0]{\@gobble}%
\providecommand \bibinfo  [0]{\@secondoftwo}%
\providecommand \bibfield  [0]{\@secondoftwo}%
\providecommand \translation [1]{[#1]}%
\providecommand \BibitemOpen [0]{}%
\providecommand \bibitemStop [0]{}%
\providecommand \bibitemNoStop [0]{.\EOS\space}%
\providecommand \EOS [0]{\spacefactor3000\relax}%
\providecommand \BibitemShut  [1]{\csname bibitem#1\endcsname}%
\let\auto@bib@innerbib\@empty
\bibitem [{\citenamefont {Aharonov}\ \emph {et~al.}(1993)\citenamefont
  {Aharonov}, \citenamefont {Davidovich},\ and\ \citenamefont
  {Zagury}}]{Aharonov}%
  \BibitemOpen
  \bibfield  {author} {\bibinfo {author} {\bibfnamefont {Y.}~\bibnamefont
  {Aharonov}}, \bibinfo {author} {\bibfnamefont {L.}~\bibnamefont
  {Davidovich}}, \ and\ \bibinfo {author} {\bibfnamefont {N.}~\bibnamefont
  {Zagury}},\ }\href {\doibase 10.1103/PhysRevA.48.1687} {\bibfield  {journal}
  {\bibinfo  {journal} {Phys. Rev. A}\ }\textbf {\bibinfo {volume} {48}},\
  \bibinfo {pages} {1687} (\bibinfo {year} {1993})}\BibitemShut {NoStop}%
\bibitem [{\citenamefont {Kempe}(2003)}]{Kempe}%
  \BibitemOpen
  \bibfield  {author} {\bibinfo {author} {\bibfnamefont {J.}~\bibnamefont
  {Kempe}},\ }\href {\doibase 10.1080/00107151031000110776} {\bibfield
  {journal} {\bibinfo  {journal} {Contemporary Physics}\ }\textbf {\bibinfo
  {volume} {44}},\ \bibinfo {pages} {307} (\bibinfo {year} {2003})},\ \Eprint
  {http://arxiv.org/abs/https://doi.org/10.1080/00107151031000110776}
  {https://doi.org/10.1080/00107151031000110776} \BibitemShut {NoStop}%
\bibitem [{\citenamefont {Venegas-Andraca}(2012)}]{VAndraca}%
  \BibitemOpen
  \bibfield  {author} {\bibinfo {author} {\bibfnamefont {S.~E.}\ \bibnamefont
  {Venegas-Andraca}},\ }\href {\doibase 10.1007/s11128-012-0432-5} {\bibfield
  {journal} {\bibinfo  {journal} {Quantum Information Processing}\ }\textbf
  {\bibinfo {volume} {11}},\ \bibinfo {pages} {1015} (\bibinfo {year}
  {2012})}\BibitemShut {NoStop}%
\bibitem [{\citenamefont {Longhi}(2009)}]{longhi}%
  \BibitemOpen
  \bibfield  {author} {\bibinfo {author} {\bibfnamefont {S.}~\bibnamefont
  {Longhi}},\ }\href {\doibase 10.1002/lpor.200810055} {\bibfield  {journal}
  {\bibinfo  {journal} {Laser \& Photonics Reviews}\ }\textbf {\bibinfo
  {volume} {3}},\ \bibinfo {pages} {243} (\bibinfo {year} {2009})}\BibitemShut
  {NoStop}%
\bibitem [{\citenamefont {M\"ulken}\ and\ \citenamefont
  {Blumen}(2011)}]{mulken}%
  \BibitemOpen
  \bibfield  {author} {\bibinfo {author} {\bibfnamefont {O.}~\bibnamefont
  {M\"ulken}}\ and\ \bibinfo {author} {\bibfnamefont {A.}~\bibnamefont
  {Blumen}},\ }\href {\doibase https://doi.org/10.1016/j.physrep.2011.01.002}
  {\bibfield  {journal} {\bibinfo  {journal} {Physics Reports}\ }\textbf
  {\bibinfo {volume} {502}},\ \bibinfo {pages} {37 } (\bibinfo {year}
  {2011})}\BibitemShut {NoStop}%
\bibitem [{\citenamefont {Naether}\ \emph {et~al.}(2012)\citenamefont
  {Naether}, \citenamefont {Meyer}, \citenamefont {St\"{u}tzer}, \citenamefont
  {T\"{u}nnermann}, \citenamefont {Nolte}, \citenamefont {Molina},\ and\
  \citenamefont {Szameit}}]{QW_Anderson_localization}%
  \BibitemOpen
  \bibfield  {author} {\bibinfo {author} {\bibfnamefont {U.}~\bibnamefont
  {Naether}}, \bibinfo {author} {\bibfnamefont {J.~M.}\ \bibnamefont {Meyer}},
  \bibinfo {author} {\bibfnamefont {S.}~\bibnamefont {St\"{u}tzer}}, \bibinfo
  {author} {\bibfnamefont {A.}~\bibnamefont {T\"{u}nnermann}}, \bibinfo
  {author} {\bibfnamefont {S.}~\bibnamefont {Nolte}}, \bibinfo {author}
  {\bibfnamefont {M.~I.}\ \bibnamefont {Molina}}, \ and\ \bibinfo {author}
  {\bibfnamefont {A.}~\bibnamefont {Szameit}},\ }\href {\doibase
  10.1364/OL.37.000485} {\bibfield  {journal} {\bibinfo  {journal} {Opt.
  Lett.}\ }\textbf {\bibinfo {volume} {37}},\ \bibinfo {pages} {485} (\bibinfo
  {year} {2012})}\BibitemShut {NoStop}%
\bibitem [{\citenamefont {Kitagawa}\ \emph {et~al.}(2012)\citenamefont
  {Kitagawa}, \citenamefont {Broome}, \citenamefont {Fedrizzi}, \citenamefont
  {Rudner}, \citenamefont {Berg}, \citenamefont {Kassal}, \citenamefont
  {Aspuru-Guzik}, \citenamefont {Demler},\ and\ \citenamefont
  {White}}]{QW_Topological_phases}%
  \BibitemOpen
  \bibfield  {author} {\bibinfo {author} {\bibfnamefont {T.}~\bibnamefont
  {Kitagawa}}, \bibinfo {author} {\bibfnamefont {M.}~\bibnamefont {Broome}},
  \bibinfo {author} {\bibfnamefont {A.}~\bibnamefont {Fedrizzi}}, \bibinfo
  {author} {\bibfnamefont {M.}~\bibnamefont {Rudner}}, \bibinfo {author}
  {\bibfnamefont {E.}~\bibnamefont {Berg}}, \bibinfo {author} {\bibfnamefont
  {I.}~\bibnamefont {Kassal}}, \bibinfo {author} {\bibfnamefont
  {A.}~\bibnamefont {Aspuru-Guzik}}, \bibinfo {author} {\bibfnamefont
  {E.}~\bibnamefont {Demler}}, \ and\ \bibinfo {author} {\bibfnamefont
  {A.}~\bibnamefont {White}},\ }\href {\doibase 10.1038/ncomms1872} {\bibfield
  {journal} {\bibinfo  {journal} {Nature Communications}\ }\textbf {\bibinfo
  {volume} {3}} (\bibinfo {year} {2012}),\ 10.1038/ncomms1872}\BibitemShut
  {NoStop}%
\bibitem [{\citenamefont {Shor}(1999)}]{shor}%
  \BibitemOpen
  \bibfield  {author} {\bibinfo {author} {\bibfnamefont {P.~W.}\ \bibnamefont
  {Shor}},\ }\href {\doibase 10.1137/S0036144598347011} {\bibfield  {journal}
  {\bibinfo  {journal} {SIAM Review}\ }\textbf {\bibinfo {volume} {41}},\
  \bibinfo {pages} {303} (\bibinfo {year} {1999})},\ \Eprint
  {http://arxiv.org/abs/https://doi.org/10.1137/S0036144598347011}
  {https://doi.org/10.1137/S0036144598347011} \BibitemShut {NoStop}%
\bibitem [{\citenamefont {Grover}(1997)}]{Grover}%
  \BibitemOpen
  \bibfield  {author} {\bibinfo {author} {\bibfnamefont {L.~K.}\ \bibnamefont
  {Grover}},\ }\href {\doibase 10.1103/PhysRevLett.79.325} {\bibfield
  {journal} {\bibinfo  {journal} {Phys. Rev. Lett.}\ }\textbf {\bibinfo
  {volume} {79}},\ \bibinfo {pages} {325} (\bibinfo {year} {1997})}\BibitemShut
  {NoStop}%
\bibitem [{\citenamefont {Childs}\ \emph {et~al.}(2003)\citenamefont {Childs},
  \citenamefont {Cleve}, \citenamefont {Deotto}, \citenamefont {Farhi},
  \citenamefont {Gutmann},\ and\ \citenamefont
  {Spielman}}]{Childs_Conf_Proc_2003}%
  \BibitemOpen
  \bibfield  {author} {\bibinfo {author} {\bibfnamefont {A.~M.}\ \bibnamefont
  {Childs}}, \bibinfo {author} {\bibfnamefont {R.}~\bibnamefont {Cleve}},
  \bibinfo {author} {\bibfnamefont {E.}~\bibnamefont {Deotto}}, \bibinfo
  {author} {\bibfnamefont {E.}~\bibnamefont {Farhi}}, \bibinfo {author}
  {\bibfnamefont {S.}~\bibnamefont {Gutmann}}, \ and\ \bibinfo {author}
  {\bibfnamefont {D.~A.}\ \bibnamefont {Spielman}},\ }\bibfield  {booktitle}
  {\emph {\bibinfo {booktitle} {Proceedings of the Thirty-Fifth Annual ACM
  Symposium on Theory of Computing}},\ }\href {\doibase 10.1145/780542.780552}
  {\ \bibinfo {series} {STOC '03},\ \bibinfo {pages} {59–68} (\bibinfo {year}
  {2003})}\BibitemShut {NoStop}%
\bibitem [{\citenamefont {Childs}\ and\ \citenamefont
  {Goldstone}(2004)}]{Childs_PRA.70.022314}%
  \BibitemOpen
  \bibfield  {author} {\bibinfo {author} {\bibfnamefont {A.~M.}\ \bibnamefont
  {Childs}}\ and\ \bibinfo {author} {\bibfnamefont {J.}~\bibnamefont
  {Goldstone}},\ }\href {\doibase 10.1103/PhysRevA.70.022314} {\bibfield
  {journal} {\bibinfo  {journal} {Phys. Rev. A}\ }\textbf {\bibinfo {volume}
  {70}},\ \bibinfo {pages} {022314} (\bibinfo {year} {2004})}\BibitemShut
  {NoStop}%
\bibitem [{\citenamefont {Schreiber}\ \emph {et~al.}(2010)\citenamefont
  {Schreiber}, \citenamefont {Cassemiro}, \citenamefont
  {Poto\ifmmode~\check{c}\else \v{c}\fi{}ek}, \citenamefont {G\'abris},
  \citenamefont {Mosley}, \citenamefont {Andersson}, \citenamefont {Jex},\ and\
  \citenamefont {Silberhorn}}]{QW_photons}%
  \BibitemOpen
  \bibfield  {author} {\bibinfo {author} {\bibfnamefont {A.}~\bibnamefont
  {Schreiber}}, \bibinfo {author} {\bibfnamefont {K.~N.}\ \bibnamefont
  {Cassemiro}}, \bibinfo {author} {\bibfnamefont {V.}~\bibnamefont
  {Poto\ifmmode~\check{c}\else \v{c}\fi{}ek}}, \bibinfo {author} {\bibfnamefont
  {A.}~\bibnamefont {G\'abris}}, \bibinfo {author} {\bibfnamefont {P.~J.}\
  \bibnamefont {Mosley}}, \bibinfo {author} {\bibfnamefont {E.}~\bibnamefont
  {Andersson}}, \bibinfo {author} {\bibfnamefont {I.}~\bibnamefont {Jex}}, \
  and\ \bibinfo {author} {\bibfnamefont {C.}~\bibnamefont {Silberhorn}},\
  }\href {\doibase 10.1103/PhysRevLett.104.050502} {\bibfield  {journal}
  {\bibinfo  {journal} {Phys. Rev. Lett.}\ }\textbf {\bibinfo {volume} {104}},\
  \bibinfo {pages} {050502} (\bibinfo {year} {2010})}\BibitemShut {NoStop}%
\bibitem [{\citenamefont {Z\"ahringer}\ \emph {et~al.}(2010)\citenamefont
  {Z\"ahringer}, \citenamefont {Kirchmair}, \citenamefont {Gerritsma},
  \citenamefont {Solano}, \citenamefont {Blatt},\ and\ \citenamefont
  {Roos}}]{QW_trapped_ion}%
  \BibitemOpen
  \bibfield  {author} {\bibinfo {author} {\bibfnamefont {F.}~\bibnamefont
  {Z\"ahringer}}, \bibinfo {author} {\bibfnamefont {G.}~\bibnamefont
  {Kirchmair}}, \bibinfo {author} {\bibfnamefont {R.}~\bibnamefont
  {Gerritsma}}, \bibinfo {author} {\bibfnamefont {E.}~\bibnamefont {Solano}},
  \bibinfo {author} {\bibfnamefont {R.}~\bibnamefont {Blatt}}, \ and\ \bibinfo
  {author} {\bibfnamefont {C.~F.}\ \bibnamefont {Roos}},\ }\href {\doibase
  10.1103/PhysRevLett.104.100503} {\bibfield  {journal} {\bibinfo  {journal}
  {Phys. Rev. Lett.}\ }\textbf {\bibinfo {volume} {104}},\ \bibinfo {pages}
  {100503} (\bibinfo {year} {2010})}\BibitemShut {NoStop}%
\bibitem [{\citenamefont {Travaglione}\ and\ \citenamefont
  {Milburn}(2002)}]{PRA.65.032310}%
  \BibitemOpen
  \bibfield  {author} {\bibinfo {author} {\bibfnamefont {B.~C.}\ \bibnamefont
  {Travaglione}}\ and\ \bibinfo {author} {\bibfnamefont {G.~J.}\ \bibnamefont
  {Milburn}},\ }\href {\doibase 10.1103/PhysRevA.65.032310} {\bibfield
  {journal} {\bibinfo  {journal} {Phys. Rev. A}\ }\textbf {\bibinfo {volume}
  {65}},\ \bibinfo {pages} {032310} (\bibinfo {year} {2002})}\BibitemShut
  {NoStop}%
\bibitem [{\citenamefont {Schmitz}\ \emph {et~al.}(2009)\citenamefont
  {Schmitz}, \citenamefont {Matjeschk}, \citenamefont {Schneider},
  \citenamefont {Glueckert}, \citenamefont {Enderlein}, \citenamefont {Huber},\
  and\ \citenamefont {Schaetz}}]{PRL.103.090504}%
  \BibitemOpen
  \bibfield  {author} {\bibinfo {author} {\bibfnamefont {H.}~\bibnamefont
  {Schmitz}}, \bibinfo {author} {\bibfnamefont {R.}~\bibnamefont {Matjeschk}},
  \bibinfo {author} {\bibfnamefont {C.}~\bibnamefont {Schneider}}, \bibinfo
  {author} {\bibfnamefont {J.}~\bibnamefont {Glueckert}}, \bibinfo {author}
  {\bibfnamefont {M.}~\bibnamefont {Enderlein}}, \bibinfo {author}
  {\bibfnamefont {T.}~\bibnamefont {Huber}}, \ and\ \bibinfo {author}
  {\bibfnamefont {T.}~\bibnamefont {Schaetz}},\ }\href {\doibase
  10.1103/PhysRevLett.103.090504} {\bibfield  {journal} {\bibinfo  {journal}
  {Phys. Rev. Lett.}\ }\textbf {\bibinfo {volume} {103}},\ \bibinfo {pages}
  {090504} (\bibinfo {year} {2009})}\BibitemShut {NoStop}%
\bibitem [{\citenamefont {D\"ur}\ \emph {et~al.}(2002)\citenamefont {D\"ur},
  \citenamefont {Raussendorf}, \citenamefont {Kendon},\ and\ \citenamefont
  {Briegel}}]{QW_optical_lattice}%
  \BibitemOpen
  \bibfield  {author} {\bibinfo {author} {\bibfnamefont {W.}~\bibnamefont
  {D\"ur}}, \bibinfo {author} {\bibfnamefont {R.}~\bibnamefont {Raussendorf}},
  \bibinfo {author} {\bibfnamefont {V.~M.}\ \bibnamefont {Kendon}}, \ and\
  \bibinfo {author} {\bibfnamefont {H.-J.}\ \bibnamefont {Briegel}},\ }\href
  {\doibase 10.1103/PhysRevA.66.052319} {\bibfield  {journal} {\bibinfo
  {journal} {Phys. Rev. A}\ }\textbf {\bibinfo {volume} {66}},\ \bibinfo
  {pages} {052319} (\bibinfo {year} {2002})}\BibitemShut {NoStop}%
\bibitem [{\citenamefont {Karski}\ \emph {et~al.}(2009)\citenamefont {Karski},
  \citenamefont {Förster}, \citenamefont {Choi}, \citenamefont {Steffen},
  \citenamefont {Alt}, \citenamefont {Meschede},\ and\ \citenamefont
  {Widera}}]{QW_trapped_atom}%
  \BibitemOpen
  \bibfield  {author} {\bibinfo {author} {\bibfnamefont {M.}~\bibnamefont
  {Karski}}, \bibinfo {author} {\bibfnamefont {L.}~\bibnamefont {Förster}},
  \bibinfo {author} {\bibfnamefont {J.-M.}\ \bibnamefont {Choi}}, \bibinfo
  {author} {\bibfnamefont {A.}~\bibnamefont {Steffen}}, \bibinfo {author}
  {\bibfnamefont {W.}~\bibnamefont {Alt}}, \bibinfo {author} {\bibfnamefont
  {D.}~\bibnamefont {Meschede}}, \ and\ \bibinfo {author} {\bibfnamefont
  {A.}~\bibnamefont {Widera}},\ }\href {\doibase 10.1126/science.1174436}
  {\bibfield  {journal} {\bibinfo  {journal} {Science (New York, N.Y.)}\
  }\textbf {\bibinfo {volume} {325}},\ \bibinfo {pages} {174} (\bibinfo {year}
  {2009})}\BibitemShut {NoStop}%
\bibitem [{\citenamefont {Dadras}\ \emph {et~al.}(2018)\citenamefont {Dadras},
  \citenamefont {Gresch}, \citenamefont {Groiseau}, \citenamefont {Wimberger},\
  and\ \citenamefont {Summy}}]{QW_BEC}%
  \BibitemOpen
  \bibfield  {author} {\bibinfo {author} {\bibfnamefont {S.}~\bibnamefont
  {Dadras}}, \bibinfo {author} {\bibfnamefont {A.}~\bibnamefont {Gresch}},
  \bibinfo {author} {\bibfnamefont {C.}~\bibnamefont {Groiseau}}, \bibinfo
  {author} {\bibfnamefont {S.}~\bibnamefont {Wimberger}}, \ and\ \bibinfo
  {author} {\bibfnamefont {G.~S.}\ \bibnamefont {Summy}},\ }\href {\doibase
  10.1103/PhysRevLett.121.070402} {\bibfield  {journal} {\bibinfo  {journal}
  {Phys. Rev. Lett.}\ }\textbf {\bibinfo {volume} {121}},\ \bibinfo {pages}
  {070402} (\bibinfo {year} {2018})}\BibitemShut {NoStop}%
\bibitem [{\citenamefont {Solenov}\ and\ \citenamefont
  {Fedichkin}(2006)}]{QW_cycle}%
  \BibitemOpen
  \bibfield  {author} {\bibinfo {author} {\bibfnamefont {D.}~\bibnamefont
  {Solenov}}\ and\ \bibinfo {author} {\bibfnamefont {L.}~\bibnamefont
  {Fedichkin}},\ }\href {\doibase 10.1103/PhysRevA.73.012313} {\bibfield
  {journal} {\bibinfo  {journal} {Phys. Rev. A}\ }\textbf {\bibinfo {volume}
  {73}},\ \bibinfo {pages} {012313} (\bibinfo {year} {2006})}\BibitemShut
  {NoStop}%
\bibitem [{\citenamefont {Krapivsky}\ \emph {et~al.}(2014)\citenamefont
  {Krapivsky}, \citenamefont {Luck},\ and\ \citenamefont
  {Mallick}}]{QW_Krapivsky1}%
  \BibitemOpen
  \bibfield  {author} {\bibinfo {author} {\bibfnamefont {P.~L.}\ \bibnamefont
  {Krapivsky}}, \bibinfo {author} {\bibfnamefont {J.~M.}\ \bibnamefont {Luck}},
  \ and\ \bibinfo {author} {\bibfnamefont {K.}~\bibnamefont {Mallick}},\ }\href
  {\doibase 10.1007/s10955-014-0936-8} {\bibfield  {journal} {\bibinfo
  {journal} {Journal of Statistical Physics}\ }\textbf {\bibinfo {volume}
  {154}},\ \bibinfo {pages} {1430} (\bibinfo {year} {2014})}\BibitemShut
  {NoStop}%
\bibitem [{\citenamefont {Krapivsky}\ \emph {et~al.}(2015)\citenamefont
  {Krapivsky}, \citenamefont {Luck},\ and\ \citenamefont
  {Mallick}}]{QW_Krapivsky2}%
  \BibitemOpen
  \bibfield  {author} {\bibinfo {author} {\bibfnamefont {P.~L.}\ \bibnamefont
  {Krapivsky}}, \bibinfo {author} {\bibfnamefont {J.~M.}\ \bibnamefont {Luck}},
  \ and\ \bibinfo {author} {\bibfnamefont {K.}~\bibnamefont {Mallick}},\ }\href
  {http://stacks.iop.org/1751-8121/48/i=47/a=475301} {\bibfield  {journal}
  {\bibinfo  {journal} {Journal of Physics A: Mathematical and Theoretical}\
  }\textbf {\bibinfo {volume} {48}},\ \bibinfo {pages} {475301} (\bibinfo
  {year} {2015})}\BibitemShut {NoStop}%
\bibitem [{\citenamefont {Bhandari}\ and\ \citenamefont
  {Durganandini}(2019)}]{Hemlata}%
  \BibitemOpen
  \bibfield  {author} {\bibinfo {author} {\bibfnamefont {H.}~\bibnamefont
  {Bhandari}}\ and\ \bibinfo {author} {\bibfnamefont {P.}~\bibnamefont
  {Durganandini}},\ }\href {\doibase 10.1103/PhysRevA.99.032313} {\bibfield
  {journal} {\bibinfo  {journal} {Phys. Rev. A}\ }\textbf {\bibinfo {volume}
  {99}},\ \bibinfo {pages} {032313} (\bibinfo {year} {2019})}\BibitemShut
  {NoStop}%
\bibitem [{\citenamefont {{Cuevas}}\ \emph {et~al.}(2011)\citenamefont
  {{Cuevas}}, \citenamefont {{Curilef}},\ and\ \citenamefont
  {{Plastino}}}]{cuevas}%
  \BibitemOpen
  \bibfield  {author} {\bibinfo {author} {\bibfnamefont {F.~A.}\ \bibnamefont
  {{Cuevas}}}, \bibinfo {author} {\bibfnamefont {S.}~\bibnamefont {{Curilef}}},
  \ and\ \bibinfo {author} {\bibfnamefont {A.~R.}\ \bibnamefont {{Plastino}}},\
  }\href {\doibase 10.1016/j.aop.2011.07.003} {\bibfield  {journal} {\bibinfo
  {journal} {Annals of Physics}\ }\textbf {\bibinfo {volume} {326}},\ \bibinfo
  {pages} {2834} (\bibinfo {year} {2011})}\BibitemShut {NoStop}%
\bibitem [{\citenamefont {Arias}\ and\ \citenamefont {Luck}(1998)}]{Toro}%
  \BibitemOpen
  \bibfield  {author} {\bibinfo {author} {\bibfnamefont {S.~D.~T.}\
  \bibnamefont {Arias}}\ and\ \bibinfo {author} {\bibfnamefont {J.~M.}\
  \bibnamefont {Luck}},\ }\href {http://stacks.iop.org/0305-4470/31/i=38/a=007}
  {\bibfield  {journal} {\bibinfo  {journal} {Journal of Physics A:
  Mathematical and General}\ }\textbf {\bibinfo {volume} {31}},\ \bibinfo
  {pages} {7699} (\bibinfo {year} {1998})}\BibitemShut {NoStop}%
\bibitem [{\citenamefont {Schonhammer}(2019)}]{schoenhammer}%
  \BibitemOpen
  \bibfield  {author} {\bibinfo {author} {\bibfnamefont {K.}~\bibnamefont
  {Schonhammer}},\ }\href {\doibase 10.1119/1.5089752} {\bibfield  {journal}
  {\bibinfo  {journal} {American Journal of Physics}\ }\textbf {\bibinfo
  {volume} {87}},\ \bibinfo {pages} {186} (\bibinfo {year} {2019})},\ \Eprint
  {http://arxiv.org/abs/https://doi.org/10.1119/1.5089752}
  {https://doi.org/10.1119/1.5089752} \BibitemShut {NoStop}%
\bibitem [{\citenamefont {Lieb}\ and\ \citenamefont
  {Robinson}(1972)}]{lieb-robinson}%
  \BibitemOpen
  \bibfield  {author} {\bibinfo {author} {\bibfnamefont {E.~H.}\ \bibnamefont
  {Lieb}}\ and\ \bibinfo {author} {\bibfnamefont {D.~W.}\ \bibnamefont
  {Robinson}},\ }\href {\doibase 10.1007/BF01645779} {\bibfield  {journal}
  {\bibinfo  {journal} {Communications in Mathematical Physics}\ }\textbf
  {\bibinfo {volume} {28}},\ \bibinfo {pages} {251} (\bibinfo {year}
  {1972})}\BibitemShut {NoStop}%
\bibitem [{\citenamefont {Bravyi}\ \emph {et~al.}(2006)\citenamefont {Bravyi},
  \citenamefont {Hastings},\ and\ \citenamefont {Verstraete}}]{bravyi2006}%
  \BibitemOpen
  \bibfield  {author} {\bibinfo {author} {\bibfnamefont {S.}~\bibnamefont
  {Bravyi}}, \bibinfo {author} {\bibfnamefont {M.~B.}\ \bibnamefont
  {Hastings}}, \ and\ \bibinfo {author} {\bibfnamefont {F.}~\bibnamefont
  {Verstraete}},\ }\href {\doibase 10.1103/PhysRevLett.97.050401} {\bibfield
  {journal} {\bibinfo  {journal} {Phys. Rev. Lett.}\ }\textbf {\bibinfo
  {volume} {97}},\ \bibinfo {pages} {050401} (\bibinfo {year}
  {2006})}\BibitemShut {NoStop}%
\bibitem [{\citenamefont {Calabrese}\ and\ \citenamefont
  {Cardy}(2006)}]{cardy2006}%
  \BibitemOpen
  \bibfield  {author} {\bibinfo {author} {\bibfnamefont {P.}~\bibnamefont
  {Calabrese}}\ and\ \bibinfo {author} {\bibfnamefont {J.}~\bibnamefont
  {Cardy}},\ }\href {\doibase 10.1103/PhysRevLett.96.136801} {\bibfield
  {journal} {\bibinfo  {journal} {Phys. Rev. Lett.}\ }\textbf {\bibinfo
  {volume} {96}},\ \bibinfo {pages} {136801} (\bibinfo {year}
  {2006})}\BibitemShut {NoStop}%
\bibitem [{\citenamefont {Lu}\ \emph {et~al.}(2016)\citenamefont {Lu},
  \citenamefont {Biamonte}, \citenamefont {Li}, \citenamefont {Li},
  \citenamefont {Johnson}, \citenamefont {Bergholm}, \citenamefont {Faccin},
  \citenamefont {Zimbor\'as}, \citenamefont {Laflamme}, \citenamefont {Baugh},\
  and\ \citenamefont {Lloyd}}]{Chiral_QWs}%
  \BibitemOpen
  \bibfield  {author} {\bibinfo {author} {\bibfnamefont {D.}~\bibnamefont
  {Lu}}, \bibinfo {author} {\bibfnamefont {J.~D.}\ \bibnamefont {Biamonte}},
  \bibinfo {author} {\bibfnamefont {J.}~\bibnamefont {Li}}, \bibinfo {author}
  {\bibfnamefont {H.}~\bibnamefont {Li}}, \bibinfo {author} {\bibfnamefont
  {T.~H.}\ \bibnamefont {Johnson}}, \bibinfo {author} {\bibfnamefont
  {V.}~\bibnamefont {Bergholm}}, \bibinfo {author} {\bibfnamefont
  {M.}~\bibnamefont {Faccin}}, \bibinfo {author} {\bibfnamefont
  {Z.}~\bibnamefont {Zimbor\'as}}, \bibinfo {author} {\bibfnamefont
  {R.}~\bibnamefont {Laflamme}}, \bibinfo {author} {\bibfnamefont
  {J.}~\bibnamefont {Baugh}}, \ and\ \bibinfo {author} {\bibfnamefont
  {S.}~\bibnamefont {Lloyd}},\ }\href {\doibase 10.1103/PhysRevA.93.042302}
  {\bibfield  {journal} {\bibinfo  {journal} {Phys. Rev. A}\ }\textbf {\bibinfo
  {volume} {93}},\ \bibinfo {pages} {042302} (\bibinfo {year}
  {2016})}\BibitemShut {NoStop}%
\bibitem [{\citenamefont {Zimbor{\'a}s}\ \emph {et~al.}(2013)\citenamefont
  {Zimbor{\'a}s}, \citenamefont {Faccin}, \citenamefont {K{\'a}d{\'a}r},
  \citenamefont {Whitfield}, \citenamefont {Lanyon},\ and\ \citenamefont
  {Biamonte}}]{TRS_quantum_transport}%
  \BibitemOpen
  \bibfield  {author} {\bibinfo {author} {\bibfnamefont {Z.}~\bibnamefont
  {Zimbor{\'a}s}}, \bibinfo {author} {\bibfnamefont {M.}~\bibnamefont
  {Faccin}}, \bibinfo {author} {\bibfnamefont {Z.}~\bibnamefont
  {K{\'a}d{\'a}r}}, \bibinfo {author} {\bibfnamefont {J.~D.}\ \bibnamefont
  {Whitfield}}, \bibinfo {author} {\bibfnamefont {B.~P.}\ \bibnamefont
  {Lanyon}}, \ and\ \bibinfo {author} {\bibfnamefont {J.}~\bibnamefont
  {Biamonte}},\ }\href {\doibase 10.1038/srep02361} {\bibfield  {journal}
  {\bibinfo  {journal} {Scientific Reports}\ }\textbf {\bibinfo {volume} {3}},\
  \bibinfo {pages} {2361} (\bibinfo {year} {2013})}\BibitemShut {NoStop}%
\bibitem [{\citenamefont {Lifshitz}(1960)}]{Lifshitz60}%
  \BibitemOpen
  \bibfield  {author} {\bibinfo {author} {\bibfnamefont {I.~M.}\ \bibnamefont
  {Lifshitz}},\ }\href@noop {} {\bibfield  {journal} {\bibinfo  {journal} {Sov.
  Phys. JETP}\ }\textbf {\bibinfo {volume} {11 1130}} (\bibinfo {year}
  {1960})}\BibitemShut {NoStop}%
\bibitem [{\citenamefont {Bertini}\ \emph {et~al.}(2016)\citenamefont
  {Bertini}, \citenamefont {Collura}, \citenamefont {De~Nardis},\ and\
  \citenamefont {Fagotti}}]{Fagotti_prl}%
  \BibitemOpen
  \bibfield  {author} {\bibinfo {author} {\bibfnamefont {B.}~\bibnamefont
  {Bertini}}, \bibinfo {author} {\bibfnamefont {M.}~\bibnamefont {Collura}},
  \bibinfo {author} {\bibfnamefont {J.}~\bibnamefont {De~Nardis}}, \ and\
  \bibinfo {author} {\bibfnamefont {M.}~\bibnamefont {Fagotti}},\ }\href
  {\doibase 10.1103/PhysRevLett.117.207201} {\bibfield  {journal} {\bibinfo
  {journal} {Phys. Rev. Lett.}\ }\textbf {\bibinfo {volume} {117}},\ \bibinfo
  {pages} {207201} (\bibinfo {year} {2016})}\BibitemShut {NoStop}%
\bibitem [{\citenamefont {Castro-Alvaredo}\ \emph {et~al.}(2016)\citenamefont
  {Castro-Alvaredo}, \citenamefont {Doyon},\ and\ \citenamefont
  {Yoshimura}}]{Doyon_PhysRevX.6.041065}%
  \BibitemOpen
  \bibfield  {author} {\bibinfo {author} {\bibfnamefont {O.~A.}\ \bibnamefont
  {Castro-Alvaredo}}, \bibinfo {author} {\bibfnamefont {B.}~\bibnamefont
  {Doyon}}, \ and\ \bibinfo {author} {\bibfnamefont {T.}~\bibnamefont
  {Yoshimura}},\ }\href {\doibase 10.1103/PhysRevX.6.041065} {\bibfield
  {journal} {\bibinfo  {journal} {Phys. Rev. X}\ }\textbf {\bibinfo {volume}
  {6}},\ \bibinfo {pages} {041065} (\bibinfo {year} {2016})}\BibitemShut
  {NoStop}%
\bibitem [{\citenamefont {Fagotti}(2017)}]{Fagotti}%
  \BibitemOpen
  \bibfield  {author} {\bibinfo {author} {\bibfnamefont {M.}~\bibnamefont
  {Fagotti}},\ }\href {\doibase 10.1103/PhysRevB.96.220302} {\bibfield
  {journal} {\bibinfo  {journal} {Phys. Rev. B}\ }\textbf {\bibinfo {volume}
  {96}},\ \bibinfo {pages} {220302} (\bibinfo {year} {2017})}\BibitemShut
  {NoStop}%
\bibitem [{\citenamefont {Doyon}\ \emph {et~al.}(2018)\citenamefont {Doyon},
  \citenamefont {Spohn},\ and\ \citenamefont {Yoshimura}}]{DOYON_2018}%
  \BibitemOpen
  \bibfield  {author} {\bibinfo {author} {\bibfnamefont {B.}~\bibnamefont
  {Doyon}}, \bibinfo {author} {\bibfnamefont {H.}~\bibnamefont {Spohn}}, \ and\
  \bibinfo {author} {\bibfnamefont {T.}~\bibnamefont {Yoshimura}},\ }\href
  {\doibase https://doi.org/10.1016/j.nuclphysb.2017.12.002} {\bibfield
  {journal} {\bibinfo  {journal} {Nuclear Physics B}\ }\textbf {\bibinfo
  {volume} {926}},\ \bibinfo {pages} {570 } (\bibinfo {year}
  {2018})}\BibitemShut {NoStop}%
\bibitem [{\citenamefont {Agrawal}\ \emph {et~al.}(2019)\citenamefont
  {Agrawal}, \citenamefont {Gopalakrishnan},\ and\ \citenamefont
  {Vasseur}}]{Agrawal_PhysRevB.99.174203}%
  \BibitemOpen
  \bibfield  {author} {\bibinfo {author} {\bibfnamefont {U.}~\bibnamefont
  {Agrawal}}, \bibinfo {author} {\bibfnamefont {S.}~\bibnamefont
  {Gopalakrishnan}}, \ and\ \bibinfo {author} {\bibfnamefont {R.}~\bibnamefont
  {Vasseur}},\ }\href {\doibase 10.1103/PhysRevB.99.174203} {\bibfield
  {journal} {\bibinfo  {journal} {Phys. Rev. B}\ }\textbf {\bibinfo {volume}
  {99}},\ \bibinfo {pages} {174203} (\bibinfo {year} {2019})}\BibitemShut
  {NoStop}%
\bibitem [{\citenamefont {Antal}\ \emph {et~al.}(1999)\citenamefont {Antal},
  \citenamefont {R\'acz}, \citenamefont {R\'akos},\ and\ \citenamefont
  {Sch\"utz}}]{antal_pre59}%
  \BibitemOpen
  \bibfield  {author} {\bibinfo {author} {\bibfnamefont {T.}~\bibnamefont
  {Antal}}, \bibinfo {author} {\bibfnamefont {Z.}~\bibnamefont {R\'acz}},
  \bibinfo {author} {\bibfnamefont {A.}~\bibnamefont {R\'akos}}, \ and\
  \bibinfo {author} {\bibfnamefont {G.~M.}\ \bibnamefont {Sch\"utz}},\ }\href
  {\doibase 10.1103/PhysRevE.59.4912} {\bibfield  {journal} {\bibinfo
  {journal} {Phys. Rev. E}\ }\textbf {\bibinfo {volume} {59}},\ \bibinfo
  {pages} {4912} (\bibinfo {year} {1999})}\BibitemShut {NoStop}%
\bibitem [{\citenamefont {Hunyadi}\ \emph {et~al.}(2004)\citenamefont
  {Hunyadi}, \citenamefont {R\'acz},\ and\ \citenamefont
  {Sasv\'ari}}]{sasvari_pre69}%
  \BibitemOpen
  \bibfield  {author} {\bibinfo {author} {\bibfnamefont {V.}~\bibnamefont
  {Hunyadi}}, \bibinfo {author} {\bibfnamefont {Z.}~\bibnamefont {R\'acz}}, \
  and\ \bibinfo {author} {\bibfnamefont {L.}~\bibnamefont {Sasv\'ari}},\ }\href
  {\doibase 10.1103/PhysRevE.69.066103} {\bibfield  {journal} {\bibinfo
  {journal} {Phys. Rev. E}\ }\textbf {\bibinfo {volume} {69}},\ \bibinfo
  {pages} {066103} (\bibinfo {year} {2004})}\BibitemShut {NoStop}%
\bibitem [{\citenamefont {Suzuki}(1971)}]{suzuki}%
  \BibitemOpen
  \bibfield  {author} {\bibinfo {author} {\bibfnamefont {M.}~\bibnamefont
  {Suzuki}},\ }\href {\doibase https://doi.org/10.1016/0375-9601(71)90218-0}
  {\bibfield  {journal} {\bibinfo  {journal} {Physics Letters A}\ }\textbf
  {\bibinfo {volume} {34}},\ \bibinfo {pages} {94 } (\bibinfo {year}
  {1971})}\BibitemShut {NoStop}%
\bibitem [{\citenamefont {Thakur}\ and\ \citenamefont
  {Durganandini}(2016)}]{pradeep2}%
  \BibitemOpen
  \bibfield  {author} {\bibinfo {author} {\bibfnamefont {P.}~\bibnamefont
  {Thakur}}\ and\ \bibinfo {author} {\bibfnamefont {P.}~\bibnamefont
  {Durganandini}},\ }\href {\doibase 10.1063/1.4948157} {\bibfield  {journal}
  {\bibinfo  {journal} {AIP Conference Proceedings}\ }\textbf {\bibinfo
  {volume} {1731}},\ \bibinfo {pages} {130051} (\bibinfo {year} {2016})},\
  \Eprint
  {http://arxiv.org/abs/https://aip.scitation.org/doi/pdf/10.1063/1.4948157}
  {https://aip.scitation.org/doi/pdf/10.1063/1.4948157} \BibitemShut {NoStop}%
\bibitem [{\citenamefont {Mahan}(2000)}]{gd_mahan}%
  \BibitemOpen
  \bibfield  {author} {\bibinfo {author} {\bibfnamefont {G.~D.}\ \bibnamefont
  {Mahan}},\ }\href@noop {} {\emph {\bibinfo {title} {Many Particle Physics,
  Third Edition}}}\ (\bibinfo  {publisher} {Plenum},\ \bibinfo {address} {New
  York},\ \bibinfo {year} {2000})\BibitemShut {NoStop}%
\bibitem [{\citenamefont {{de Andrada e Silva}}(1992)}]{current}%
  \BibitemOpen
  \bibfield  {author} {\bibinfo {author} {\bibfnamefont {E.~A.}\ \bibnamefont
  {{de Andrada e Silva}}},\ }\href {\doibase 10.1119/1.17084} {\bibfield
  {journal} {\bibinfo  {journal} {American Journal of Physics}\ }\textbf
  {\bibinfo {volume} {60}},\ \bibinfo {pages} {753} (\bibinfo {year}
  {1992})}\BibitemShut {NoStop}%
\bibitem [{Note1()}]{Note1}%
  \BibitemOpen
  \bibinfo {note} {We previously)~\protect \citep {Hemlata} classified fronts
  as an ordinary front if $v'(q) \not =0$ while the $k$th ($k \geq 1$) order
  extremal front is that for which the first non-zero derivative of the group
  velocity is the $(k+1)^{th}$ derivative ($f^{(n)}(q) =\protect \frac
  {d^nf(q)}{dq^n}$): $v^{(1)}(q) = v^{(2)}(q) = \protect \cdots =
  v^{(k)}(q)=0;\hskip 1em\relax v^{(k+1)}(q) \not =0$ \protect \vspace
  {0pt}}\BibitemShut {NoStop}%
\bibitem [{\citenamefont {Stephan}(2019)}]{SciPost_Phys_6}%
  \BibitemOpen
  \bibfield  {author} {\bibinfo {author} {\bibfnamefont {J.-M.}\ \bibnamefont
  {Stephan}},\ }\href {\doibase 10.21468/SciPostPhys.6.5.057} {\bibfield
  {journal} {\bibinfo  {journal} {SciPost Phys.}\ }\textbf {\bibinfo {volume}
  {6}},\ \bibinfo {pages} {57} (\bibinfo {year} {2019})}\BibitemShut {NoStop}%
\bibitem [{\citenamefont {Haberman}(2013)}]{haberman}%
  \BibitemOpen
  \bibfield  {author} {\bibinfo {author} {\bibfnamefont {R.}~\bibnamefont
  {Haberman}},\ }\href {https://books.google.co.in/books?id=hGNwLgEACAAJ}
  {\emph {\bibinfo {title} {Applied Partial Differential Equations: With
  Fourier Series and Boundary Value Problems}}},\ Featured Titles for Partial
  Differential Equations\ (\bibinfo  {publisher} {PEARSON},\ \bibinfo {year}
  {2013})\BibitemShut {NoStop}%
\bibitem [{\citenamefont {Novo}\ and\ \citenamefont
  {Ribeiro}(2020)}]{novo2020floquet}%
  \BibitemOpen
  \bibfield  {author} {\bibinfo {author} {\bibfnamefont {L.}~\bibnamefont
  {Novo}}\ and\ \bibinfo {author} {\bibfnamefont {S.}~\bibnamefont {Ribeiro}},\
  }\href@noop {} {\  (\bibinfo {year} {2020})},\ \Eprint
  {http://arxiv.org/abs/2012.00448} {arXiv:2012.00448 [quant-ph]} \BibitemShut
  {NoStop}%
\end{thebibliography}%

\end{document}